\documentclass[10pt]{iopart}
\usepackage{iopams}  
\usepackage{graphicx}
\usepackage{color}
\usepackage{cite}
\usepackage{url}
\begin{document}

\review{A review of metasurfaces: physics and applications}

\author{Hou-Tong Chen$^1$, Antoinette J Taylor$^2$ and Nanfang Yu$^3$}

\address{$^1$Center for Integrated Nanotechnologies, Los Alamos National Laboratory, Los Alamos, NM 87545, USA}

\address{$^2$Associate Directorate for Chemistry, Life, and Earth Sciences, Los Alamos National Laboratory, Los Alamos, NM 87545, USA}

\address{$^3$Department of Applied Physics and Applied Mathematics, Columbia University, New York, NY 10027, USA}
\eads{\mailto{chenht@lanl.gov}, \mailto{ttaylor@lanl.gov}, \mailto{ny2214@columbia.edu}}

\vspace{10pt}
\begin{indented}
\item[]\today
\end{indented}

\begin{abstract}
Metamaterials are composed of periodic subwavelength metal/dielectric structures that resonantly couple to the electric and/or magnetic components of the incident electromagnetic fields, exhibiting properties that are not found in nature. This class of micro- and nano-structured artificial media have attracted great interest during the past 15 years and yielded ground-breaking electromagnetic and photonic phenomena. However, the high losses and strong dispersion associated with the resonant responses and the use of metallic structures, as well as the difficulty in fabricating the micro- and nanoscale 3D structures, have hindered practical applications of metamaterials. Planar metamaterials with subwavelength thickness, or metasurfaces, consisting of single-layer or few-layer stacks of planar structures, can be readily fabricated using lithography and nanoprinting methods, and the ultrathin thickness in the wave propagation direction can greatly suppress the undesirable losses. Metasurfaces enable a spatially varying optical response (e.g., scattering amplitude, phase, and polarization), mold optical wavefronts into shapes that can be designed at will, and facilitate the integration of functional materials to accomplish active control and greatly enhanced nonlinear response. This paper reviews recent progress in the physics of metasurfaces operating at wavelengths ranging from microwave to visible. We provide an overview of key metasurface concepts such as anomalous reflection and refraction, and  introduce metasurfaces based on the Pancharatnam-Berry phase and Huygens' metasurfaces, as well as their use in wavefront shaping and beam forming applications, followed by a discussion of polarization conversion in few-layer metasurfaces and their related properties. An overview of dielectric metasurfaces reveals their ability to realize unique functionalities coupled with Mie resonances and their low ohmic losses. We also describe metasurfaces for wave guidance and radiation control, as well as active and nonlinear metasurfaces. Finally, we conclude by providing our opinions of opportunities and challenges in this rapidly developing research field.
\end{abstract}

%
%
%
%
\newpage
\tableofcontents

\ioptwocol

\section{Introduction}

Optical devices control and manipulate light by altering its amplitude, phase, and polarization states in a desired manner, which result in steering the beam propagation direction, shaping the wavefront (e.g., focusing), and imparting information for applications such as sensing, imaging and communication. Conventional optical components are based on refraction, reflection, absorption, and/or diffraction of light, and light manipulation is achieved via propagation through media of given refractive indices, which can be engineered to control the optical path of light beams. In this way, phase, amplitude, and polarization changes are accumulated through propagation in optical components based on refraction and reflection, such as lenses, waveplates, and optical modulators. Ancient people already used ice lenses to focus sunlight and start fires~\cite{Wu_Lu_Gong_Guo_2015_AAPPS}, one example of controlling light propagation. They still prevail in today's optical laboratories and many consumer-based optical products, but are bulky and heavy, unsuitable for the increasingly demanding requirements of integration and miniaturization in modern electromagnetic and photonic systems. The propagation effect is also used in transformation optics~\cite{Pendry_2006_Science,Leonhardt_2006_Science}, which utilizes optical materials structured on a subwavelength scale to produce spatially varying refractive indices that can range from positive to negative.

Metamaterials are composed of periodic subwavelength metal/dielectric structures (i.e., meta-atoms) that resonantly couple to the electric or magnetic or both components of the incident electromagnetic fields, exhibiting effective electric (represented by electric permittivity $\epsilon$) and/or magnetic (represented by magnetic permeability $\mu$) response not found in nature. This class of micro- and nano-structured artificial media have attracted great interest during the past 15 years and yielded ground-breaking electromagnetic and photonic phenomena~\cite{Cai_Shalaev_Book,Engheta_Ziolkowski_Book}. Their electromagnetic properties are mainly determined by the subwavelength structures together with the integrated functional materials, therefore, producing the desirable electromagnetic response and device functionalities by structural engineering. The initial overwhelming interest in metamaterials lies in the realization of simultaneously negative electric and magnetic responses and, thereby, negative refractive index~\cite{Smith_2000_PRL,Shelby_Smith_2011_Science}, which can be used to accomplish superresolution in optical imaging~\cite{Pendry_2000_PRL_Superlens,Fang_Zhang_2005_Science}. The capability of tailoring inhomogeneous and anisotropic refractive index resulted in electromagnetic invisibility~\cite{Schurig_Smith_2006_Science}, another hallmark accomplishment using metamaterials. These promising potential applications are, however, hindered in practice due to the high losses and strong dispersion associated with the resonant responses and the use of metallic structures. 

Another challenge in metamaterials is the difficult micro- and nano-fabrication of the required three-dimensional (3D) structures~\cite{Soukoulis_2011_NatPhoton}, as permittivity, permeability and refractive index are essentially properties of bulk materials. Planar metamaterials, however, can be readily fabricated using existing technologies such as lithography and nanoprinting methods, driving many metamaterial researchers to focus on single-layer or few-layer stacks of planar structures that are more accessible particularly in the optical regime. They are called metasurfaces and can be considered as the two-dimensional (2D) equivalent of bulk metamaterials. Because the subwavelength thickness introduces minimal propagation phase, the effective permittivity, permeability and refractive index are of less interest in metasurfaces. In contrast, of significant importance are the surface or interface reflection and transmission resulting from the tailored surface impedance, including their amplitude, phase, and polarization states. The ultrathin thickness in the wave propagation direction can greatly suppress the undesirable losses by using appropriately chosen materials and metasurface structures. Overall, metasurfaces can overcome the challenges encountered in bulk metamaterials while their interactions with the incident waves can be still sufficiently strong to obtain very useful functionalities. For this reason, we envision that metasurfaces will dominate the general field of metamaterials research given their high potential in applications.  

Metasurfaces diminish our dependence on the propagation effect by introducing abrupt changes in optical properties~\cite{Yu_Capasso_2011_Science,Ni_Shalaev_2012_Science,Sun_Zhou_2012_NatMater}. At microwave and terahertz (THz) frequencies, one can take advantage of subwavelength metallic resonators such as split-ring resonators (SRRs)~\cite{Pendry_1999_IEEE,Chen_2007_OE} and a variety of elements typically used in frequency selective surfaces~\cite{Munk_2000_FSS}. Abrupt and controllable changes of optical properties are achieved by engineering the interaction between light and an array of optical scatterers called ``optical antennas''~\cite{Bharadwaj_Novotny_2009_AdvOptPhoton,Novotny_2011_NatPhoton}, which can take a variety of forms, including metallic or dielectric micro/nano-particles~\cite{Svirko_Osipov_2001_APL,Zou_Fumeaux_2013_OE_Dielectric_Resonator_Nanoantennas}, apertures formed in metallic films~\cite{Walther_2012_AdvMater,Lin_Capasso_2013_Science}, and their multi-layer structures~\cite{Cheng_Tian_2015_AdvMater}. The most critical feature of metasurfaces is that they provide degrees of freedom in designing spatial inhomogeneity over an optically thin interface. Arrays of antennas with subwavelength separation between adjacent elements can have spatially varying structural features or material compositions. Thus, metasurfaces are able to introduce a spatially varying electromagnetic or optical response (i.e., scattering amplitude, phase, and polarization), and mold wavefronts into shapes that can be designed at will. 

As metasurfaces comprise a rapidly growing field of research, there have been a few good review articles focusing on different areas~\cite{Holloway_Smith_2012_IEEE,Yu_Capasso_2013_IEEE,Kildishev_Shalaev_2013_Science,Yu_Capasso_2014_NatMater,Walia_2015_ApplPhysRev,Minovich_Kivsha_2015_LPR,Genevet_Capasso_2015_RepProPhys}.  Here we provide our perspective on this research field by reviewing the progress during the past few years, where metasurfaces are broadly defined as planar metamaterial structures with subwavelength thickness, operating at wavelengths ranging from microwave to visible. The paper is organized as described below. In section~\ref{Anomalous_reflection_and_refraction} we overview the concept and provide demonstrations of anomalous reflection and refraction, which have largely stimulated and reformed worldwide research interest in metasurfaces.
In section~\ref{Phase_Gradient_Beam_Forming} we introduce metasurfaces based on the Pancharatnam-Berry phase and Huygens' metasurfaces, as well as their use in wavefront shaping and beam forming applications. In section~\ref{Polarization_Conversion} we review polarization conversion in few-layer metasurfaces and their related properties. This section is followed by an overview in section~\ref{Dielectric_Metasurfaces} of dielectric metasurfaces that not only reduce the ohmic losses in metallic metasurfaces but also realize some other unique properties and functionalities. In section~\ref{Wave_Guidance_Radiation} we describe metasurfaces for wave guidance and radiation control. We also summarize active and nonlinear metasurfaces in section~\ref{Active_Metasurfaces}, and in the last section we provide concluding remarks and an outlook on future research directions.

\section{Anomalous reflection and refraction} \label{Anomalous_reflection_and_refraction}

\subsection{Generalized laws of reflection and refraction}

When a plane electromagnetic wave encounters a boundary between two homogeneous media with different refractive indices, it is split into a reflected beam that propagates back to the first medium and a transmitted beam that proceeds into the second medium. The reflection and transmission coefficients and their directions are determined by the continuity of field components at the boundary, and are given by the Fresnel equations and Snell's law, respectively. If we add to the interface an array of subwavelength resonators of negligible thickness forming a metasurface, the reflection and transmission coefficients will be then dramatically changed because the boundary conditions are modified by the resonant excitation of an effective current within the metasurface. The reflection and transmission waves carry a phase change that can vary from $-\pi$ to $\pi$, depending on the wavelength of the incident wave relative to the metasurface resonance. When the resonators are anisotropic, the polarization state may be also altered. When the phase change is uniform along the interface, the directions of reflection and refraction are unaltered; in contrast, one of the merits provided by metasurfaces is that we can create spatial phase variation with subwavelength resolution to effectively control the direction of wave propagation and the shape of wavefront.   

\begin{figure}[htp]
\centerline{\includegraphics[width=2.5in]{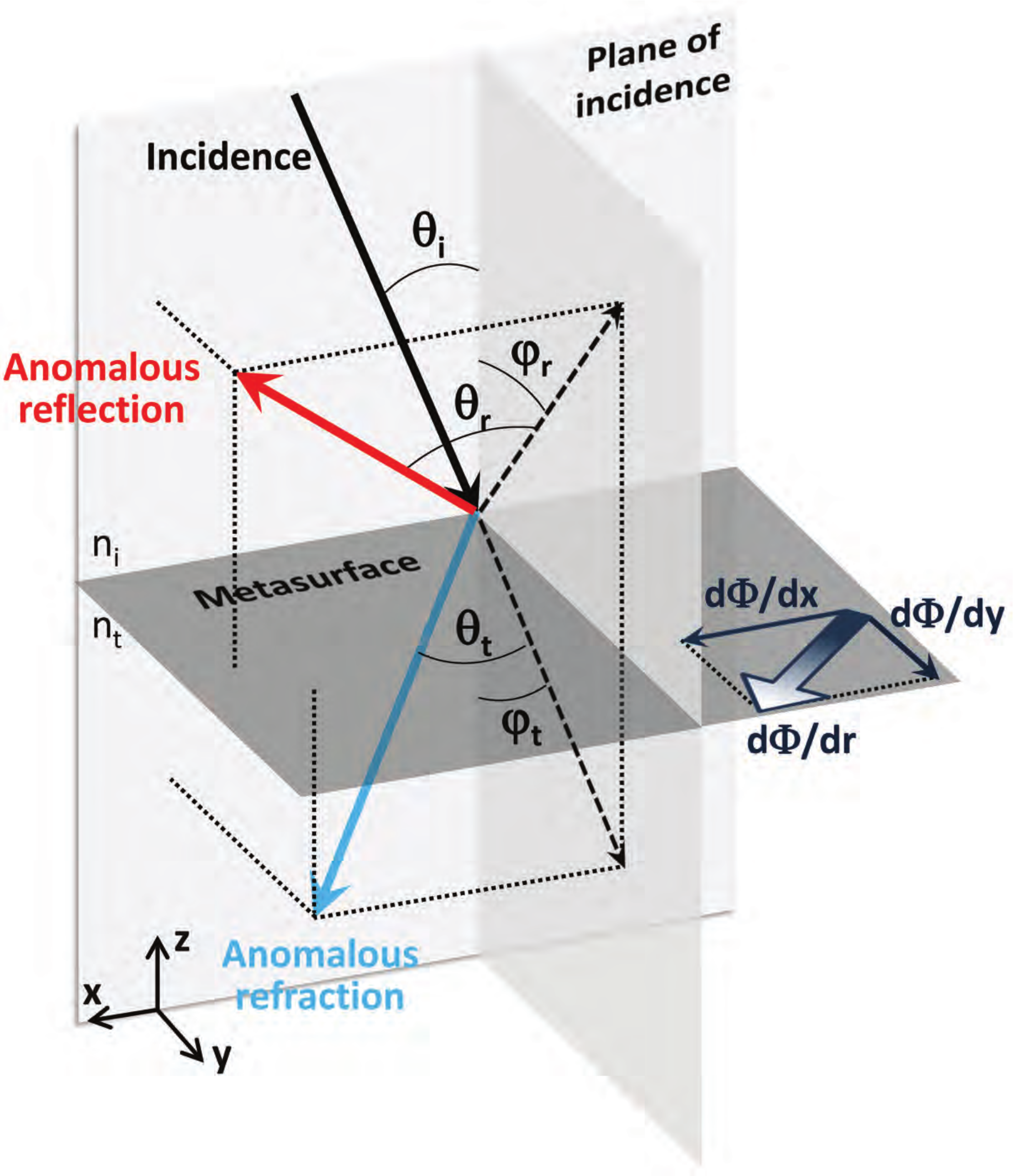}} \caption{A gradient of interfacial phase jump $\rmd \Phi/\rmd r$ provides an effective wavevector along the interface that can bend transmitted and reflected light into arbitrary directions.}
\label{Figure_1}
\end{figure}
We can understand quantitatively the control of wave propagation direction using Fermat's principle, which states that the route for the propagation of a light beam should be stationary in the total accumulated phase with respect to small variations of the route. Now we consider a specific case where a metasurface introduces a spatial distribution of phase jumps due to electromagnetic scattering at its constitutive antennas. The actual optical path in the presence of these phase jumps should be stationary in the total accumulated optical phase. This law of stationary phase ensures that wavelets starting from a source point with the same initial phase will arrive at the point of destination in phase after reflecting from or transmitting through the metasurface, and thus constructively interfere, which makes the route a physical path of optical power. A set of generalized laws of refraction and reflection can be derived from Fermat's principle of stationary phase~\cite{Yu_Capasso_2011_Science,Aieta_Capasso_2012_NL_Out_of_Plane,Ni_Shalaev_2012_Science}:
\begin{eqnarray}
\left\{ \begin{array}{l l}
n_\mathrm{t} \sin(\theta_\mathrm{t}) - n_\mathrm{i} \sin(\theta_\mathrm{i}) = \frac{1}{k_0} \frac{\rmd \Phi}{\rmd x}\\
\cos(\theta_\mathrm{t}) \sin(\varphi_\mathrm{t}) = \frac{1}{n_\mathrm{t} k_0} \frac{\rmd \Phi}{\rmd y}
\end{array}\right.
\end{eqnarray}
\begin{eqnarray}
\left\{ \begin{array}{l l}
\sin(\theta_\mathrm{r}) - \sin(\theta_\mathrm{i}) = \frac{1}{n_\mathrm{i} k_0} \frac{\rmd \Phi}{\rmd x}\\
\cos(\theta_\mathrm{r}) \sin(\varphi_\mathrm{r}) = \frac{1}{n_\mathrm{r} k_0} \frac{\rmd \Phi}{\rmd y}
\end{array}\right.
\end{eqnarray}
where the definition of angles is shown in \fref{Figure_1}, and $\rmd \Phi/\rmd x$ and $\rmd \Phi/\rmd y$ are, respectively, the components of the phase gradient parallel and perpendicular to the plane of incidence. Looking at the problem from an alternative point of view, the interfacial phase gradient functions as an effective wavevector along the interface, and is imparted to the transmitted and reflected waves. The above generalized laws can thus be derived by considering the conservation of wavevector along the interface. These generalized laws indicate that the transmitted and reflected light beams can be bent in arbitrary directions in their respective half space, depending on the direction and magnitude of the interfacial phase gradient, as well as the refractive indices of the surrounding media.

\subsection{Demonstration of generalized optical laws}

To experimentally demonstrate the generalized laws, one has to devise miniature scatterers that are able to controllably change the phase of the scattered waves and to place such scatterers into an array, forming an artificial interface. The scattering amplitudes should be the same for all scatterers and the spacing between neighboring scatterers in the array should be much less than the wavelength. These conditions ensure that the superposition of spherical waves emanating from the scatterers gives rise to refracted and reflected waves with planar wavefronts, following Huygens' principle. 

One approach to design the phase response of scatterers is to use antenna dispersion. That is, for a fixed electromagnetic wavelength and a variation of antenna geometries, or for a fixed antenna geometry and a variation of excitation wavelengths, there is an associated phase variation of the waves scattered from the antenna (there are also associated amplitude and polarization changes that can be utilized or otherwise managed for designing metasurfaces). For example, when a beam of light impinges on a metallic optical antenna, the optical energy is coupled into surface electromagnetic waves propagating back and forth along the antenna surface. These are accompanied by charge oscillations inside the antenna. These coupled surface electromagnetic waves and oscillating charges are known as surface plasmons. For a fixed excitation wavelength, the antenna resonance occurs when the antenna length $L_\textrm{res} \approx \lambda/2$, where $\lambda$ is the surface plasmon wavelength; under this condition the incident electromagnetic wave is in phase with the excited antenna current. When the antenna length is smaller or larger than $L_\textrm{res}$, the current leads or lags the driving field. Therefore, the phase of the antenna current and the phase of electromagnetic waves created by the oscillating current (i.e., scattered waves from the antenna) can be controlled by choosing the appropriate antenna length. The tuning range of phase is up to $\pi$ if a single antenna resonance is involved. Multiple independent resonances, coupled antenna resonances, or geometric effects (see discussion of the Pancharatnam-Berry phase in section~\ref{Section_PBPhase}) are able to extend the phase response to cover the entire $2\pi$ range, which is necessary for complete control of the wavefront. In addition to metallic antennas, dielectric ones are also able to introduce phase variations to the scattered light associated with Mie resonances (i.e., establishment of standing wave patterns in the dielectric antennas, see the discussion of dielectric metasurfaces in section~\ref{Dielectric_Metasurfaces}).

\begin{figure*}[htp]
\centerline{\includegraphics[width=5in]{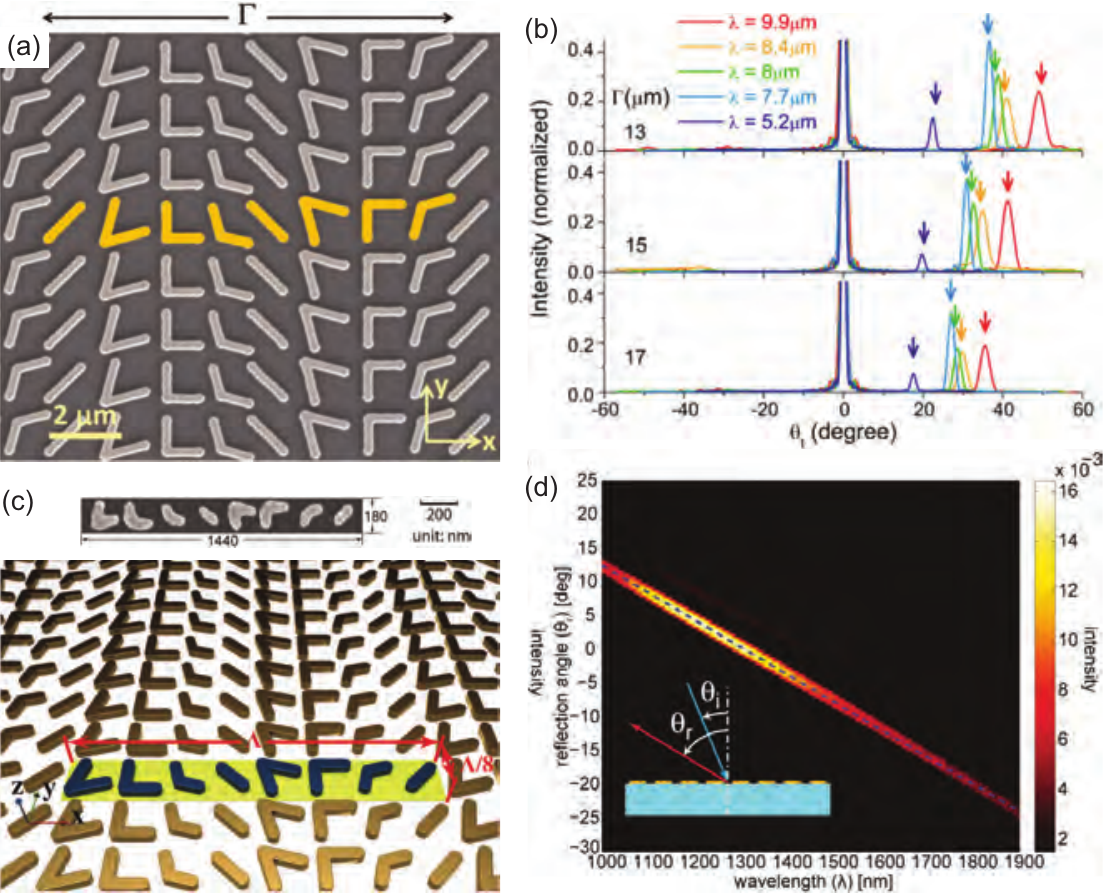}} \caption{(a) SEM image of a mid-infrared metasurface consisting of an array of V-shaped gold optical antennas patterned on a silicon wafer, with the unit cell highlighted and $\Gamma = 11$~$\mu$m. It creates a constant gradient of phase jump along the metasurface for the control of the propagation direction of light transmitted through or reflected from the metasurface. (b) Under normal incidence, measured far-field intensity profiles show the ordinary (co-polarized) and anomalous (cross-polarized) refraction generated by metasurfaces like the one shown in (a) and with different interfacial phase gradients (from $2\pi/13$-$\mu$m to $2\pi/17$-$\mu$m). The far-field profiles are normalized with respect to the intensity of the ordinary beams located at $\theta_\mathrm{t} = 0^\circ$. The arrows indicate the calculated angular positions of the anomalous refraction according to $\theta_\mathrm{t} = -\arcsin(\lambda/\Gamma)$. (c) A metasurface used to demonstrate generalized laws of reflection and refraction in the near-infrared. Upper panel depicts one unit cell of the fabricated structure and lower panel reveals a schematic of the metasurface. (d) Measured far-field intensity profiles of the metasurface in (c) showing reflection angle $\theta_r$ versus wavelength for cross-polarized light with $65^\circ$ incidence angle. (a) used with permission from~\cite{Yu_Capasso_2011_Science}, (b) used with permission from~\cite{Yu_Capasso_2012_NL}, (c) and (d) reproduced with permission from~\cite{Ni_Shalaev_2012_Science}.
}
\label{Figure_2}
\end{figure*}

Generalized optical laws were first demonstrated using V-shaped optical antennas in the mid-infrared spectral range~\cite{Yu_Capasso_2011_Science} and later confirmed in the near-infrared~\cite{Ni_Shalaev_2012_Science} (see \fref{Figure_2}). Such optically anisotropic antennas support two plasmonic eigenmodes with different resonant properties. The geometry and orientation of antennas in the array are properly chosen so that for an incident wave at around 8~$\mu$m wavelength, over a wide range of incident angles, and polarized along the \textit{x}-axis (see \fref{Figure_2}(a)), the component of the scattered wave polarized along the \textit{y}-axis has an incremental phase of $\pi/4$ between adjacent V-antennas in the unit cell of the metasurface. The amplitude of the component polarized along the \textit{y}-axis is also tuned to be uniform across the antenna array. The antenna spacing is between one tenth and one fifth of the free space wavelength. The metasurface creates anomalously refracted and reflected beams satisfying the generalized laws over a wide wavelength range, with negligible spurious beams and optical background, as shown in \fref{Figure_2}(b) and (d). The broadband performance is due to the fact that the two eigenmodes supported by the V-antennas form a broad effective resonance over which the scattering efficiency is nearly constant and the phase response is approximately linear~\cite{Yu_Capasso_2012_NL,Kats_Capasso_2012_PNAS}. The scalability of metasurfaces allows the extension of this concept to other frequency ranges, e.g., broadband anomalous refraction was also demonstrated at THz frequencies using C-shaped metallic resonators~\cite{Zhang_Zhang_2013_AdvMater}

\begin{figure*}[htp]
\centerline{\includegraphics[width=5in]{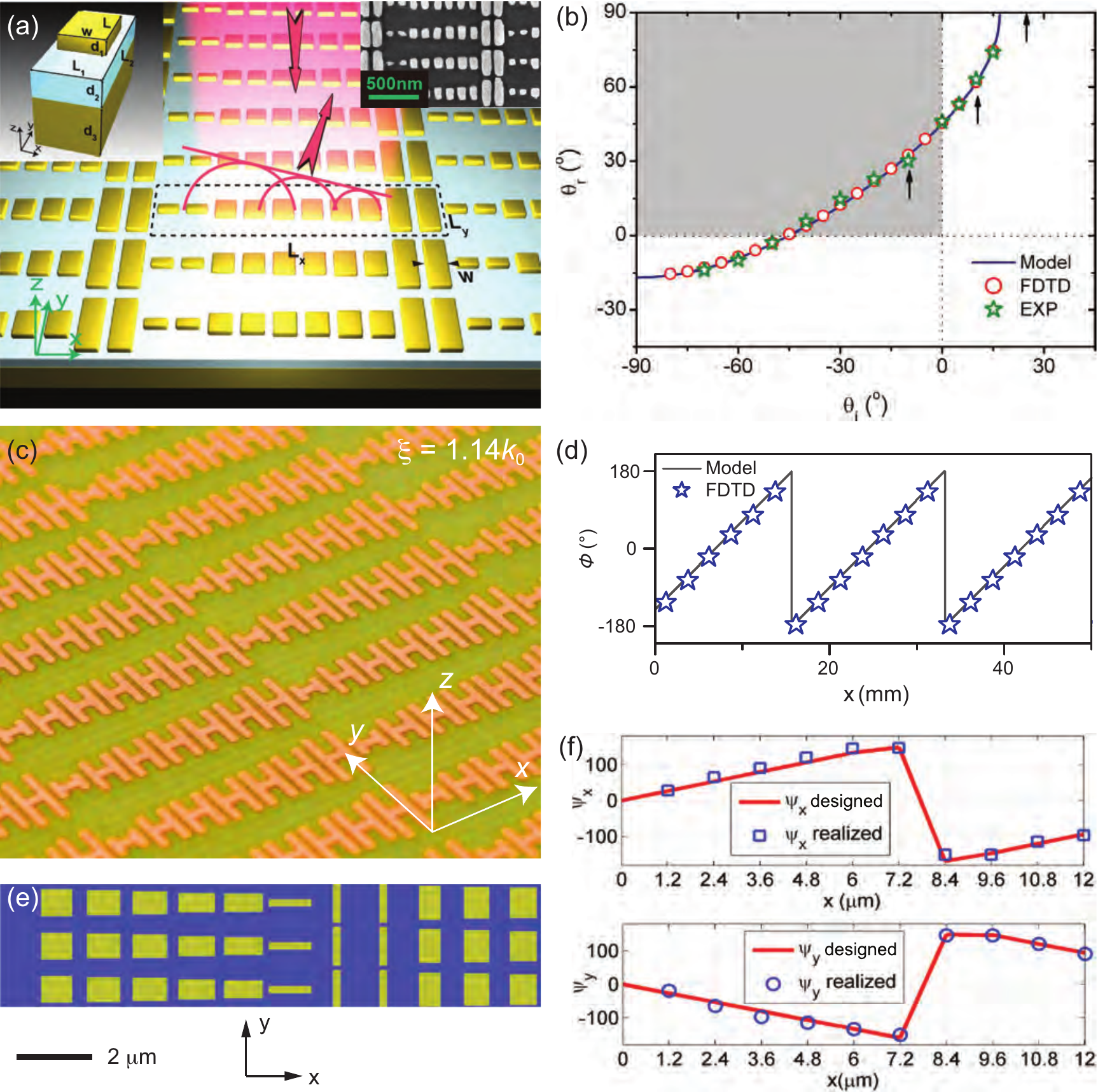}} \caption{(a) Schematic of a near-infrared reflect-array metasurface consisting of gold patch antennas separated from a gold back plane by a MgF$_2$ spacer with subwavelength thickness. Left inset shows a basic building block, and right inset is an SEM image of part of the metasurface. (b) Anomalous reflections at different incident angles for the metasurface shown in (a). The shaded quadrant indicates ``negative'' reflection. (c) Photograph of a fabricated microwave reflect-array consisting of H-antennas separated from a metallic back plane by a dielectric spacer. The reflect-array introduces an interfacial phase gradient $\xi = 1.14 k_0$, where $k_0$ is the wavevector of the incident beam corresponding to a wavelength of 2~cm. (d) Scattering phase profile from the metasurface in (c) showing the phase gradient along the \textit{x}-direction. (e) Schematic of part of a birefringent reflect-array metasurface working at $\lambda = 8.06$~$\mu$m. (f) Depending on the polarization of the incident light, the phase gradient is either positive or negative along the \textit{x}-direction for the metasurface in (e), leading to polarization-dependent anomalous reflection. (a) and (b) reproduced with permission from~\cite{Sun_Tsai_2012_NL}, (c) and (d) used with permission from~\cite{Sun_Zhou_2012_NatMater}, (e) and (f) used with permission from~\cite{Farmahini_2013_OL}.
}
\label{Figure_3}
\end{figure*}
The generalized law of reflection has also been demonstrated using reflect-arrays~\cite{Sun_Tsai_2012_NL,Sun_Zhou_2012_NatMater,Farmahini_2013_OL}, which consist of metallic antennas separated from a back metallic plane by a thin layer of dielectric material (see \fref{Figure_3}). Such reflect-array metasurfaces are inspired by initial work on microwave and millimeter wave reflect-array antennas~\cite{Pozar_1997_IEEE,Gagnon_2013_IEEE}. \Fref{Figure_3}(a) shows a near-infrared reflect-array metasurface based on patch antennas and \fref{Figure_3}(c) shows one that is based on H-shaped antennas for microwaves. The essence of reflect-arrays is to use antennas coupled with their dipolar images in the back mirror to achieve a phase coverage of $2\pi$. Ideally, all incident power will be coupled into anomalous reflection, which will have the same polarization as that of the incident light; the transmission and specular reflection will be absent. Experimentally demonstrated efficiency in generating anomalous reflection in reflect-array metasurface is as high as 80\%, significantly higher than the initial proof-of-principle demonstrations in \fref{Figure_2}, which are based on a single antenna layer, rely on polarization rotation to achieve the $2\pi$ phase coverage, and have an efficiency of 10-20\%. 

\Fref{Figure_3}(b) shows three regimes of operation for a reflect-array metasurface: negative angle of reflection ($\theta_\mathrm{r}$ and $\theta_\mathrm{i}$ of different signs), angle of incidence and angle of reflection of the same sign but not equal to each other, and coupling of incident light into evanescent waves propagating on the metasurface ($\theta_\mathrm{r}$ beyond $90^\circ$). In the last case, the interaction between the incident light and the metasurface leads to a lateral wavevector that is larger than the free space wavevector; as a result, no reflection exists and the incident optical power can only be coupled into surface waves. The work shown in \fref{Figure_3}(c) and (d)  confirms the existence of such surface waves by experimentally measuring their near-field characteristics. A number of variations of the reflect-array metasurface have been also demonstrated. For example, birefringent reflect-array metasurfaces that steer incident light into different directions according to its polarization state have been demonstrated in simulations~\cite{Farmahini_2013_OL} (see \fref{Figure_3}(e) and (f)).

\section{Arbitrary phase gradient and beam forming} \label{Phase_Gradient_Beam_Forming}

\subsection{Metasurfaces based on Pancharatnam-Berry phase} \label{Section_PBPhase}

\begin{figure*}[htp]
\centerline{\includegraphics[width=6in]{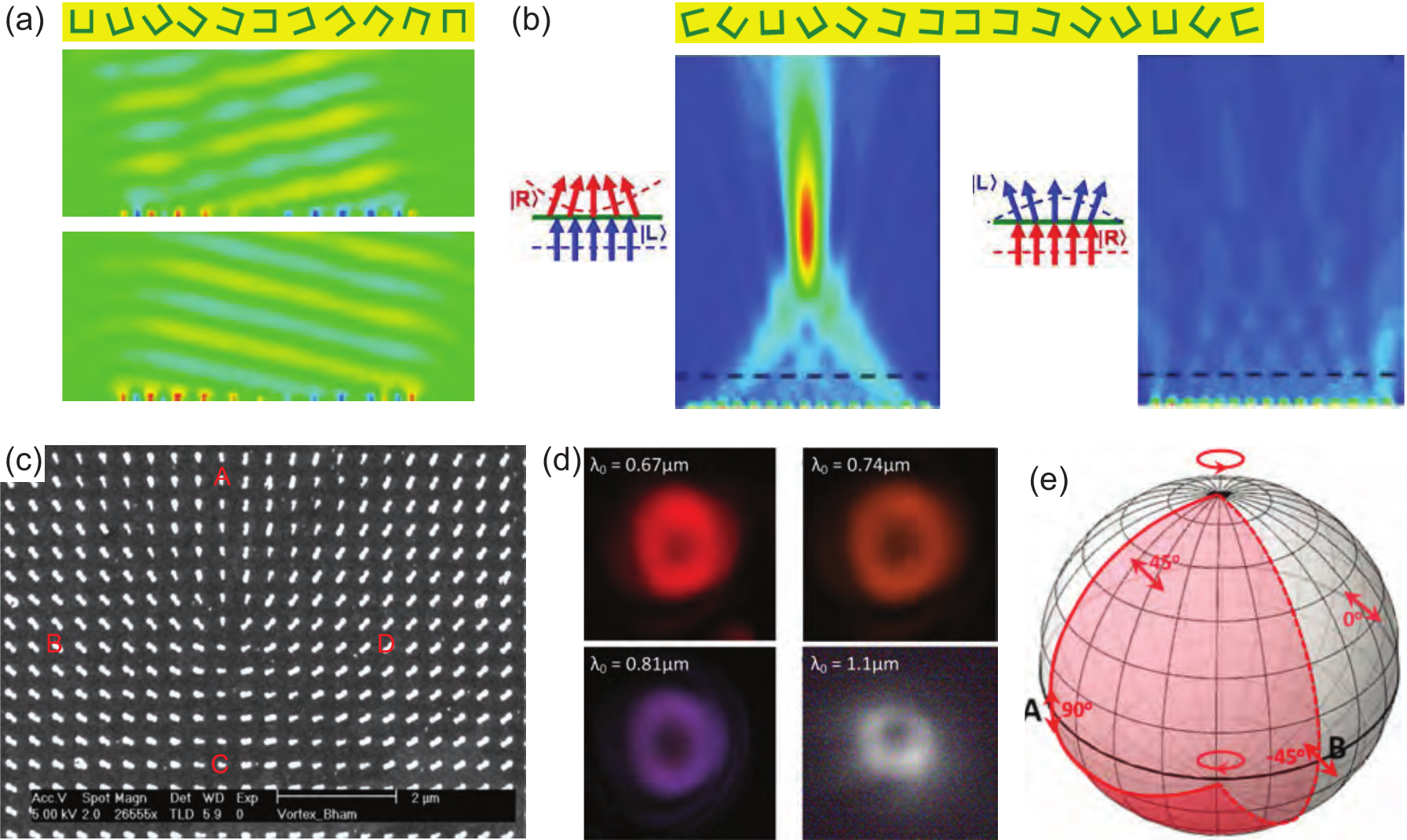}} \caption{(a) Upper panel: schematic of the super-unit-cell of a metasurface consisting of an array of identical U-shaped apertures with gradually increasing rotation angles. Lower panels: simulation rusults showing that the metasurface in (a) bends a circularly polarized incident beam under normal incidence into left or right direction according to the handedness of the incident beam. (b) Upper panel: super-unit-cell of a planar cylindrical lens consisting of an array of identical U-apertures with different orientations. Lower panels: schematics and simulation results showing that the lens focuses right-handed circularly polarized transmission component when the incident light is left-handed circularly polarized, and that the same lens defocuses left-handed circularly polarized transmission component when the incident light is right-handed circularly polarized. (c) SEM image of a metasurface consisting of an array of gold rod antennas with identical geometry but spatially varying orientations, which is designed for generating an optical vortex beam with $L = 1$ (incidence: right-handed circularly polarized; detection: left-handed circularly polarized). (d) Measured intensity distribution of vortex beams generated by the metasurface in (c) at different wavelengths from 670 to 1100~nm. (e) Pioncar\'{e} sphere used to derive the phase difference between scattered waves of left-handed circular polarization from rod antennas located at points A and B in (c), with right-handed circularly polarized incident light. (a) and (b) reproduced with permission from~\cite{Kang_Li_2012_OE}, (c) and (d) reproduced with permission from~\cite{Huang_Zhang_2012_NL}.
}
\label{Figure_4}
\end{figure*}

In the previous examples, variations in phase or amplitude response are introduced by the dispersion of antenna resonance. A completely different approach to introducing phase jumps is to use the so-called Pancharatnam-Berry phase~\cite{Pancharatnam_1956_PIASSA,Berry_1984_PRSLA}. The latter is associated with polarization change and can be created by using anisotropic, subwavelength scatterers with identical geometric parameters but spatially varying orientations. The recent development of metasurfaces based on Pancharatnam-Berry phase has been largely following the innovative early works by Hasman and co-workers~\cite{Hasman_2005_Progress_in_Optics}, who used continuous or discrete subwavelength gratings to control the polarization states for the generation of vector beams and manipulation of wavefronts. The easiest way to reveal the relation between polarization and phase is to use Jones calculus~\cite{Menzel_2010_PRA,Armitage_2014_PRB}. In general, the Jones matrix of an anisotropic scatterer can be written as
\begin{eqnarray}
\hat{M} = \hat{R}(-\alpha)
\left( \begin{array}{c c}
t_\mathrm{o} & 0\\
0 & t_\mathrm{e}
 \end{array}
 \right) \hat{R}(\alpha),
\end{eqnarray}
where $t_\mathrm{o}$ and $t_\mathrm{e}$ are, respectively, the coefficients of forward scattering for incident light linearly polarized along the two principal axes of the anisotropic scatterer,
\begin{eqnarray}
\hat{R}(\alpha) = \left( \begin{array}{c c}
\cos(\alpha) & \sin(\alpha)\\
-\sin(\alpha) & \cos(\alpha)
 \end{array}
 \right)
\end{eqnarray}
is the rotation matrix and $\alpha$ is the rotation angle. Given an incident wave of right/left circular polarization $\mathbf{E}^\mathrm{R/L}_\mathrm{I}$, the scattered light from the anisotropic scatterer in the forward direction $\mathbf{E}^\mathrm{R/L}_\mathrm{T}$ can then be written as~\cite{Kang_Li_2012_OE}: 
\begin{eqnarray}
\mathbf{E}^\mathrm{R/L}_\mathrm{T} &= \hat{M} \cdot \mathbf{E}^\mathrm{R/L}_\mathrm{I} \nonumber\\
&= \frac{t_\mathrm{o} + t_\mathrm{e}}{2} \mathbf{E}^\mathrm{R/L}_\mathrm{I}  + \frac{t_\mathrm{o} - t_\mathrm{e}}{2} \exp(i m 2\alpha) \mathbf{E}^\mathrm{L/R}_\mathrm{I}.
\end{eqnarray}
The first term represents circularly polarized scattered waves with the same handedness as the incident light, and the second term represents circularly polarized scattered waves with opposite handedness and an additional Pancharatnam-Berry phase of $m2\alpha$, where $m$ is `$-$' for right-handed and `$+$' for left-handed circularly polarized incident light. The second component can be selected in experiments by using a quarter-wave plate and a polarizer. Its phase can cover the entire 2$\pi$ range if the anisotropic scatterer is rotated from 0 to $180^\circ$. Based on this principle, a phase-gradient metasurface has been demonstrated to steer light into different directions depending on the handedness of the incident circular polarization (see \fref{Figure_4}(a))~\cite{Kang_Li_2012_OE}. The unit cell of the metasurface consists of U-shaped aperture antennas with an incremental angle of rotation between adjacent elements, with the total rotation angle being $180^\circ$ within the unit cell. Similar U-shaped aperture antennas have been used to create a planar lens, which either functions as a focusing or diverging lens depending on the handedness of the incident circular polarization~\cite{Kang_Li_2012_OE}, as shown in \fref{Figure_4}(b). A broadband phase plate generating optical vortex beams has been demonstrated by using an array of rod antennas with different orientations (\fref{Figure_4}(c) and (d))~\cite{Huang_Zhang_2012_NL}. A bi-layer metallic aperture metasurface was also demonstrated to accomplish the simultaneous manipulation of polarization and phase of the transmitted light~\cite{Li_Tian_2015_AFM}.  

The metasurfaces based on the Pancharatnam-Berry phase work for circularly polarized incident light and control the component of the circularly polarized transmission with the opposite handedness. A major advantage of the approach based on the Pancharatnam-Berry phase is ultra-broadband performance: given a certain antenna geometry, the magnitude of the phase jump is only a function of the orientation angle of the antenna and the sign of the phase jump is determined by the handedness of the incident circularly polarized light; there is no wavefront distortion resulting from antenna dispersion. The operating bandwidth is limited on the long-wavelength side by reduced scattering efficiency and on the short-wavelength side by the requirement that the wavelength has to be at least several times larger than the spacing between scatterers (i.e., metasurface regime). In early demonstrations of broadband metasurfaces based on Pancharatnam-Berry phase, the presence of scattered waves that do not carry Pancharatnam-Berry phase inevitably decreases their efficiency. In a new generation of metasurfaces, Luo \textit{et al.} were able to suppress these scattered components and created metasurfaces based on Pancharatnam-Berry phase with efficiency approaching unity~\cite{Luo_Zhou_2015_AOM}. They demonstrated two different metasurfaces that separate a linearly polarized incident microwave into a left-handed circularly polarized beam and a right-handed circularly polarized beam over a frequency range of 11--14 GHz, within which the linearly polarized background is very weak. The metasurface design is based on rigorous Jones matrix analyses that provide a set of criteria for achieving 100\% efficiency \cite{Luo_Zhou_2015_AOM}.

A completely different perspective to understand the operation of metasurfaces based on the Pancharatnam-Berry phase results from tracing the evolution of polarization in the Poincar\'{e} sphere~\cite{Pancharatnam_1956_PIASSA,Bomzon_Hasman_2001_OL,Bomzon_Hasman_2002_OL,Li_Li_2013_NL}. The phase difference between the scattered waves from any two points on the metasurface is equal to the solid angle enclosed by their corresponding traces in the Poincar\'{e} sphere divided by two~\cite{Bomzon_Hasman_2001_OL}. For example, the solid red trace in \fref{Figure_4}(e) corresponds to light passing through point A in \fref{Figure_4}(c): The trace starts at north pole of the Poincar\'{e} sphere representing right-handed circularly polarized incident light; the trace passes a point on the equator that represents linear polarization in the vertical direction because the antenna at point A on the metasurface preferentially scatters vertically polarized waves; the trace ends at the south pole because left-handed circularly polarized transmission is selectively monitored. Similarly, the dashed red trace in \fref{Figure_4}(e) corresponds to light passing through point B on the metasurface shown in \fref{Figure_4}(c). The solid angle enclosed by the two traces is $\pi$; therefore the phase difference between left-handed circularly polarized light scattering from points A and B on the metasurface is $\pi/2$. Similar analyses show that the phase difference between points A and C is $\pi$ and between A and D is $3\pi/2$. Therefore, the metasurface in \fref{Figure_4}(c) introduces a constant phase gradient in the azimuthal direction and the phase variation is $2\pi$ during one circle around the central point of the metasurface. The metasurface thus imprints a spiral phase distribution to the transmitted wavefront, creating a vortex beam with orbital angular momentum of $L = 1$.

\subsection{Huygens' surfaces}
\begin{figure*}[htp]
\centerline{\includegraphics[width=6in]{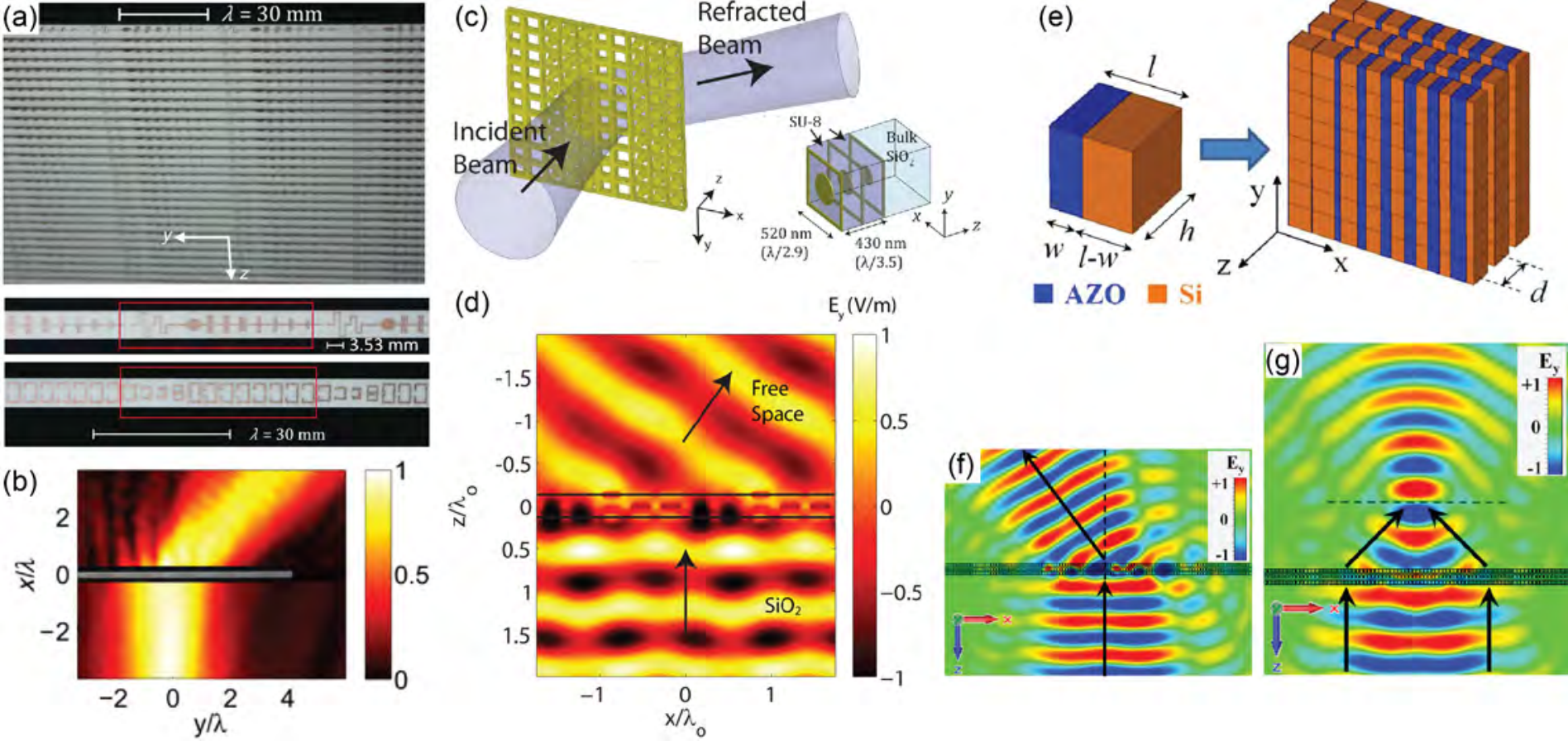}} \caption{(a) Upper panel: photograph of a fabricated microwave metasurface that can redirect an incident beam with nearly 100\% efficiency into a refracted beam. It is made of a stack of identical circuit board stripes, the top and bottom sides of which are printed with copper traces. Bottom panel: one unit cell of the metasurface consists of capacitively and inductively loaded traces to realize desired electric sheet reactance (on the top side of each stripe) and capacitively loaded loops to realize desired magnetic sheet reactance (on the bottom side of each stripe). (b) Measured magnetic field distribution of the beam-deflecting metasurface in (a). (c) Schematic of an optically thin, isotropic Huygens' metasurface that efficiently refracts a normally incident beam at telecommunication wavelengths. Inset: schematic of a unit cell. (d) Simulated electric field distribution of a beam deflector based on the metasurface in (c).(e) Left panel: basic building block of a metasurface made of plasmonic (AZO: aluminum-doped zinc oxide) and dielectric (silicon) materials, with $l = 250$~nm and $h = 250$~nm. Right panel: metatransmit-array made of three stacked metasurfaces with center-center distance of $d = \lambda_0/8 = 375$~nm. (f) and (g) Simulated electric field distributions of a beam deflector and a flat lens based on the metatransmit-array shown in (e). (a) and (b) reproduced with permission from~\cite{Pfeiffer_Grbic_2013_PRL}, (c) and (d) reproduced with permission from~\cite{Pfeiffer_Grbic_2014_NL}, (e)--(g) used with permission from~\cite{Monticone_Alu_2013_PRL}.
}
\label{Figure_5}
\end{figure*}
To boost the efficiency of a metasurface in controlling the transmitted light, one has to match its impedance with that of free space. Complete elimination of reflection can be realized by controlling the surface electric and magnetic polarizabilities, $\alpha_\mathrm{e}$ and $\alpha_\mathrm{m}$, of the metasurfaces so that~\cite{Pfeiffer_Grbic_2013_PRL}
\begin{eqnarray}
\sqrt{\alpha_\mathrm{m}/\alpha_\mathrm{e}} = \eta_0,
\label{Eq_Impedance}
\end{eqnarray}
where $\eta_0$ is the impedance of the surrounding media. The effective electric and magnetic surface currents, which are proportional to $\alpha_\mathrm{e}$ and $\alpha_\mathrm{m}$, respectively, change the boundary conditions at the metasurface and lead to the new scattered wavefronts. The complex transmission coefficient of the metasurface is~\cite{Pfeiffer_Grbic_2013_PRL}
\begin{eqnarray}
T = \frac{2 - j \omega \alpha_\mathrm{e} \eta_0}{2 + j \omega \alpha_\mathrm{e} \eta_0}.
\label{Eq_Transmission}
\end{eqnarray}

If $\alpha_\mathrm{e}$ is predominantly real, one can vary $\alpha_\mathrm{e}$ and $\alpha_\mathrm{m}$ simultaneously at each point on the metasurface to ensure that the waves transmitted through the metasurface acquire a phase jump anywhere from $-\pi$ to $+\pi$ according to (\ref{Eq_Transmission}) and that the transmission efficiency is close to unity by satisfying (\ref{Eq_Impedance}) everywhere on the metasurface. The above design concept has been implemented in the microwave spectral region by using spatially varying copper traces supporting both electric and magnetic polarization currents (\fref{Figure_5}(a))~\cite{Pfeiffer_Grbic_2013_PRL}. A transmission efficiency of 86\% was experimentally demonstrated in a beam deflector shown in \fref{Figure_5}(b). Although the demonstrations are in the microwave regime, the concepts can be adapted to the optical regime and one example is shown in \fref{Figure_5}(c) and (d)~\cite{Pfeiffer_Grbic_2014_NL}. Another metasurface that is impedance matched to free space and able to fully control the phase of the transmitted light was proposed in a recent paper~\cite{Monticone_Alu_2013_PRL}. It is designed based on optical nano-circuit concepts and is comprised of three planarized arrays stacked together, as shown in \fref{Figure_5}(e), where the building blocks of the array are subwavelength components made of metallic and dielectric materials with different filling ratios and function as LC nano-circuit elements. A beam deflector and a flat lens with high transmission efficiency were demonstrated in simulations, as shown in \fref{Figure_5}(f) and (g), by engineering the effective surface impedance of the metasurface via tuning of the filling ratios.

\subsection{Wavefront shaping and beam forming}

\begin{figure}[htp]
\centerline{\includegraphics[width=2.5in]{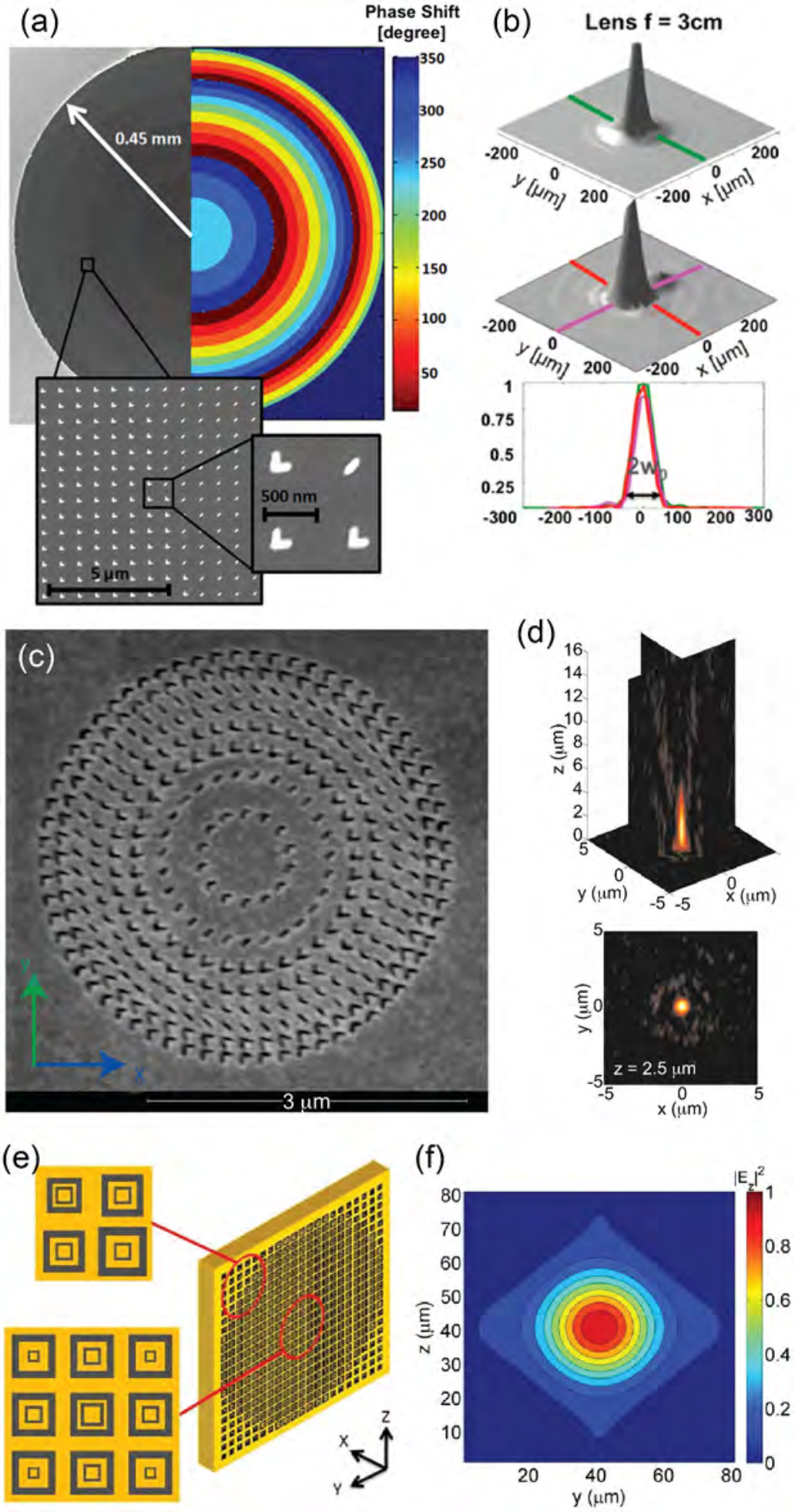}} \caption{(a) Left panel: SEM image of a fabricated metasurface lens with 3 cm focal length, consisting of an array of V-antennas. Right panel: phase profile of the lens discretized according to the phase responses of eight constituent antennas. Insets: zoom-in view of fabricated antennas. (b) 3D plots of the simulated (top panel) and measured (middle panel) and 1D plots (cross-sectional planes along the lines) of intensity distribution of the lens in (a) on the focal plane. (c) SEM image of a planar plasmonic metalens consisting of V-shaped apertures and with a focal length $f = 2.5$~$\mu$m at an operational wavelength of 676~nm. (d) Intensity distributions for two cross-sectional planes (top panel) cutting through the center of the metalens in (c), and on the focal plane of the metalens (bottom panel). (e) Schematic of a metasurface lens consisting of an array of $21 \times 21$ scatterers each made of two silver concentric loops (gray). The scatterers are placed on both sides of the substrate (yellow). (f) Simulated intensity distribution on the focal plane of the metasurface lens in (e). (a) and (b) used with permission from~\cite{Aieta_Capasso_2012_NL_Flat_Lens}, (c) and (d) used with permission from~\cite{Ni_Shalaev_2013_LSA}, and (e) and (f) used with permission from~\cite{Memarzadeh_2011_OL}. 
}
\label{Figure_6}
\end{figure}

Metasurfaces provide us with an unprecedented opportunity to design the wavefronts of light at will, as we have seen from the descriptions of anomalous reflection/refraction and beam focusing in the previous sections. \Fref{Figure_6} further shows a few planar devices based on metasurfaces. To realize flat lenses, a metasurface should impose a phase profile 
\begin{eqnarray}
\varphi_\mathrm{L}(x,y) = \frac{2\pi}{\lambda}\left(\sqrt{x^2 + y^2 + f^2} - f \right)
\label{Eq_Lens}
\end{eqnarray}
to convert incident planar wavefronts into spherical ones, which converge at a distance $f$ from the lenses. The optical wavefronts in transmission or reflection remain spherical as long as the incident plane wave impinges normal to the flat lenses. It is therefore straightforward to achieve high numerical-aperture (NA) focusing without spherical aberration. Flat lenses at telecom wavelengths have been experimentally demonstrated using V-shaped antennas (see \fref{Figure_6}(a) and (b))~\cite{Aieta_Capasso_2012_NL_Flat_Lens}. The efficiency of these flat lenses is, however, rather small (i.e., ~1\% of the incident optical power is focused) because of the use of only a single scatterer layer, the small surface filling factor, and focusing only the component of the scattered light that is cross-polarized with respect to the incident polarization. At THz and microwave frequencies, high-performance planar components can benefit from few-layer metasurfaces, which have enabled highly efficient and ultra-broadband polarization conversion and anomalous refraction~\cite{Grady_Chen_2013_Science}, and highly efficient reflect-array metasurface lenses~\cite{Li_Zhou_2012_OL}. V-shaped apertures allow similar control of scattering polarization, amplitude and phase as in their complementary V-antennas according to Babinet's principle; they have been used to demonstrate flat lenses to focus visible light (\fref{Figure_6}(c) and (d))~\cite{Ni_Shalaev_2013_LSA} and THz waves~\cite{Jiang_Zhang_2013_OE}, with one of the advantages being significant suppression of the background light. 

Based on the Pancharatnam-Berry phase, U-shaped and other nano aperture antennas have been used to create a planar lens, which either functions as a focusing or diverging lens depending on the handedness of the incident circular polarization (see \fref{Figure_4}(b))~\cite{Kang_Li_2012_OE,Chen_Zentgraf_2012_NatCommun}. A flat lens design at telecom wavelengths with potentially high efficiency has been demonstrated in simulations (\fref{Figure_6}(e) and (f))~\cite{Memarzadeh_2011_OL}. The design uses concentric loop antennas placed on both sides of a substrate to enhance scattering efficiency and to increase the range of phase coverage. Flat lenses working in the near-infrared with high efficiency have been demonstrated experimentally by using reflect-arrays of patch antennas~\cite{Pors_2013_NL_Flat_Mirrors}. Note that except for spherical aberration, monochromatic aberrations are still present in the above demonstrated flat lenses. For example, when incident light is not perpendicular to the lenses, the transmitted or reflected wavefront is no longer spherical because its phase distribution is that of (\ref{Eq_Lens}) plus a linear phase distribution introduced by the non-normal incidence angle. Flat lenses also have chromatic aberration, although one can design antennas with multiple resonances to eliminate it by engineering their dispersion~\cite{Khorasaninejad_Capasso_2015_NL} (see discussions in section~\ref{Dielectric_Metasurfaces}). 

It is of particular interest to focus Gaussian beams to non-diffracting and long focal depth Bessel beams that are traditionally generated using axicons and have been widely used in microscopy imaging. A metasurface approach to generate a Bessel beam (axicons)~\cite{Aieta_Capasso_2012_NL_Flat_Lens,Pfeiffer_Grbic_2014_PhysRevAppl} has been demonstrated by creating a linear phase gradient along the radial direction of the metasurface. An arbitrary spatially varying phase profile can be created in the azimuthal direction. A V-shaped antenna based metasurface has been used to create a vortex beam (i.e., Laguerre-Gaussian modes) from a Gaussian beam~\cite{Yu_Capasso_2011_Science,Genevet_2012_APL}, resulting in optical singularity at the beam center and a helicoidal equal-phase wavefront carrying orbital angular momentum. Using a phase profile based on the Pancharatnam-Berry phase, a broadband phase plate generating optical vortex beams has been demonstrated using an array of rod antennas with different orientations (see \fref{Figure_4}(c) and (d))~\cite{Huang_Zhang_2012_NL}. 

\begin{figure*}[htp]
\centerline{\includegraphics[width=6in]{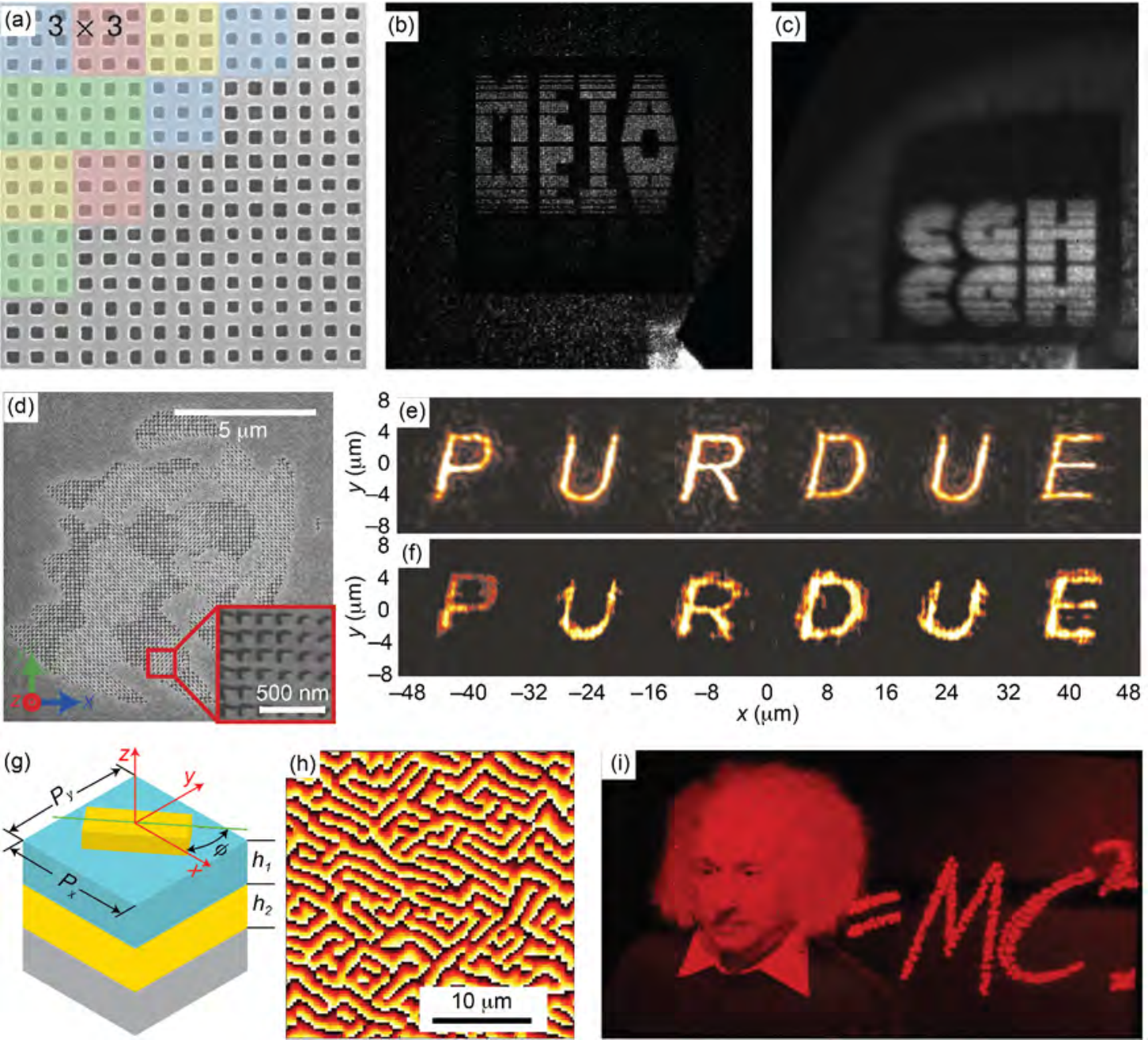}} \caption{Metasurface holograms. (a) SEM image of part of a metasurface hologram consisting of nanoaperture antennas. Different colors represent pixels with distinctive transmission coefficients. (b) and (c) Transmitted light intensity of the metasurface in (a) recorded in the far-field at $\lambda_1 = 905$~nm and $\lambda_2 = 1385$~nm, respectively. (d) SEM image of a fabricated metasurface for generating a holographic image of the letter ``\textit{P}''. Inset: zoomed-in view of the hologram. (e) Simulated and (f) measured holographic image created by the metasurface holograms similar to (d) with an eight-level phase modulation and a two-level amplitude modulation. (g) One-pixel cell structure of a nanorod-based hologram. The nanorod can rotate in the \textit{x-y} plane with an orientation angle $\phi$ to create different Pancharatnam?Berry phase delays. (h) 16-level phase distribution of the nanorod-based hologram ($100 \times 100$ pixels shown). (i) Experimentally obtained image in the far field created by the nanorod-based hologram at 632.8~nm wavelength. (a)-(c) used with permission from~\cite{Walther_2012_AdvMater}, (d)-(f) used with permission from~\cite{Ni_Shalaev_2013_NatCommun}, and (g)-(i) used with permission from~\cite{Zheng_Zhang_2015_NatNano}.
}
\label{Figure_7}
\end{figure*}
The most complex and general wavefront shaping is to create a holographic image in the far-field. Metasurfaces provide the degrees of freedom to engineer the local amplitude, phase, and polarization response on an interface, and thus are a good platform to realize all types of computer-generated holograms (CGHs) (e.g., binary holograms, phase-only holograms, amplitude and phase modulation holograms). \Fref{Figure_7}(a) shows a metasurface consisting of arrays of nanoaperture antennas that produce a spatially varying transmission coefficient~\cite{Walther_2012_AdvMater}. By utilizing the dispersion of aperture antennas, the metasurface was designed to operate as two distinctive binary transmission holograms at two different wavelengths, $\lambda_1 = 905$~nm and $\lambda_2 = 1385$~nm. It creates a word ``META'' shown in \fref{Figure_7} (b) at $\lambda_1 = 905$~nm and a word ``CGH'' shown in \fref{Figure_7} (c) at $\lambda_2 = 1385$~nm in the far-field. In another metasurface hologram, V-shaped aperture antennas shown in \fref{Figure_7}(d) were used to introduce an eight-level phase distribution and a two-level amplitude distribution~\cite{Ni_Shalaev_2013_NatCommun}. The amplitude and phase distributions approximated the required near-field amplitude and phase distributions on the metasurface plane, so that a certain holographic image was obtained in the far-field, as shown in \fref{Figure_7}(f). Additionally, a reflect-array metasurface that introduced a 16-level Pancharatnam-Berry phase has been demonstrated to create complex holographic images in the far-field (\fref{Figure_7}(g), (h) and (i))~\cite{Zheng_Zhang_2015_NatNano}. The antenna-orientation-controlled Pancharatnam-Berry phase combined with the reflect-array design led to broadband performance and high efficiency of the hologram. Experimentally demonstrated efficiency reaches 80\% at $\lambda = 825$~nm and the hologram operates between 630~nm and~1,050 nm. 

\section{Polarization conversion} \label{Polarization_Conversion}

Polarization state is an intrinsic property of electromagnetic waves, and the conversion between polarization states is very often highly desirable (or even necessary) for many modern electromagnetic and photonic applications. For instance, in advanced communication and sensing, converting linear polarization to circular polarization makes a beam resistant to environmental variation, scattering and diffraction. During recent years, conversion among polarization states using metasurfaces has attracted increasing interest due to their design flexibility and compactness. The accompanied capability of tuning a phase delay spanning the entire $2\pi$ range over a broad bandwidth and with a deep subwavelength resolution could potentially address some critical issues in the development of flat optics.

Highly symmetric simple meta-atoms can be advantageous in maintaining polarization states. Breaking the symmetry can, however, provide additional degrees of freedom to achieve customized functionality that enables the manipulation of polarization states. Through tailoring the two eigenmodes corresponding to orthogonal linear polarizations, it is possible to have equal transmission magnitude but a relative phase delay $\Delta \phi$ at a specific frequency. Narrowband polarization conversions between linear and circular polarization states ($\Delta \phi = \pi/2$, quarter wave plates), or linear polarization rotation ($\Delta \phi = \pi$, half wave plates) have been realized using single-layer metasurfaces~\cite{Strikwerda_Averitt_2009_OE,Khoo_Crozier_2011_OL,Zhao_Alu_2011_PRB,Cong_Zhang_2012_NJP,Zhu_Wu_2013_OE,Wang_Hong_2015_OE} or multi-layer cascading metasurfaces~\cite{Chin_Cui_2008_APL,Li_Zhu_2010_APL,Peralta_OHara_2009_OE} operating from microwave to optical frequencies. However, the efficiency is limited, in general, up to 50\% with a bandwidth comparable to a meanderline quarter wave plate~\cite{Young_1973_IEEE,Strikwerda_Averitt_2009_OE}. The low level of polarization conversion efficiency can be addressed by the implementation of few-layer metasurfaces. 

Following the Jones matrix description~\cite{Menzel_2010_PRA,Armitage_2014_PRB} the transmission of linearly polarized incident fields $(E_x, E_y)$ through a metasurface can be described as 
\begin{eqnarray}
\left( \begin{array}{c c}
E_\mathrm{x}^\mathrm{t}\\
E_\mathrm{y}^\mathrm{t}
\end{array} \right) &= 
\left( \begin{array}{c c}
T_\mathrm{xx} & T_\mathrm{xy} \nonumber\\
T_\mathrm{yx} & T_\mathrm{yy}
\end{array} \right) \left( \begin{array}{c c}
E_\mathrm{x}^\mathrm{i}\\
E_\mathrm{y}^\mathrm{i}
\end{array} \right) \\
& = \hat{T}_\mathrm{lin} \left( \begin{array}{c c}
E_\mathrm{x}^\mathrm{i}\\
E_\mathrm{y}^\mathrm{i}
\end{array} \right), \label{JonesMatrix}
\end{eqnarray} 
For circularly polarized incident fields, it becomes
\begin{eqnarray}
\left( \begin{array}{c c}
E^\mathrm{t}_+\\
E^\mathrm{t}_-
\end{array} \right) &= \left( \begin{array}{c c}
T_{++} & T_{+-}\\
T_{-+} & T_{--}
\end{array} \right)   \left( \begin{array}{c c}
E^\mathrm{i}_+\\
E^\mathrm{i}_-
\end{array} \right)\nonumber \\
&= \hat{T}_\mathrm{circ} \left( \begin{array}{c c}
E^\mathrm{i}_+\\
E^\mathrm{i}_-
\end{array} \right),
\end{eqnarray}
where $T_{\pm\pm} = \frac{1}{2}(T_\mathrm{xx}+T_\mathrm{yy}) \pm \frac{i}{2}(T_\mathrm{xy}-T_\mathrm{yx})$ and $T_{\pm\mp} = \frac{1}{2}(T_\mathrm{xx}-T_\mathrm{yy}) \mp \frac{i}{2}(T_\mathrm{xy}+T_\mathrm{yx})$. Under normal incidence and in general, $x$ and $y$ directions do not necessarily coincide with the structure's principal axes. There are a few properties of Jones matrices that are related to metasurface structural symmetries:
\begin{enumerate}
\item All components in the Jones matrices could be different if the metasurface lacks reflection or rotational symmetries; 
\item If the metasurface structure has a mirror symmetry, $T_\mathrm{xy} = T_\mathrm{yx}$ and $T_{++} = T_{--}$, and if the incident linear polarization is further parallel or perpendicular to the symmetry plane, $T_\mathrm{xy} = T_\mathrm{yx} = 0$;
\item For metasurface structures with a $C_4$ or $C_3$ rotational symmetry with respect to the $z$-axis, we have $T_\mathrm{xx} = T_\mathrm{yy}$, $T_\mathrm{yx} = -T_\mathrm{xy}$, and $T_{+-} = T_{-+}$.  
\end{enumerate}
When designing metasurfaces for polarization conversion between the same kinds (\textit{x} and \textit{y} linear polarizations or left- and right-handed circular polarizations), we need to maximize the off-diagonal components of the Jones matrices. For the conversion between linear and circular polarizations, the metasurfaces need to enable $\pi/2$ phase difference between the orthogonal components.

\subsection{Linear-to-circular polarization conversion}

An antenna array backed with a ground plane has been widely exploited at microwave frequencies to enhance the radiation efficiency and beam directionality. This configuration also enhances the polarization conversion in reflection for anisotropic subwavelength metallic resonator arrays. Early work at microwave frequencies demonstrated that narrowband conversion to various polarization states, including linear-to-circular polarization and linear polarization perpendicular to the incident one, is possible depending on the structural parameters, incident angle, and frequency~\cite{Hao_Zhou_2007_PRL}. It has also been shown that a pair of perpendicularly oriented and detuned electric dipoles (e.g., rectangular, elliptical, squeezed cross resonators, \textit{etc.}) can be used to manipulate polarization states including the construction of quarter-wave plates operating in reflection at optical wavelengths~\cite{Pors_2011_OL,Wang_Wei_2012_APL}. This type of structure is similar to those widely used in metamaterial perfect absorbers~\cite{Watts_Padilla_2012_AM}, where the Fabry-P\'{e}rot-like interference plays an important role~\cite{Chen_2012_OE}. 

\begin{figure}[htp]
\centerline{\includegraphics[width=3.0in]{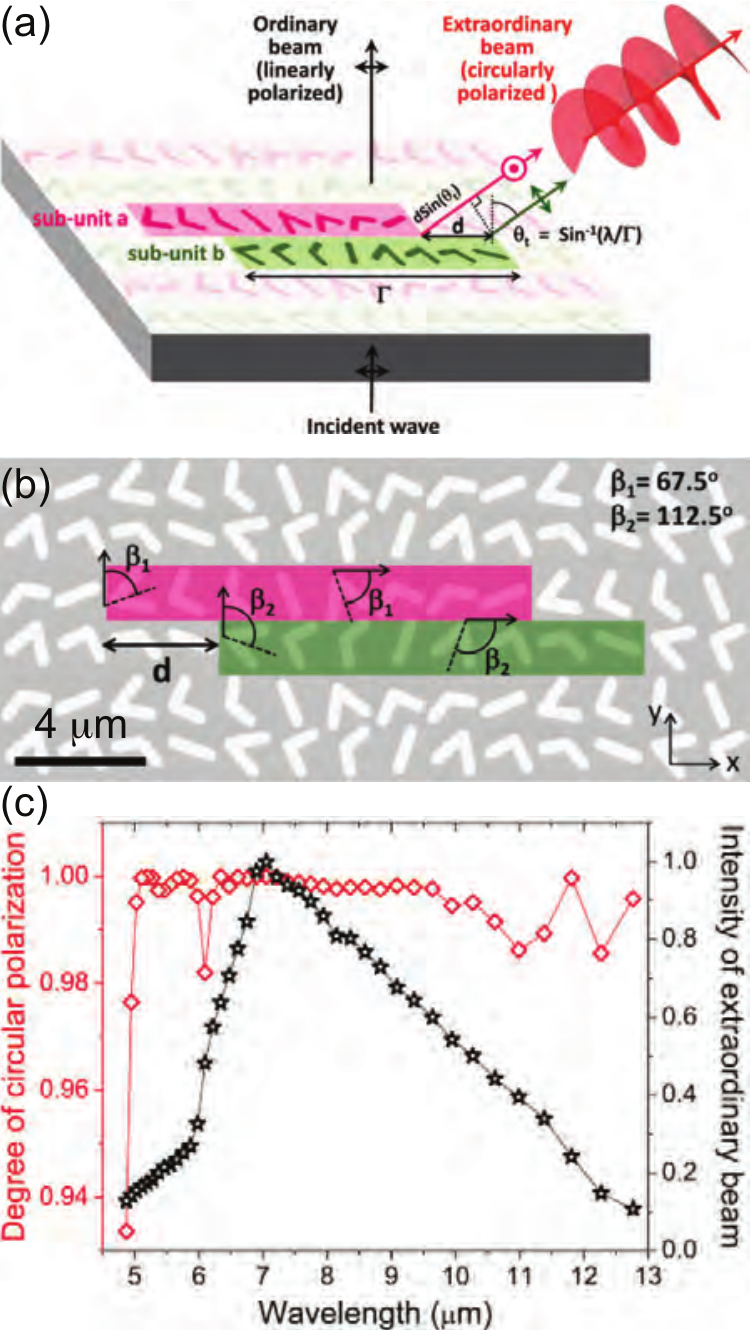}} \caption{(a) Schematic of a metasurface quarter-wave plate, with the unit cell of the metasurface consisting of two subunits (pink and green). Each subunit contains eight V-antennas. Upon excitation by linearly polarized incident light, the subunits generate two copropagating waves with equal amplitudes, orthogonal linear polarizations, and a $\pi/2$ phase difference (when offset $d = \Gamma/4$), which produce a circularly polarized anomalous refraction that is separated from the normal beam. (b) SEM image of a portion of the fabricated metasurface quarter-wave plate with a footprint of $230 \times 230$~$\mu$m$^2$ to accommodate the plane-wave like incident light. Antenna orientation angles are indicated by $\beta_1$ and $\beta_2$, and dashed lines represent the antenna symmetry axes. (c) Calculated degree of circular polarization and intensity of the anomalously refracted beam as a function of wavelength, showing the broadband and high efficiency properties of the quarter-wave plate. Used with permission from~\cite{Yu_Capasso_2012_NL}.
}
\label{Figure_8}
\end{figure}
New device functionalities could be realized by controlling spatial distribution of polarization response using metasurfaces. \Fref{Figure_8} show a metasurface-based quarter-wave plate~\cite{Yu_Capasso_2012_NL} that generates high-quality circularly polarized light (degree of circular polarization or ellipticity $>0.97$) over a broad wavelength range ($\lambda = 5$ to $12~\mu$m) (\fref{Figure_8}(c)). The unit cell of the metasurface comprises two subunits (colored pink and green in \fref{Figure_8}(a) and (b)). Upon excitation by linearly polarized incident light, the subunits generate two co-propagating waves with equal amplitudes, orthogonal linear polarizations, and a $\pi/2$ phase difference (when offset $d = \Gamma/4$), which produce a circularly polarized anomalously refracted beam that bends away from the surface normal.

\begin{figure}[htp]
\centerline{\includegraphics[width=2.5in]{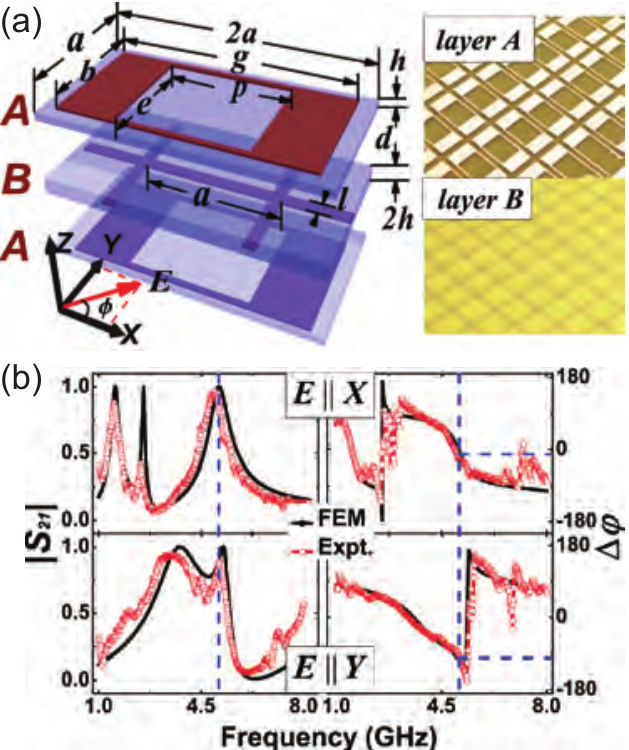}} \caption{(a) Unit cell of a tri-layer \textit{ABA}-type narrowband microwave linear-to-circular polarization converter, with the transmission amplitude and phase shown in (b) at two orthogonal directions. The operation frequency is indicated by the dashed vertical line. Used with permission from~\cite{Sun_Zhou_2011_OL}.}
\label{Figure_9}
\end{figure}

\begin{figure}[htp]
\centerline{\includegraphics[width=3.2in]{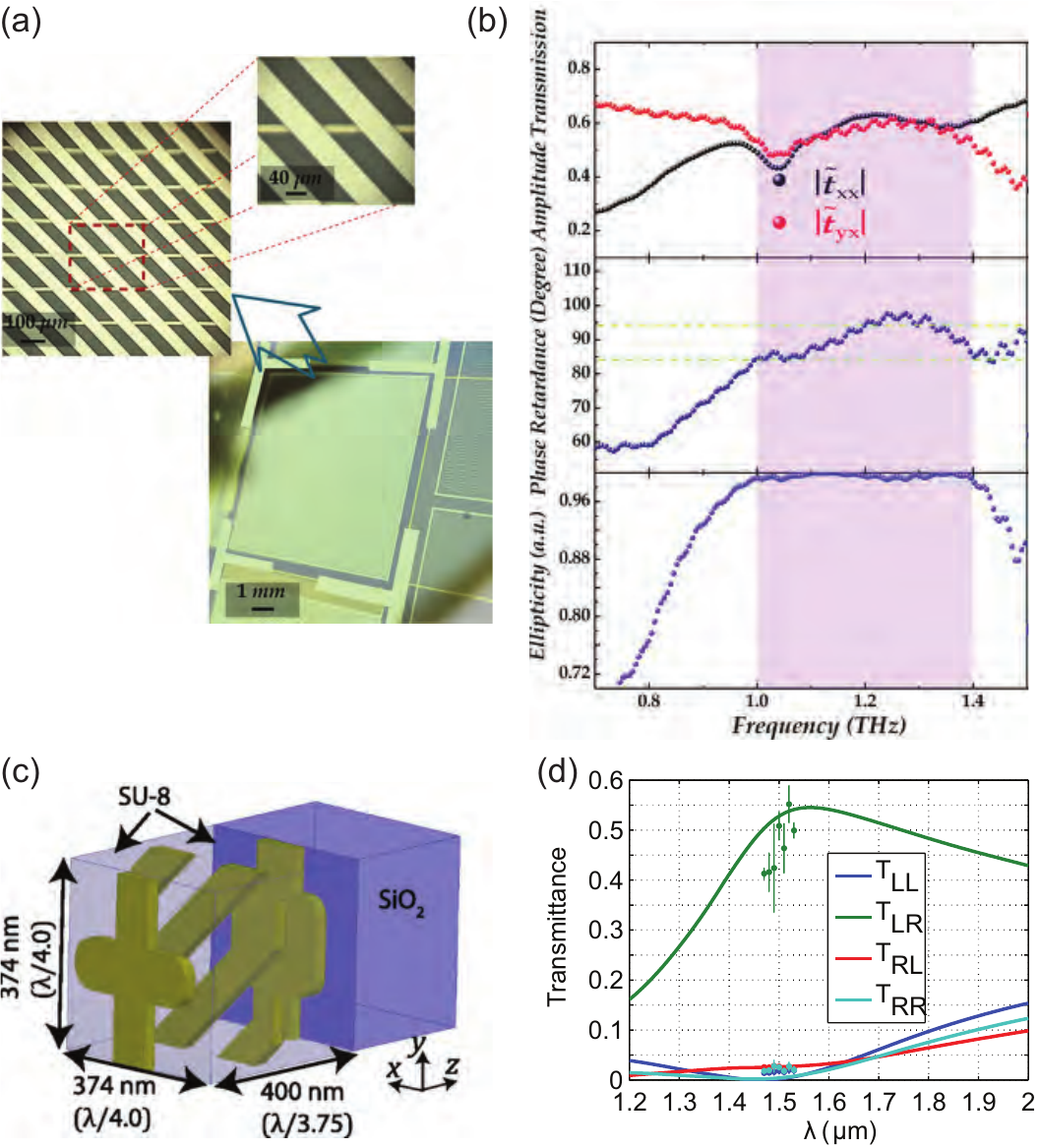}} \caption{(a) Optical images with different levels of zooming for a fabricated bi-layer THz metasurface embedded within a polyimide film. (b) Experimentally measured transmission amplitude, phase retardation, and ellipticity under horizontally polarized incidence, for the sample shown in (a). (c) Unit cell of a tri-layer metasurface for circular-to-circular polarization conversion operating at the near infrared, with simulated and measured transmittance shown in (d). (a) and (b) used with permission from~\cite{Cong_Zhang_2014_LPR}, (c) and (d) used with permission from~\cite{Pfeiffer_Grbic_2014_PRL}.}
\label{Figure_10}
\end{figure}

By increasing the number of layers to two or three, the near field or Fabry-P\'{e}rot-like coupling can significantly enhance the efficiency of linear-to-circular polarization conversion as well as the operation bandwidth. This property is realized in the few-layer metasurface structures illustrated in \fref{Figure_9} and \fref{Figure_10}. An \textit{ABA}-type, anisotropic tri-layer metasurface, shown in \fref{Figure_9}(a), has enabled narrowband, highly efficient linear-to-circular polarization conversion in transmission at microwave frequencies~\cite{Sun_Zhou_2011_OL}. Here layer \textit{A} is an electric metasurface with periodically arranged resonant microstructures, while layer \textit{B} is a metallic mesh. There are two mechanisms that are responsible for transparency. The first one is the electromagnetic wave tunneling~\cite{Zhou_2005_PRL} (a mechanism that is essentially equivalent to Fabry-P\'{e}rot-like resonance~\cite{Chen_2010_PRL_ARC}), and the second one is the extraordinary optical transmission (EOT) of layer \textit{B} that is mediated by the periodic structure of layer \textit{A}~\cite{Sun_Zhou_2011_OL}. Through structural tailoring, these two transparency bands, corresponding to the two orthogonal linear polarization directions (\textit{x} and \textit{y}), can overlap and, at the same time, have a phase difference of $\pi/2$, as shown in \fref{Figure_9}(b) at the frequency indicated by the dashed vertical line. This means that an incident electromagnetic wave linearly polarized at $45^\circ$ has been transformed to a circularly polarized one, with a conversion efficiency greatly exceeding any single-layer metasurface.  

Bi-layer metasurfaces have enabled high-efficiency and broadband conversions from linear to circular polarizations~\cite{Cong_Zhang_2014_LPR,Li_Zhang_2015_JAP}. A bi-layer metasurface comprised of stacked and twisted metallic wire grids shown in \fref{Figure_10}(a) was developed to operate at THz frequencies~\cite{Cong_Zhang_2014_LPR}.  For normal incidence and linearly polarized light in the \textit{x} direction, the first wire grid is aligned at $45^\circ$ with respect to \textit{x} direction. The wire grid is designed such that the transmission amplitude of orthogonal components $|t_\mathrm{xx}|$ and $|t_\mathrm{xy}|$ are approximately constant and equal, while the linear phase retardance is frequency dependent. This frequency dependent phase retardance is compensated through tailoring the geometric parameters of the second wire grid, which also has simultaneously high transmission coefficients $|t_\mathrm{xx}|$ and $|t_\mathrm{yy}|$. The metallic grids were embedded within a polyimide film so there are 4 interfaces: front air/polyimide, front wire grid, back wire grid, and back polyimide/air. Through combining the multiple reflections due to these interfaces and the dispersion of specially designed wire grids, the overall output of the two orthogonal \textit{x} and \textit{y} components have approximately equal amplitude and a phase delay of about $90^\circ$, resulting in circularly polarized transmission over a relatively broad bandwidth from 0.98 to 1.36 THz where the ellipticity is about 0.99, as shown in \fref{Figure_10}(b). Circular-to-circular polarization conversion was demonstrated employing a tri-layer metasurface designed through the approach developed by Pfeiffer and Grbic~\cite{Pfeiffer_Grbic_2014_PRAppl}, with the unit cell illustrated in \fref{Figure_10}(c). The measured and simulated Jones matrix of the metasurface~\cite{Pfeiffer_Grbic_2014_PRL}, shown in \fref{Figure_10}(d), reveals a high transmittance of $50\%$ for right-handed to left-handed circular polarization conversion, while all other components in the Jones matrix are below $2.5\%$, suggesting an extinction ration of $\sim20:1$ at the designed wavelength of 1.5~$\mu$m. It was also observed that the circular-to-circular polarization conversion extends over a quite broad wavelength range.  

\subsection{Linear polarization rotation}

Planar chiral response can yield optical activity, rotating the direction of linear polarization. While the polarization rotation power may significantly exceed naturally occurring materials per unit thickness, typically it is insufficient to obtain the desirable $90^\circ$ polarization rotation. Increasing the number of layers can yield half wave rotation; however, in general, this approach cannot sustain the polarization rotation power through increasing the number of layers by simple stacking, due to the near-field coupling or interference of the multireflections. In the past, efficient linear polarization conversion still employed anisotropic properties of metamaterials. 

\begin{figure}[htp]
\centerline{\includegraphics[width=3.2in]{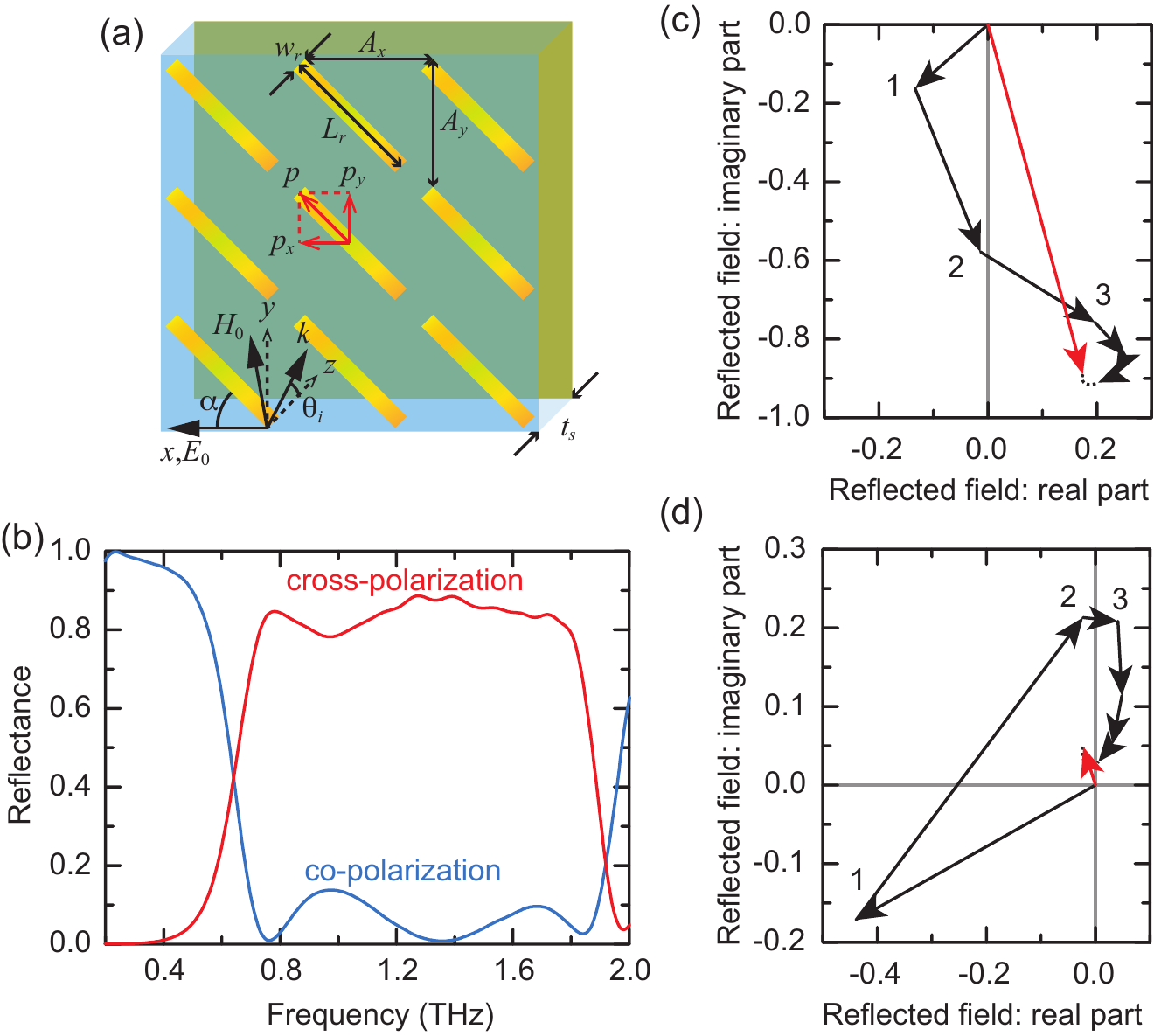}} \caption{Metasurface broadband polarization conversion in reflection. (a) Schematic metasurface structure. The incidence angle $\theta_i = 25^\circ$, and the incident electric field $E_0$ is linearly polarized in the \textit{x} direction with an angle $\alpha = 45^\circ$ with respect to the cut-wire orientation. (b) Experimentally measured co- and cross-polarized reflectance. (c) Cross- and (d) co-polarized multiple reflections theoretically calculated at 0.76 THz, revealing the constructive and destructive
interferences, respectively. Used with permission from~\cite{Grady_Chen_2013_Science}.}
\label{Figure_11}
\end{figure}  
A simple structure is shown in \fref{Figure_11}(a) where an array of cut-wires was separated from the ground plane by a polyimide spacer~\cite{Grady_Chen_2013_Science}. Under normal incidence, the incident \textit{x} polarized THz waves were converted to \textit{y} polarized waves in reflection with a conversion efficiency higher than 80\% over an ultrabroad bandwidth, as shown in \fref{Figure_11}(b). The co-polarized reflection approaches zero at several individual frequencies where the destructive interference conditions~\cite{Chen_2012_OE,Grady_Chen_2013_Science} are largely satisfied, as illustrated in \fref{Figure_11}(c) and (d), in which the superposition seems to be responsible for the observed broadband performance. Following this concept, a variety of metasurface structures, mostly at microwave frequencies, have been demonstrated to accomplish multi-band and ultra broadband linear polarization conversion in reflection~\cite{Chen_Li_2014_JAP}; even at visible wavelengths the high efficiency can be still largely maintained according to the simulation results in ~\cite{Dai_Wang_2014_OE}. The observed linear polarization rotation is consistent with an earlier contribution using a similar structure to control optical polarization in a reflection geometry~\cite{Hao_Zhou_2009_PRA}, while the bandwidth was much improved and different theoretical models were used. In order to avoid the increasing metallic loss in the optical frequency range, dielectric metasurfaces for linear polarization conversion in reflection were also demonstrated, based on the same principle~\cite{Yang_Valentine_2014_NL} (see discussions in section~\ref{Dielectric_Metasurfaces}).  

\begin{figure*}[htp]
\centerline{\includegraphics[width=6in]{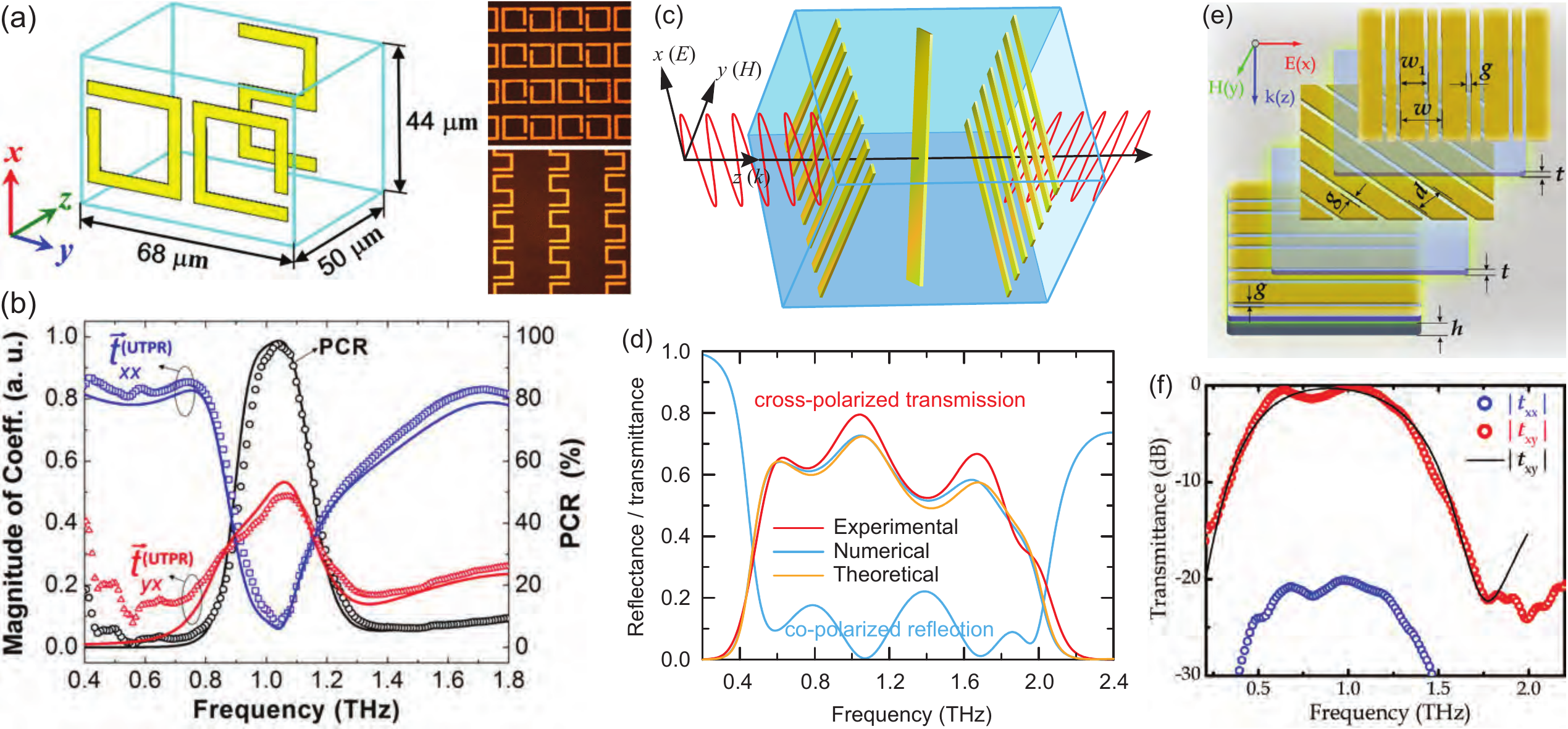}} \caption{(a) Unit cell (left panel) and optical images (right panel) of a bi-layer polarization rotator, and (b) measured co- and cross-polarized transmission coinciding with the simulated results, together with the polarization conversion ratio. (c) Schematic of the unit cell of a tri-layer metasurface linear polarization converter and (d) cross-polarized transmittance obtained through experimental measurements, numerical simulations, and theoretical calculations, together with the numerically simulated
co-polarized reflectance. (e) Schematic of a tri-layer metasurface polarization rotator consisting of three metallic gratings, and (f) experimental transmittance spectra. (a) and (b) used with permission from~\cite{Chiang_Yen_2013_APL}, (c) and (d) used with permission from~\cite{Grady_Chen_2013_Science}, (e) and (f) used with permission from~\cite{Cong_Zhang_2013_APL}.}
\label{Figure_12}
\end{figure*}  

It is more desirable to have linear polarization converters operating in the transmission mode. There have been a few bi-layer or tri-layer metasurfaces demonstrated to realize cross polarization conversion operating at a narrow single band or multiple bands~\cite{Ye_He_2010_APL,Mutlu_Ozbay_2012_APL,Shi_Jiang_2014_APL}, where the polarization rotation is insensitive to the azimuthal angle of the incident polarization due to the use of structures with four-fold rotational symmetry. \Fref{Figure_12}(a) shows the anisotropic unit cell of a bi-layer metasurface for $90^\circ$ rotation of the THz linear polarization, consisting of a front array of asymmetric split-ring resonators (ASRR) for polarization conversion and a rear array of S-shaped resonators (SR) for polarization selection~\cite{Chiang_Yen_2013_APL}. For the ASRR metasurface, the incident \textit{x}-polarized THz waves induces currents and forms a net electric dipole in the \textit{y}-direction, providing both \textit{x}- and \textit{y}-polarized components in reflection and transmission. The SR metasurface, however, exhibits negligible polarization conversion; it was tailored to have a resonance frequency coinciding with the ASRRs for \textit{x}-polarized waves, allowing only \textit{y}-polarized waves to pass through and blocking the \textit{x}-polarized waves. Due to the dispersion of the metasurfaces and through carefully optimizing the PET spacer thickness, a Fabry-P\'erot resonance occurs within the ultrathin polarization rotator, which can enhance the polarization conversion efficiency exceeding that of the ASRR metasurface alone. Although numerical simulations predict a conversion efficiency of 50\% (cross-polarized transmission magnitude 0.71) and a polarization-conversion ratio (PCR) up to 99.9\% using lossless PET spacer, the experimental values realized are 23\% (magnitude 0.48) and 97.7\%, respectively, at 1.04 THz as shown in \fref{Figure_12}(b), due to the significant loss within the PET spacer~\cite{Chiang_Yen_2013_APL}.   

Increasing the conversion efficiency and/or bandwidth becomes particularly interesting when metasurfaces are used to realize a new class of flat optical components where the transmission phase can be simultaneously controlled. An intriguing example for linear polarization rotation is a tri-layer THz metasurface demonstrated by Chen and co-workers~\cite{Grady_Chen_2013_Science}. It consists of a pair of identical gratings that are aligned in orthogonal directions, and an array of cut-wires tilted at an azimuthal angle of $45^\circ$, as shown in \fref{Figure_12}(c). The front grating is transparent when the incident THz field is linearly polarized along the \textit{x} direction. As it continues to propagate and excite the cut-wires, the scattering results in both \textit{x} and \textit{y} polarized components. For forward scattering, the back grating allows the newly generated \textit{y} polarized component to pass through while blocking the \textit{x} polarized component; for back scattering, the front grating reflects the \textit{y} polarized component and allows the \textit{x} polarized component to pass. This process continues due to a multireflection process within this multi-layer structure. When the thicknesses of the polyimide spacer layers are carefully tuned, a constructive interference enhances the polarization conversion and a destructive interference of the co-polarized reflections largely reduces the reflection loss (insertion loss) at multiple frequencies, as shown in \fref{Figure_12}(d), a mechanism similar to metamaterial antireflection coatings~\cite{Chen_2010_PRL_ARC} and perfect absorbers~\cite{Chen_2012_OE}. The back grating also guarantees a purely \textit{y} polarized output -- there is practically no co-polarized transmission. The overall result is that the \textit{x} polarized incident THz waves can be completely converted to its orthogonal \textit{y} polarization, over a bandwidth exceeding 2 octaves and with a conversion efficiency up to 80\%. Simply by scaling, a variety of similar structures~\cite{Li_Xu_2014_EL,Liu_Wang_2015_OptCommun} were employed in the microwave and infrared frequency ranges to demonstrate broadband, high-efficiency linear polarization rotators. Furthermore, in the structure shown in \fref{Figure_12}(c), the transmission phase can be finely tuned to span an entire $2\pi$ range and with subwavelength resolution through replacing the cut-wires with a variety of anisotropic resonators with varying geometric dimensions~\cite{Grady_Chen_2013_Science}. Combining this property and the high polarization conversion efficiency promises great potential in wavefront control, resulting in a new class of practical flat optical devices.

A similar broadband THz polarization rotator was demonstrated by Cong \textit{et al.}~\cite{Cong_Zhang_2013_APL}, where the middle cut-wire array was replaced by a wire grating, as schematically shown in \fref{Figure_12}(e). The formation of a Fabry-P\'{e}rot cavity makes this metasurface structure perform in remarkable contrast to cascading wire polarizers with consecutive $45^\circ$ rotation. The latter does rotate the incident linear polarization by $90^\circ$ but allows only up to $25\%$ power transmission. This metasurface showed a conversion efficiency up to $85\%$, and the output waves exhibit extremely clean cross linear polarization over a broad bandwidth, as shown in \fref{Figure_12}(f), although the transmission phase cannot be controlled.  

\subsection{Asymmetric transmission}

By reducing the structural symmetry and converting between polarization states, metasurfaces have yielded a polarization sensitive and asymmetric transmission with respect to the direction of wave propagation~\cite{Fedotov_Zheludev_2006_PRL}. Asymmetric polarization conversion and transmission were observed in planar chiral metasurfaces for circularly polarized incident fields with $T_{\pm\mp}^\mathrm{f} \neq T_{\mp\pm}^\mathrm{f}$ and $T_{\pm\mp}^\mathrm{f} \neq T_{\pm\mp}^\mathrm{b}$, where the superscripts ``f'' and ``b'' denote the forward and backward propagation directions, respectively, though $T_{\pm\pm}^{f} = T_{\pm\pm}^{b}$ and $T_{\pm\mp}^{f} = T_{\mp\pm}^{b}$ as required by Lorentz Reciprocity Lemma. The planar chiral metasurfaces are more transparent to a circularly polarized wave from one side than from the other side, with an experimentally measured transmission difference up to $40\%$ at microwave~\cite{Fedotov_Zheludev_2006_PRL} and $15\%$ at visible~\cite{Schwanecke_Zheludev_2008_NL,Fedotov_Zheludev_2007_NL} frequencies. This effect is caused by the different efficiencies of polarization conversion in the opposite propagation directions for lossy metasurfaces, in remarkable contrast to the optical activity and Faraday effect. It implies that when circularly polarized light passes through the metasurface and then retraces its path after reflection from a mirror, the final polarization state will be different from that of the initial state~\cite{Drezet_Ebbesen_2008_OE}.

\begin{figure*}[htp]
\centerline{\includegraphics[width=5in]{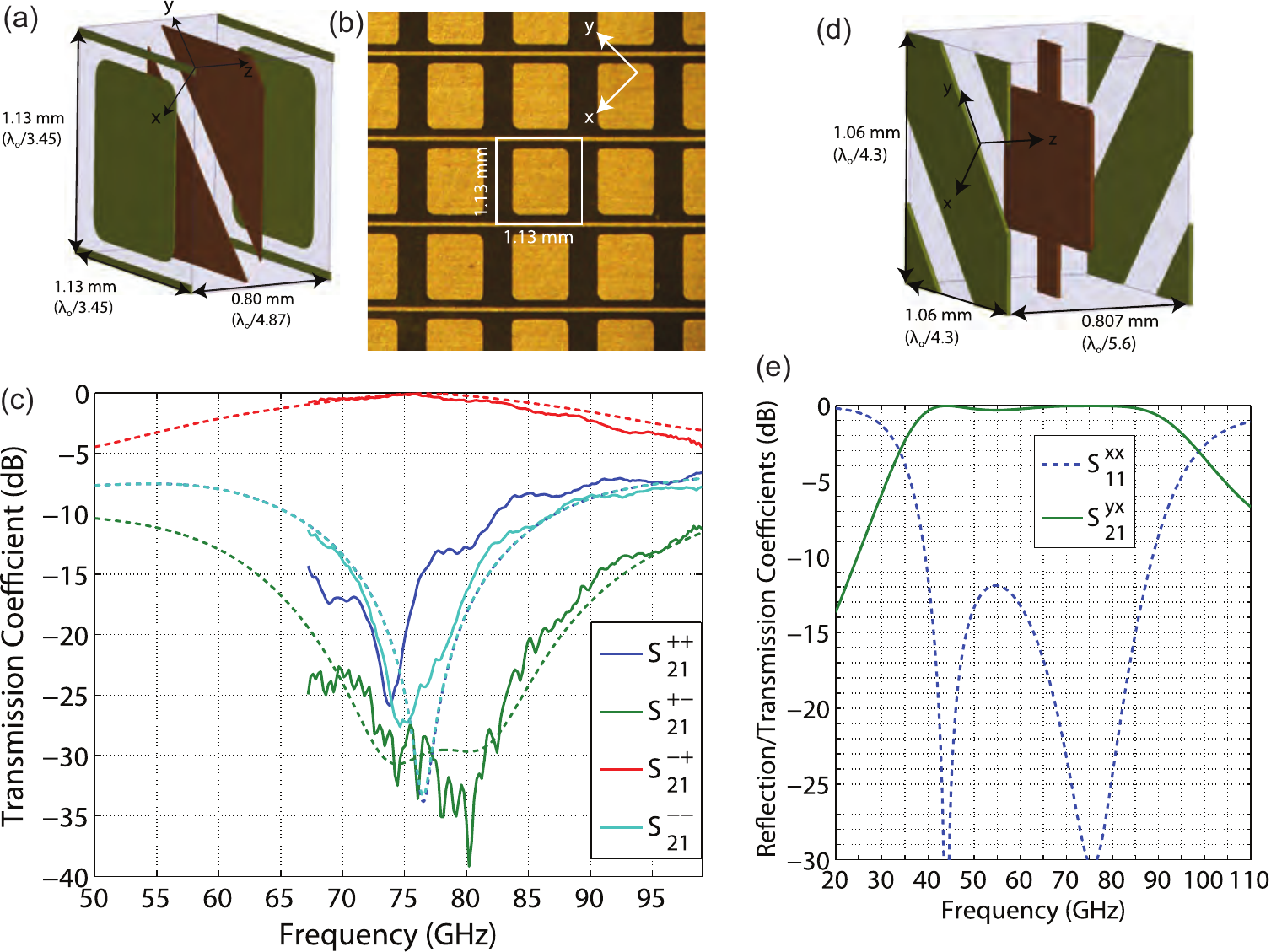}} \caption{(a) Schematic of a tri-layer metasurface unit cell and (b) optical image of its top metallic sheet, which exhibits asymmetric transmission of circularly polarized millimeter waves with transmission coefficients shown in (c). Solid curves: measured data; dashed curves: simulated data. (d) The unit cell of a tri-layer metasurface which enables (e) broadband and highly asymmetric transmission of linearly polarized millimeter waves. Used with permission from~\cite{Pfeiffer_Grbic_2014_PRAppl}.}
\label{Figure_13}
\end{figure*}

Bi-layer and multi-layer metasurfaces can increase the polarization conversion and consequently enhance the transmission asymmetry. Pfeiffer and Grbic recently presented systematic methods to analyze and synthesize bianisotropic metasurfaces realized by cascading anisotropic, patterned metallic sheets. This design approach starts with the desirable S-parameters and solves for the necessary admittances of the metallic sheets. Once the required sheet admittances are known, the theory of frequency-selective surface and full-wave numerical simulations are used for their physical realization. One such metasurface exhibiting strongly asymmetric transmission of circularly polarized millimeter waves is shown in \fref{Figure_13}(a) and (b)~\cite{Pfeiffer_Grbic_2014_PRAppl}. As shown in \fref{Figure_13}(c), the $S_{21}$ parameter (i.e., transmission) is below $-10$ dB for $++$, $+-$, and $--$, and it is above $-0.8$ dB for $-+$, resulting in an asymmetric response of 0.99 over a bandwidth of $20\%$ at the designed millimeter wavelengths. Similar behaviors were observed in tri-layer metasurfaces operating at near infrared wavelengths~\cite{Pfeiffer_Grbic_2014_PRL}, as shown in \fref{Figure_10}(c) and (d). 

A variety of bi-layer metasurfaces have been also reported to exhibit asymmetric transmission for linearly polarized incident light~\cite{Menzel_Lederer_2010_PRL,Huang_2012_PRB}. Further developments showed that bi-layer metasurface structures can be used to demonstrate increased bandwidth of the asymmetric transmission in the near infrared~\cite{Li_Tian_2014_APL,Li_Tian_2015_Plasmonics}. It was shown that the interlayer alignment could have very little effect on the asymmetric transmission~\cite{Li_Tian_2014_APL}, which indicates that the near-field coupling is negligible. This advantageous property is particularly useful in the optical regime where the interlayer alignment is challenging. In order to take full advantage of the asymmetric transmission, it is necessary to suppress other components and only obtain a high contrast asymmetric component (e.g., $t_\mathrm{yx}$) within the Jones transmission matrix~\cite{Han_2011_APL,Shi_Cui_2014_PRB,Mutlu_Ozbay_2012_PRL}. Tri-layer metasurfaces have demonstrated the best performance in both the efficiency and bandwidth. Excellent examples include the ultra-broadband THz linear polarization rotator shown in \fref{Figure_12}(c)-(f), which exhibits a bandwidth over two octaves~\cite{Grady_Chen_2013_Science,Cong_Zhang_2013_APL}. Another tri-layer metasurface is shown in \fref{Figure_13}(d) and (e), which demonstrates highly efficient, broadband asymmetric transmission of linearly polarized millimeter waves~\cite{Pfeiffer_Grbic_2014_PRAppl}. The simulated results show that a 1-dB transmission bandwidth of 2.43:1 for the desired polarization is achieved, and that the rejection of the unwanted polarization exceeds 30 dB in this band. 

\section{Dielectric metasurfaces} \label{Dielectric_Metasurfaces}

The majority of metasurface research has focused on using subwavelength metallic structures, where ohmic losses pose a severe issue, particularly in the optical frequency range, limiting the performance of arguably any desirable functions. Low-loss, high-refractive-index dielectric materials have received much attention during recent years partially due to their ability in addressing the efficiency issue in metallic metasurfaces. Furthermore, the capability of tuning the magnetic and electric resonances through tailoring the geometry and spacing of dielectric resonators enables device functionalities beyond metallic metasurfaces. 

\subsection{Dielectric resonators}

Dielectric resonators can be traced back to the discussions by Richtmyer~\cite{Richtmyer_1939_JAP}. Due to the excitation of the resonant modes as well as their leaky nature, dielectric resonators can serve as radiative antennas, as developed theoretically and experimentally in the 1980's by Long \textit{et al.} at microwave frequencies~\cite{Long_1983_IEEE}. Increasing the dielectric constant $\epsilon$ can significantly reduce the required size $d$ of the resonators, which is related to the free space resonant wavelength $\lambda_0$ by $d \sim \lambda_0 / \sqrt{\epsilon}$. However, increasing the dielectric constant also reduces the radiation efficiency and narrows the operational bandwidth, which is inversely related to the dielectric constant. Typical values of the dielectric constant used range from 8 to 100 in order to balance the compactness, radiation efficiency and bandwidth requirements. Very often dielectric resonators are mounted on top of a metal ground plane, which improves the radiation efficiency and acts as an electrical symmetry plane to improve the compactness. Early work in resonant dielectric antennas at microwave frequencies has been summarized in review articles~\cite{Mongia_Bhartia_1994_IJMMWCAE,Petosa_Ittipiboon_2010_IEEE}. 

\begin{figure}[htp]
\centerline{\includegraphics[width=3.2in]{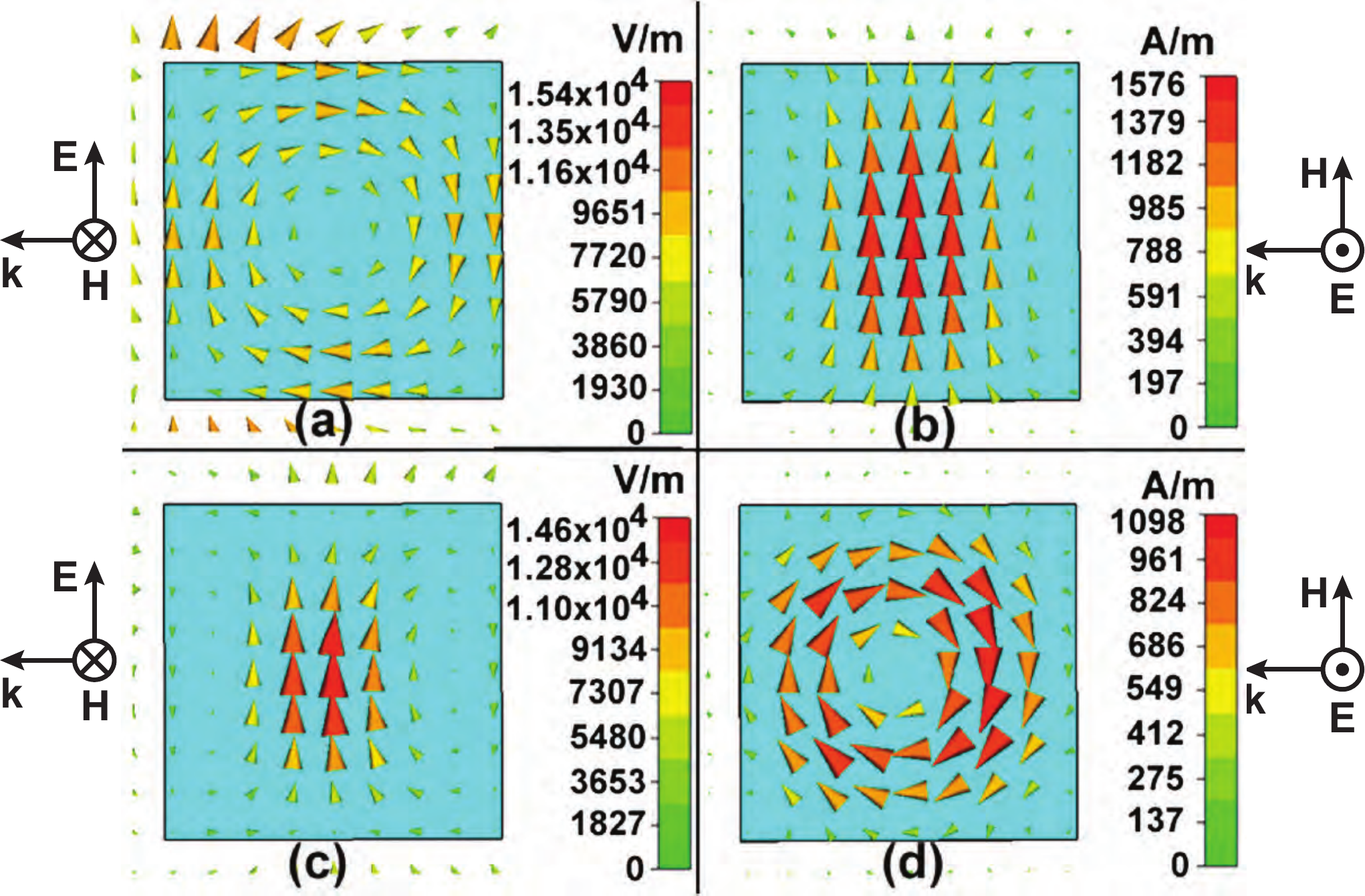}} \caption{Electric and magnetic modes in a cubic dielectric resonator. (a) and (b) Magnetic dipole resonance mode, showing electric field (a) and magnetic field (b) distributions. (c) and (d) Electric dipole resonance mode, showing electric field (c) and magnetic field (d) distributions. The incident fields are indicated in the insets. Reproduced with permission from~\cite{Zhao_2008_PRL}.}
\label{Figure_14}
\end{figure}

In the optical regime, low loss dielectric particles support strong electric and magnetic scattering known as Mie resonances, which can be decomposed into a multipole series. The modes are determined by the particle size and structural properties~\cite{Wheeler_2005_PRB,Ahmadi_2008_PRB,Evlyukhin_2010_PRB,Evlyukhin_2011_PRB}, in contrast to metallic particles where the resonance scattering is dominated by the electric resonances. In most dielectric resonators of regular shapes such as spheres, cubes, cylindrical disks and rods, the lowest resonant mode is the magnetic dipole resonance and the second lowest mode is the electric dipole resonance~\cite{Evlyukhin_2011_PRB,Kuznetsov_2012_SciRep}. \Fref{Figure_14} shows the fundamental magnetic and electrical dipole modes for a cubic dielectric resonator~\cite{Zhao_2008_PRL}. The magnetic resonance mode originates from the excitation of circulating displacement currents, resulting in the strongest magnetic polarization at the center, similar to the case of magnetic resonant response in metallic SRRs. The contribution from other higher order modes can be ignored as the coefficients of these modes are often orders of magnitude lower~\cite{Vynck_2009_PRL}.

Subwavelength dielectric resonators can be used as the basic building blocks of metamaterials and metasurfaces, as first suggested by O'Brien and Pendry to obtain magnetic activity in dielectric composites~\cite{OBrien_Pendry_2002_JPCM}. A class of Mie resonance-based dielectric metamaterials have been consequently demonstrated, with some early work reviewed in~\cite{Zhao_2009_MaterialsToday}, where high dielectric constant materials are used to create subwavelength resonators for the realization of negative electric and magnetic responses. Ferroelectric barium strontium titanate (BST or Ba$_{0.5}$Sr$_{0.5}$TiO$_3$) was used to demonstrate dielectric metamaterials because of its high dielectric constant ($\sim600$) at microwave frequencies. Left-handed behavior was observed in prisms formed by an array of periodic or random subwavelength BST rods~\cite{Peng_2007_RPL}, and negative magnetic response was also observed in a bulk metamaterial consisting of an array of subwavelength BST cubes~\cite{Zhao_2008_PRL}. In the optical frequency range, materials used to form dielectric metamaterials include tellurium (Te) cubes on barium fluoride (BaF$_2$)~\cite{Ginn_Brener_2012_PRL}, cubic ($\beta$) phase silicon carbide (SiC) whiskers on zinc selenide (ZnSe)~\cite{Schuller_Brongersma_2007_PRL,Schuller_Brongersma_2009_NatPhoton}, in the mid-infrared; silicon cylindrical nano disks embedded within silicon dioxide~\cite{Staude_Kivshar_2013_NL} in the near infrared; silicon nano spheres on glass~\cite{Evlyukhin_2012_NL} and titanium dioxide cylindrical disks on silver~\cite{Zou_Fumeaux_2013_OE_Dielectric_Resonator_Nanoantennas} at visible frequencies. 

\begin{figure}[htp]
\centerline{\includegraphics[width=3in]{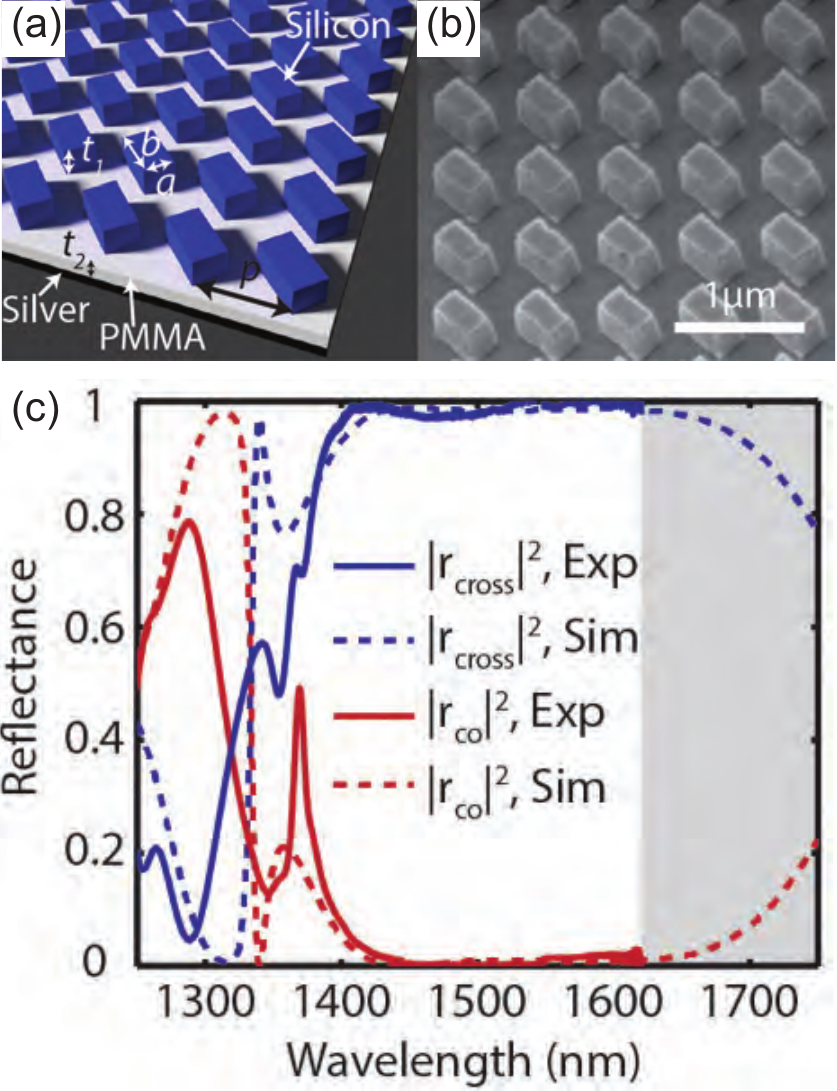}} \caption{Dielectric metasurface for broadband polarization conversion in reflection. (a) Schematic and (b) SEM image of the dielectric metasurface structure. (c) Experimentally measured (solid lines) and numerically simulated (dotted lines) co- and cross-polarized reflectance. Used with permission from~\cite{Yang_Valentine_2014_NL}.}
\label{Figure_15}
\end{figure}

The loss reduction enabled by dielectric metasurfaces becomes clear when functioning as a linear polarization rotator as shown in \Fref{Figure_15}, where an array of anisotropic (rectangular) silicon resonators is separated from a metal ground plane by a thin layer of PMMA. In experiments, linear polarization conversion with more than 98\% conversion efficiency was demonstrated over a 200 nm bandwidth in the near infrared~\cite{Yang_Valentine_2014_NL}, as shown in \fref{Figure_15}(c). This result exemplifies the significant loss reduction enabled by the use of dielectric metasurfaces instead of metallic resonators shown in \fref{Figure_11}, particularly in the infrared and visible frequency ranges.

In general, dielectric resonators offer only up to $\pi$ phase variation in transmission when the electric and magnetic resonances are at different frequencies. By overlapping the electric and magnetic dipole resonances through varying the geometry of dielectric resonators, however, it is possible to achieve a phase variation covering the entire $2\pi$ range~\cite{Cheng_Mosallaei_2014_OL}. This was experimentally verified even without satisfying the condition of equal electric and magnetic resonance width~\cite{Decker_Kivshar_2015_AOM}. In cylindrical dielectric disks, the tuning parameter could be the disk height, diameter, and period (spacing). The spacing between resonators further facilitates the tuning of resonance coupling~\cite{Evlyukhin_2010_PRB}, which affects the dispersion of the scattering phase resulting from the different transverse electric and transverse magnetic modes, and also enables electromagnetically induced transparency in dielectric metasurfaces with an ultra high quality factor~\cite{Yang_Valentine_2014_NatCommun}. 

\subsection{Directional scattering}

In 1983 Kerker \textit{et al.} discussed electromagnetic scattering by magnetic spheres. It was shown that back scattering can be reduced to zero by spheres with equal permeability $\mu$ and permittivity $\epsilon$~\cite{Kerker_1983_JOSA}. In such a situation the particle exhibits equal electric and magnetic multipole coefficients, resulting in destructive interference in the backward propagating direction and constructive interference in the forward propagating direction. The magnetic Mie resonance overcomes the absence of magnetic materials at optical frequencies and enables the investigation of directional optical scattering using dielectric metasurfaces. The complete cancellation of back scattering was also theoretically predicted in~\cite{Evlyukhin_2010_PRB} at an off-resonance frequency in an array of silicon nano spheres where the electric and magnetic polarizabilities have equal values. Such a phenomenon corresponds to a `Huygens' secondary source, and was experimentally demonstrated using nonmagnetic dielectric spherical and cylindrical scatters with moderate dielectric constants at microwave~\cite{Geffrin_2012_NatCommun} and visible~\cite{Fu_2013_NatCommun,Person_Novotny_2013_NL} frequencies. 

\begin{figure}[htp]
\centerline{\includegraphics[width=3.2in]{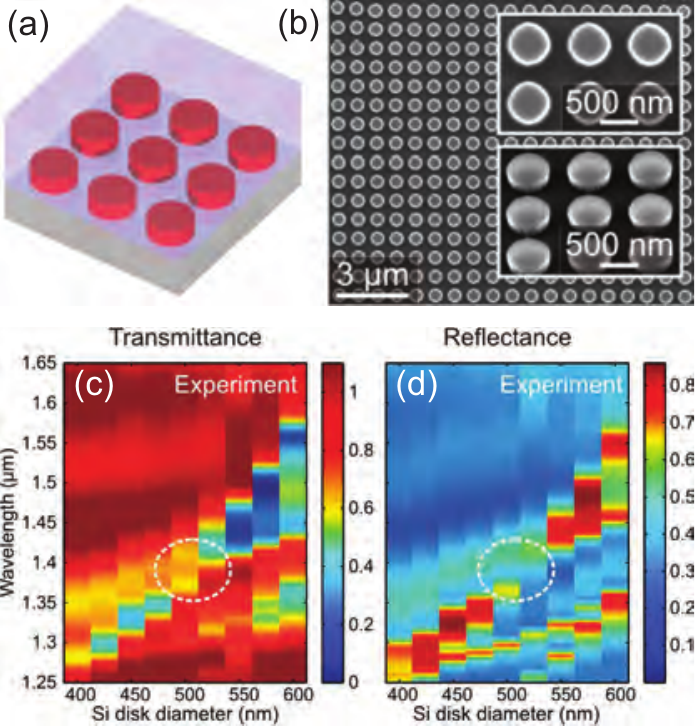}} \caption{(a) Schematic of silicon nanodisks embedded into a low-index (SiO$_2$) medium. (b) SEM image of the fabricated silicon nanodisks before embedding them into SiO$_2$. The insets show the close-up top and oblique views. (c) Optical transmittance and (d) reflectance spectra of the fabricated sample, where the white dashed ellipses indicate the regions where the back scattering is significantly reduced. Used with permission from~\cite{Staude_Kivshar_2013_NL}. }
\label{Figure_16}
\end{figure}

The resonant directional scattering is more interesting because of the large field enhancement and concentration. Resonant response usually accompanies large back scattering, which makes it more feasible for dielectric metasurfaces to operate in a reflection configuration~\cite{Zou_Fumeaux_2013_OE_Dielectric_Resonator_Nanoantennas,Yang_Valentine_2014_NL}. This enabled the demonstration of broadband dielectric metasurface mirrors~\cite{Slovick_2013_PRB,Moitra_Valentine_2014_APL,Moitra_Valentine_2015_ACSPhotonics} and optical magnetic mirrors~\cite{Liu_Brener_2014_Optica,Headland_Sriram_2015_AdvMater}, without reflection phase reversal in the latter. Using geometric shapes other than spherical or cubic dielectric resonators, one could have more degrees of freedom to tune independently the frequencies of electric and magnetic resonances to realize resonant directional scattering. This is exemplified by the closer electric and magnetic dipole resonances when squeezing the silicon spheres in the $z$-direction, which results in a larger forward-to-backward scattering ratio~\cite{Fu_2013_NatCommun}. An array of silicon cylindrical nano disks, as shown in \fref{Figure_16}(a) and (b), was used to demonstrate resonant directional scattering in the visible wavelength range~\cite{Staude_Kivshar_2013_NL}. By varying the diameter of the silicon disks, it was observed that the electric and magnetic resonances overlap, resulting in enhanced forward scattering and cancellation of backward scattering, as shown in \fref{Figure_16}(c) and (d). 

An ideal dielectric Huygens' metasurface requires overlapping electric and magnetic dipole resonances of equal resonance strength and width in order to completely cancel the reflection and obtain near unity transmission~\cite{Decker_Kivshar_2015_AOM}. High transmittance of $55\%$ at resonance was experimentally demonstrated in the near-infrared using a silicon metasurface consisting of an array of cylindrical resonators embedded within an SiO$_2$ environment, where the condition of equal width of the electric and magnetic resonances was not yet satisfied~\cite{Decker_Kivshar_2015_AOM}. By tuning the dielectric constant of the environment and the geometric dimensions of the resonators, it is possible to achieve spectral overlap and equal width of the resonances. Post-fabrication active tuning of the resonances is attractive for this purpose as well. For instance, a layer of liquid crystals was added on top of the silicon nano disks, providing temperature-dependent refractive indices when the liquid crystals were switched between the nematic and isotropic phases~\cite{Sautter_Kivshar_2013_ACSNano}. It was shown that the electric resonance has a larger tuning range because of extended fringing fields outside the resonators, while the magnetic resonance has smaller tuning capability because of the better confined field distribution within the dielectric resonators. Reconfigurable directional scattering can be also accomplished using metasurfaces consisting of semiconducting resonator arrays through injection of free charge carriers by optical excitation~\cite{Iyer_2015_ACSPhotonics}.

\subsection{Beam forming and wavefront control enabled by dielectric metasurfaces}

\begin{figure}[htp]
\centerline{\includegraphics[width=3.2in]{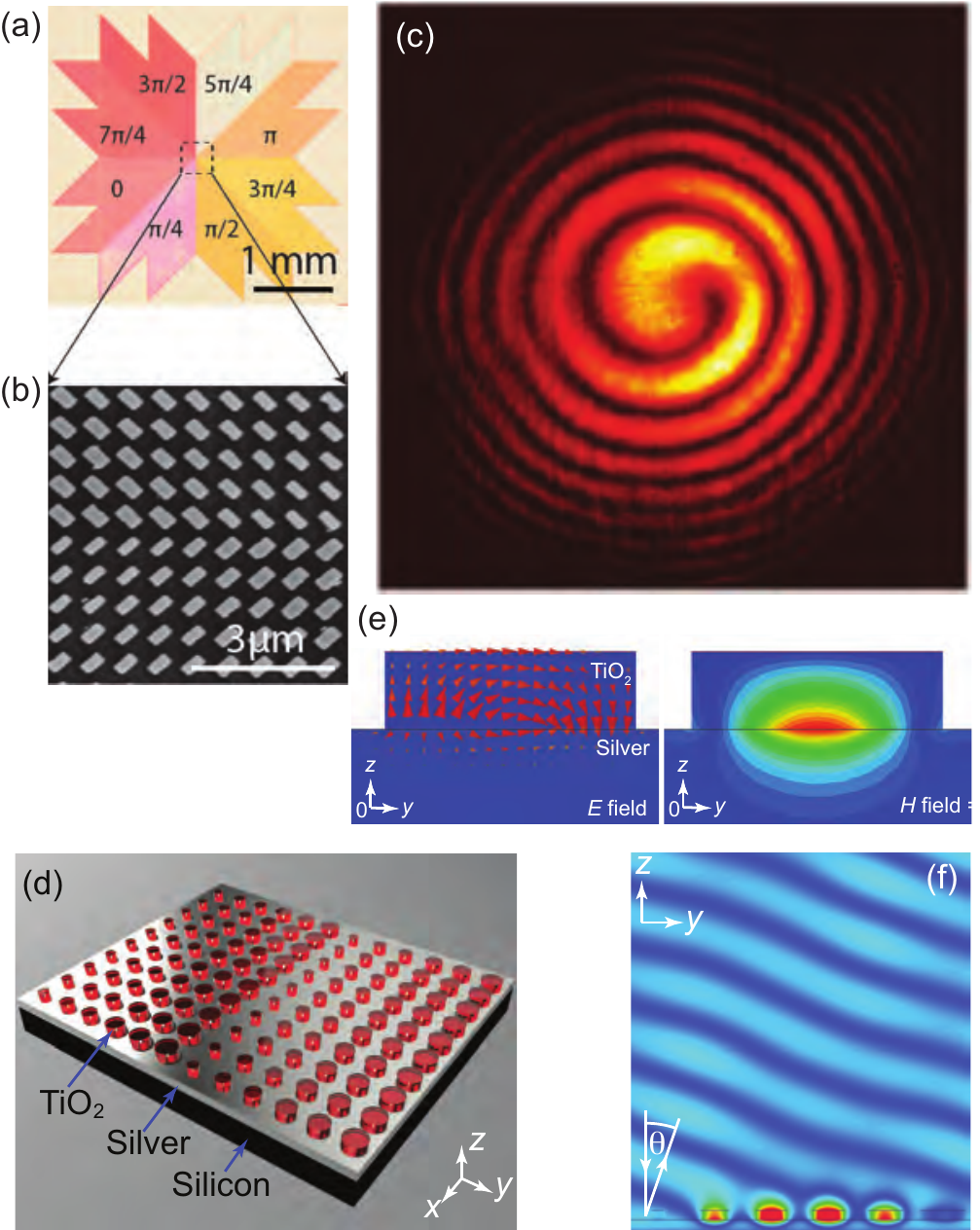}} \caption{(a) A phase profile in the azimuthal direction with an increment of $\pi/4$, created using (b) silicon rectangular resonators on top of a metal mirror with a PMMA spacer, and enabling the formation of a near infrared optical vortex beam. The pattern in (c) is the interference between the vortex beam and a reference Gaussian beam. (d) Schematic of part of a reflect-array metasurface consisting of dielectric resonators patterned on a metallic substrate and operating at $\lambda = 633$~nm. (e) Simulated electric and magnetic field distributions in a dielectric resonator antenna. (f) Simulation showing that at zero-degree angle of incidence the metasurface in (d) generates a reflected wave propagating along $20^\circ$ direction from the surface normal. (a)-(c) used with permission from~\cite{Yang_Valentine_2014_NL}, (d)-(f) used with permission from~\cite{Zou_Fumeaux_2013_OE_Dielectric_Resonator_Nanoantennas}.}
\label{Figure_17}
\end{figure}

Similar to metasurfaces consisting of plasmonic metallic resonators, wavefront control and beam forming can be accomplished using dielectric metasurfaces. By varying the dimensions of the rectangular silicon resonators shown in \fref{Figure_15}, a phase variation can span the entire $2\pi$ range. This enables the generation of a near infrared optical vortex beam in reflection with high efficiency when a phase gradient profile was created in the azimuthal direction using 8 elements of different sizes~\cite{Yang_Valentine_2014_NL}, as shown in \fref{Figure_17}(a-c). The use of a PMMA spacer layer between the silicon resonators and a metallic back plane not only provides the desirable interference resulting from the Fabry-P\'{e}rot-like multiple reflections, but also effectively prevents the incident light from coupling to surface waves. This is in remarkable contrast to the situation where the dielectric resonators are directly mounted onto the metallic surface~\cite{Zou_Fumeaux_2013_OE_Dielectric_Resonator_Nanoantennas}. In the latter work, a linear phase gradient at wavelength of 633 nm was created by using six TiO$_2$ cylindrical resonators of various diameters sitting on top of a silver plane (\fref{Figure_17}(d)), demonstrating a deflection from the specular reflection by the expected $20^\circ$ (\fref{Figure_17}(f))~\cite{Zou_Fumeaux_2013_OE_Dielectric_Resonator_Nanoantennas}. It was shown that considerable dissipation occurs within the TiO$_2$ resonators, partially because this configuration can also function as a metamaterial absorber~\cite{Liu_Zhou_2013_APL}. Even more optical power is coupled to surface waves, which was described in a recent theoretical proposal of directional launching of surface waves~\cite{Zou_2015_PRB}.

\begin{figure}[t]
\centerline{\includegraphics[width=3.2in]{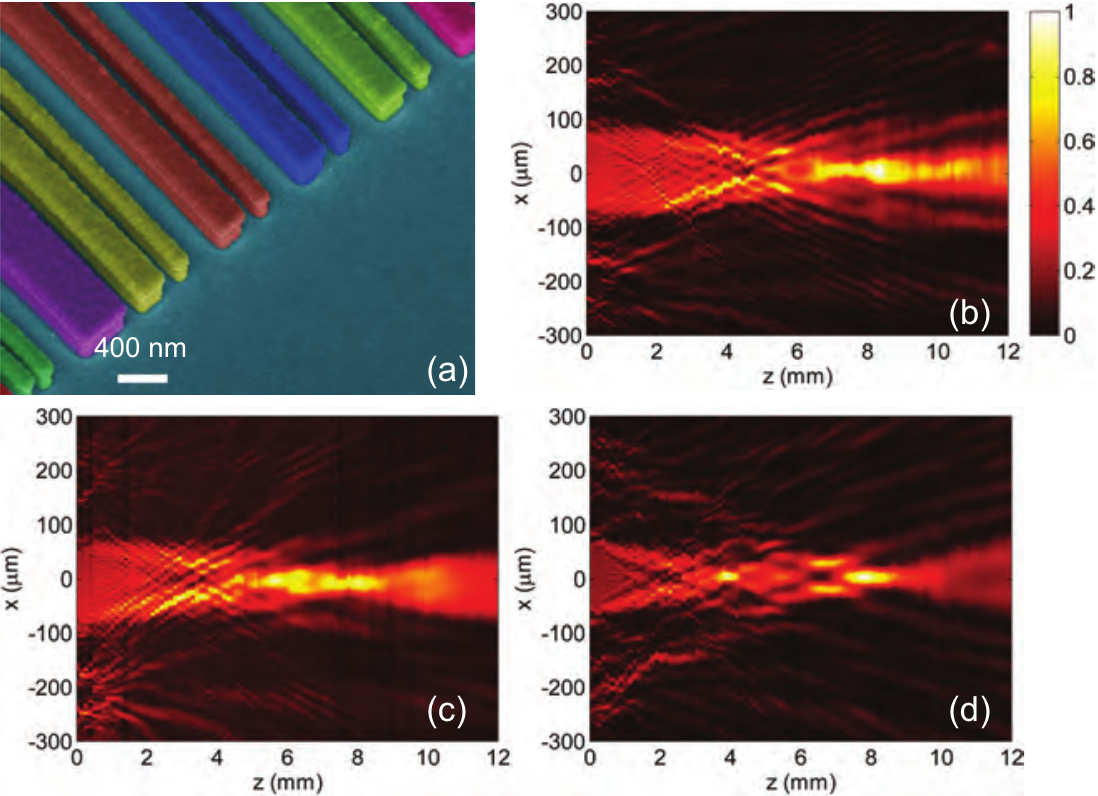}} \caption{Multi-wavelength dielectric metasurface cylindrical lens. (a) False colored side-view SEM image of the metasurface lens. Each unit cell is identified by a different color. (b)-(d) Measured intensity distributions in the plane perpendicular to the silicon ridges at wavelengths (b) 1300 nm, (c) 1550 nm, and (d) 1800 nm. Used with permission from~\cite{Khorasaninejad_Capasso_2015_NL} }
\label{Figure_18}
\end{figure}

A metasurface that converts a Gaussian beam into a vortex beam was demonstrated; it consists of four quadrants with a phase increment of $\pi/2$ and each quadrant consists of an array of cylindrical silicon nano disks of the same geometry but different separations between adjacent disks~\cite{Chong_Kivshar_2015_NL}. By varying the diameter of silicon nanoposts to control the phase profile, a high-efficiency lens was demonstrated with measured focusing efficiency in transmission up to 82\%~\cite{Arbabi_Faraon_2015_NatCommun}. Through varying the geometric dimensions and coupling strength between dielectric resonators, it is possible to create the required phase profiles to simultaneously control the wavefront at multiple wavelengths. This approach was exploited in the demonstration of a multi-wavelength dielectric metasurface lens operating near telecommunication wavelengths~\cite{Aieta_Capasso_2015_Science,Khorasaninejad_Capasso_2015_NL}. To achieve equal focal lengths at different wavelengths, the metasurface lens imparts a wavelength dependent phase contribution to compensate for the dispersive accumulated propagation phase. This is achieved by designing the dispersive phase response of coupled dielectric ridge patterned on a fused silica (SiO$_2$) substrate, as shown in \fref{Figure_18}(a). It creates a phase profile that realizes the same focal length for wavelengths at 1300, 1550, and 1800 nm as shown in \fref{Figure_18}(b-d). The focusing efficiency, defined as the ratio of power at the beam focal waist and the input power, is still rather low, in part due to the reflection loss. Few-layer metasurfaces introduced in previous sections could potentially address this issue of impedance mismatch and improve the focusing efficiency. For wavelengths other than these specific values, the operation of the lens follows the normal dispersion curves, which indicates that a dielectric metasurface lens that eliminates chromatic aberration over a broad range of wavelengths is still challenging to accomplish.  

\begin{figure}[htp]
\centerline{\includegraphics[width=3.2in]{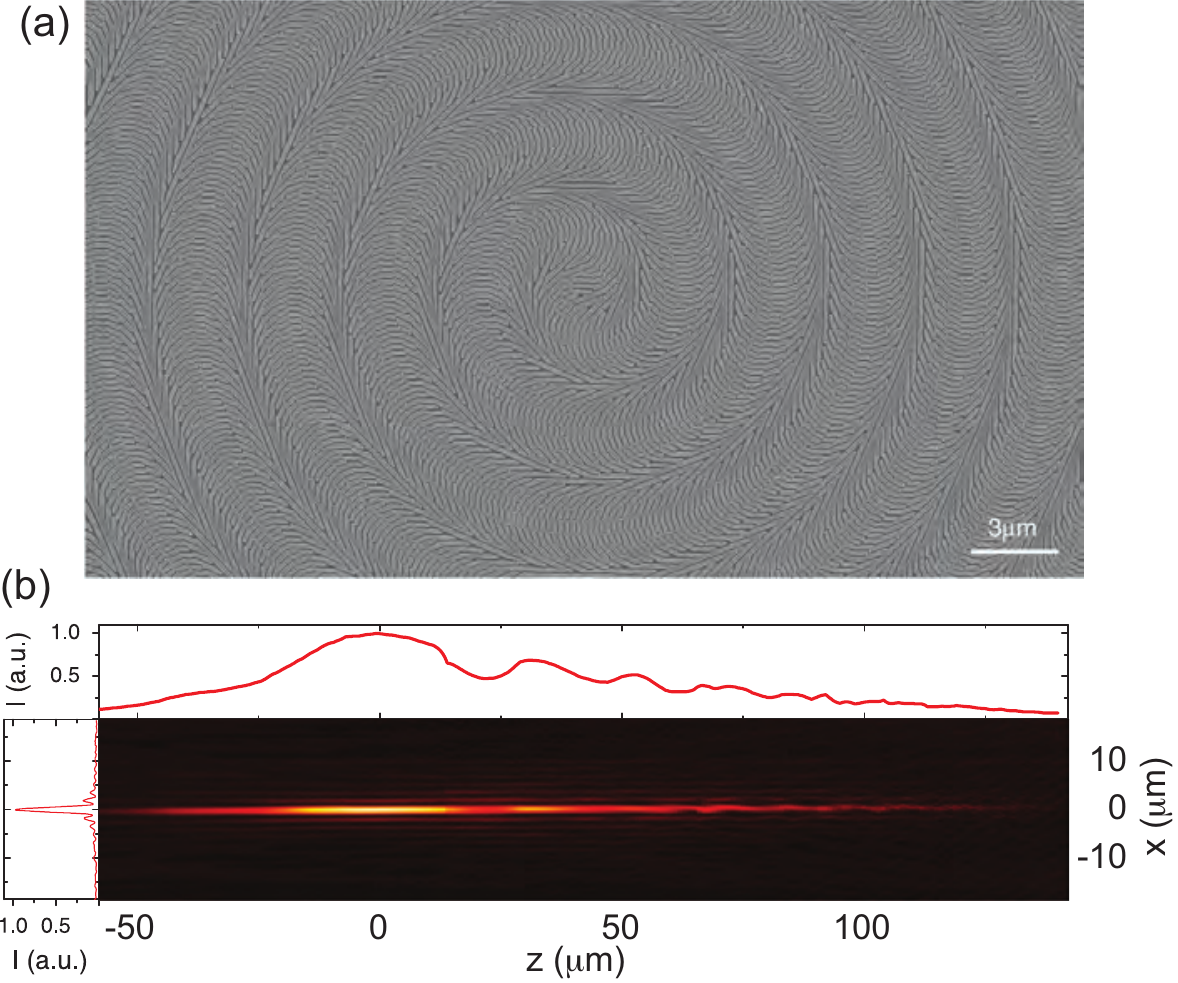}} \caption{(a) SEM image of a fabricated dielectric metasurface axicon consisting of silicon nanobeams. (b) Measured intensity profile of the nondiffractive Bessel beam generated behind the axicon in (a) in the \textit{xz} plane. Used with permission from~\cite{Lin_Brongersma_2014_Science}.}
\label{Figure_19}
\end{figure}

An alternative approach to create a spatially-varying phase profile is through the use of Pancharatnam-Berry phase~\cite{Bomzon_Hasman_2002_OL}. The key is the conversion between left- and right-handed circular polarization states via different routes on the Poincar\'{e} sphere. The required polarization control can be achieved by the excitation of electric and magnetic resonances in dielectric resonators. Using silicon nanobeams with appropriate geometric dimensions, it was shown that the incident circularly polarized light is partially converted into circularly polarized light with opposite handedness with an imparted Pancharatnam-Berry phase depending on the orientation of the silicon nanobeams~\cite{Lin_Brongersma_2014_Science}. The nanobeam metasurface exhibits anomalous refraction when forming a constant phase gradient. Linearly polarized incident light is split into right- and left-handed circularly polarized beams that propagate in different directions. Transmission spatial phase profiles have been also experimentally demonstrated, functioning as lenses for focusing and axicons for creating a Bessel beam (see \fref{Figure_19})~\cite{Lin_Brongersma_2014_Science}. 

\begin{figure*}[htp]
\centerline{\includegraphics[width=6in]{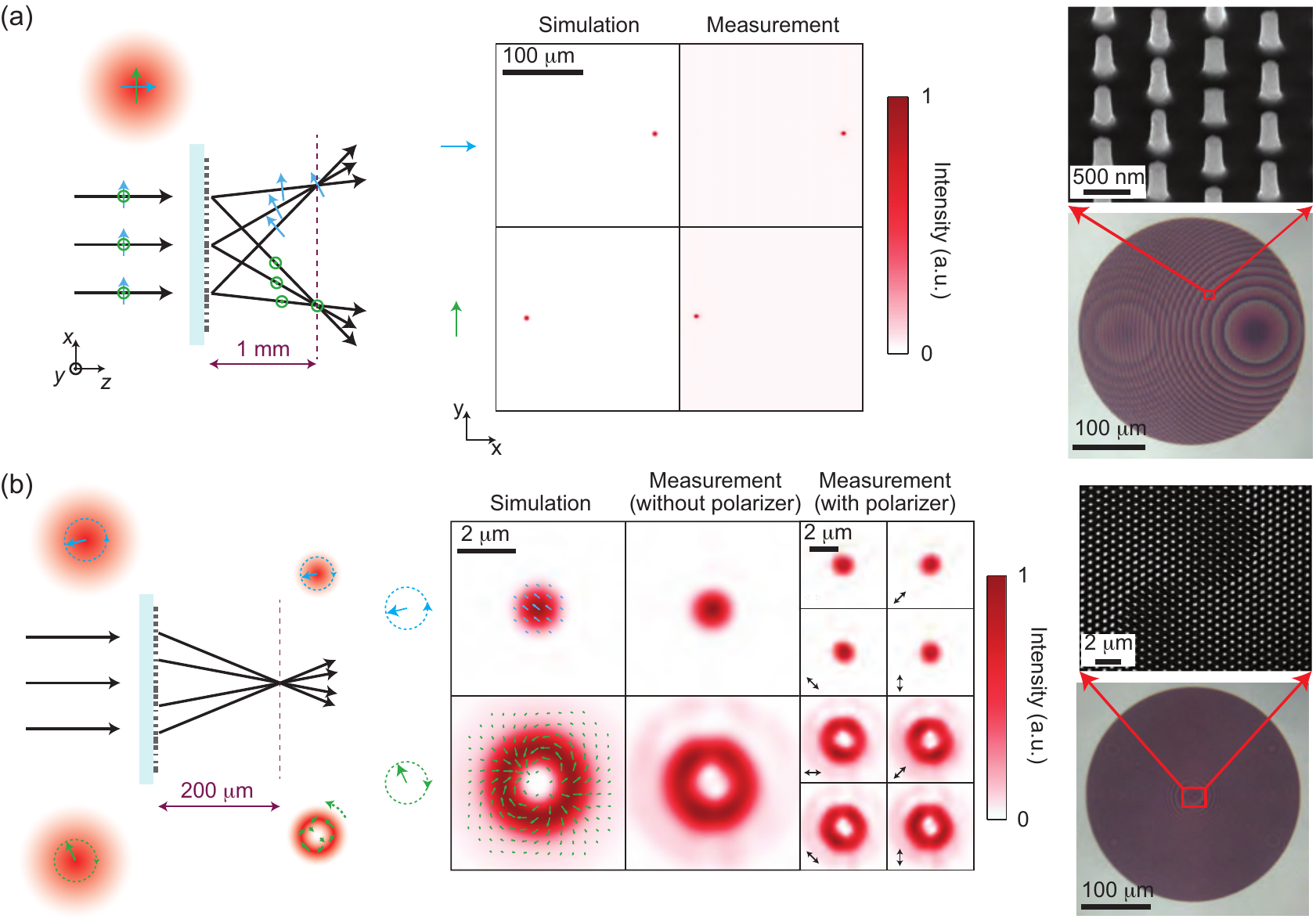}} \caption{(a) Dielectric metasurface that separates the \textit{x}- and \textit{y}-polarized incident light, deflecting and focusing them to two different spots. (b) Dielectric metasurface that focuses the incident circularly polarized light to a diffraction-limited spot or a doughnut-shaped spot depending on its handedness. Left column: schematic illustration of the devices; Mid-column: simulated and experimental results; Right column: SEM images of the dielectric metasurfaces. Used with permission from~\cite{Arbabi_Faraon_2015_NatNano}.}
\label{Figure_20}
\end{figure*} 
It is essential to realize simultaneous and complete control of polarization and phase with subwavelength resolution and high transmission. In the optical regime plasmonic metasurfaces partially accomplish this goal with limited efficiency~\cite{Li_Tian_2015_AFM}. In a recent paper from Faraon's group, a dielectric metasuface platform was demonstrated based on elliptical high-contrast dielectric nanoposts that provide complete control of transmissive polarization and phase with measured efficiency ranging from 72\% to 97\%, achieved through varying the ellipticity, size, as well as orientation of the nanoposts~\cite{Arbabi_Faraon_2015_NatNano}. It was shown that most free space high-performance transmissive optical elements can be realized, such as lenses, wave plates, beamsplitters, holograms and arbitrary vector beam generators. Two examples are illustrated in \fref{Figure_20} for incident polarization-dependent focusing.

\section{Metasurfaces for wave guidance and radiation} \label{Wave_Guidance_Radiation}
\begin{figure*}[htp]
\centerline{\includegraphics[width=6in]{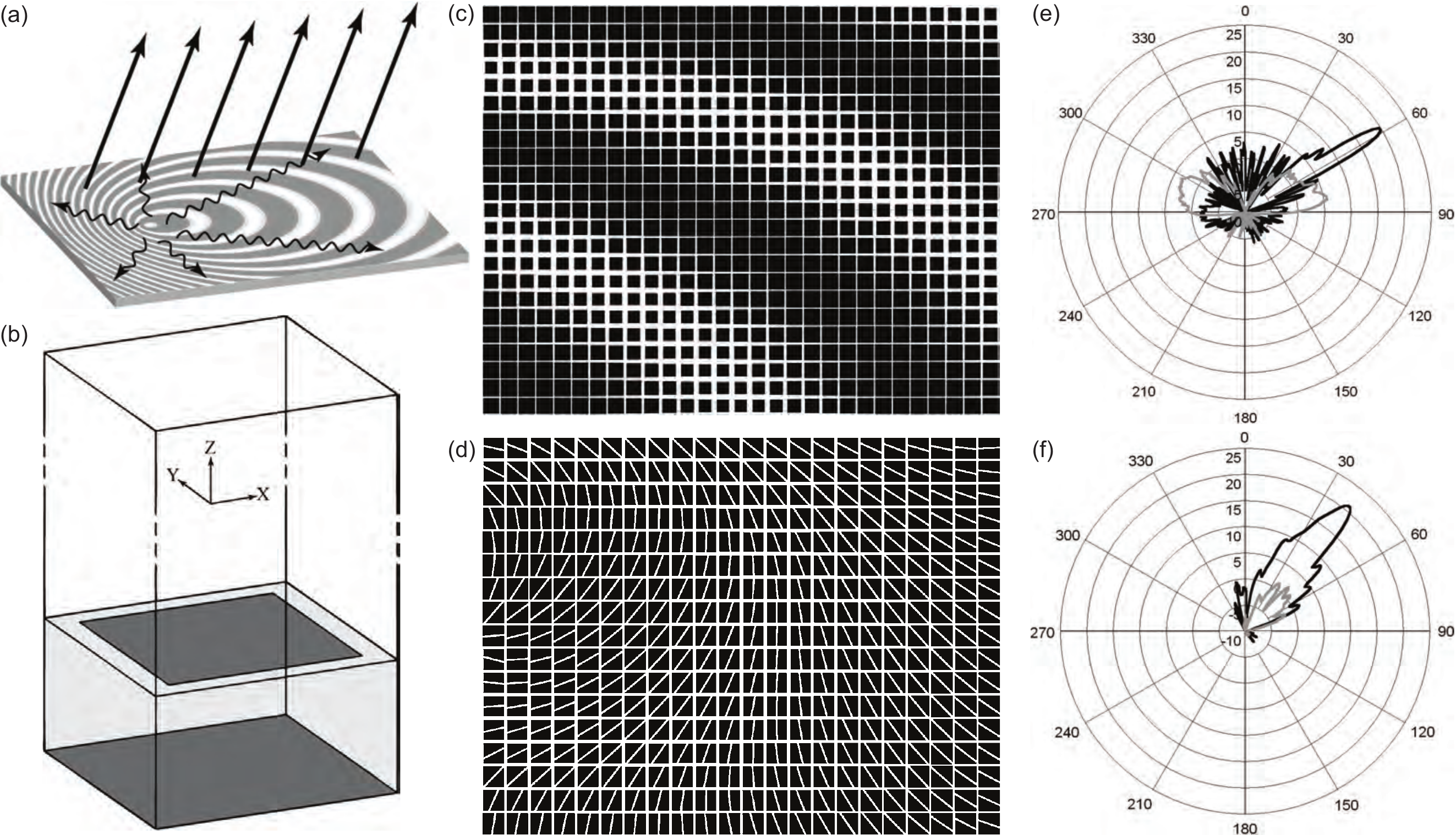}} \caption{(a) Schematic showing the concept of holographic leaky wave antenna. Surface waves (undulating arrows) are excited on a metasurface impedance surface, and are scattered by variations in the surface impedance to produce the desired radiation (straight arrows). (b) Unit cell of the impedance surface consisting of a patch antenna patterned on a metal grounded layer of insulator. (c) A section of the designed scalar impedance surface that scatters a cylindrical surface wave produced by a point source into a plane wave propagating along $60^\circ$ from the surface normal. (d) A section of the designed tensor impedance surface that scatters a cylindrical surface wave produced by a point source into a plane wave propagating along $45^\circ$ from the surface normal. (e) Black and gray curves are, respectively, radiation patterns of a monopolar antenna placed on the scalar holographic impedance surface and on a smooth metal surface. (f) Black and gray curves show, respectively, measured radiation patterns with left-handed and right-handed circular polarization produced by a monopolar antenna placed on the tensor holographic impedance surface. Used with permission from~\cite{Fong_Sievenpiper_2010_IEEE}. }
\label{Figure_21}
\end{figure*} 

In the previous sections we mainly focus on the physics and applications of metasurfaces in controlling waves that propagate in free space. The present section reviews the emerging research on using metasurfaces to control guided waves and to couple between guided waves and waves propagating in free space. Because of the spatial inhomogeneity of metasurfaces, they do not support any eigen guided modes. That is, waves propagating along metasurfaces are at a transient state and are constantly evolving. Thus, metasurfaces are most suitable for realizing mode conversions. By designing the in-plane effective wavevector using metasurface structures, one is able to realize conversion between two different guided modes or between a guided mode and a mode propagating in the free space. 

There are a couple of major differences between mode conversion using metasurfaces and using conventional grating-based mode convertors:

(1) Metasurfaces can be designed to provide a unidirectional phase gradient, or a unidirectional effective wavevector. The latter leads to an asymmetric coupling between modes: electromagnetic energy is transferred preferentially from one mode to the other, while the inverse process can be highly inefficient. Such asymmetric electromagnetic energy transfer between modes is maintained even when the conventional phase matching condition is not strictly satisfied (i.e., phase gradient $\mathrm{d}\Phi/\mathrm{d}r$ not equal to the difference in wave number between two modes, $\beta_1 - \beta_2$). This property ensures that mode conversion can be realized over a broad spectral range and won't be greatly affected by small structural changes to the metasurfaces. On the contrary, conventional grating couplers provide positive and negative reciprocal lattice vectors, $\pm 2\pi/\Lambda$, where $\Lambda$ is the grating period. The coupling between two modes is symmetric, and thus the phase matching condition, $\beta_1 - \beta_2 = 2\pi/\Lambda$, has to be strictly satisfied to ensure that electromagnetic energy is transferred from one mode to the other. 

(2) The spacing between adjacent constituent elements in a metasurface is subwavelength. Therefore, metasurfaces are able to modify the wavevector of a guided wave adiabatically. The absence of abrupt variation of wavevectors prevents scattering of electromagnetic energy into free space or into the substrate. Grating couplers, however, have a periodicity comparable to the wavelength. Guided waves are likely to be scattered, which makes in-plane confinement of electromagnetic energy a challenge.

\subsection{Coupling between free space and surface waves}
The pioneering work on using metasurfaces to control guided waves was conducted by Sievenpiper and colleagues in the microwave spectral range~\cite{Fong_Sievenpiper_2010_IEEE}. They used the concept of holography to design impedance surfaces that convert a given surface wave into a freely propagating wave with desired far-field radiation pattern and polarization. The impedance surface is essentially a hologram, which is the interference pattern between a reference beam and an object beam, and carries information of the phase, amplitude and polarization of the desired object beam. The object beam is reconstructed when the reference beam impinges on the hologram. In Sievenpiper's implementation, a source antenna produces the reference beam in the form of a surface wave, $E_\mathrm{surf}$, and the object beam is the desired wave, $E_\mathrm{rad}$, propagating in the half space above the surface (\fref{Figure_21}(a)); microwave holograms are created according to the interference pattern produced by the two waves and consist of a square lattice of dissimilar sub-wavelength conductive patches on a metal-grounded dielectric substrate. Both scalar and tensor forms of the impedance surfaces were experimentally demonstrated.

Surface impedance provides an appropriate language to characterize the properties of the metasurface. It is defined as the ratio between the electric and magnetic fields near the surface. For transverse magnetic (TM) waves (i.e., magnetic field transverse to the propagation direction) that propagate in the \textit{x}-direction, the surface impedance is $Z(x,y) = E_\mathrm{x}(x,y)/H_\mathrm{y}(x,y)$. The surface magnetic field is proportional to the
surface current, which is provided by the electromagnetic source. For example, a monopole antenna produces a cylindrical distribution of surface current. The function of the impedance surface is to translate this surface current to a distribution of electromagnetic waves on the surface, which matches the desired radiative wave. 

In Sievenpiper and colleagues' work, square patch antennas (\fref{Figure_21}(b)) were used to construct scalar impedance surfaces and square patches with an additional slice were used for tensor impedance surfaces. The three independent terms in the impedance tensor, $Z_\mathrm{xx}$, $Z_\mathrm{xy} = Z_\mathrm{yx}$ and $Z_\mathrm{yy}$, are controlled by the three degrees of freedom in antenna design: the slice width, its orientation angle, and the gap between neighboring square patches. In the case of scalar impedance surfaces, the value of surface impedance of patch antennas was determined by the following procedure:
\begin{enumerate}
\item Calculating dispersion relation of surface waves propagating on a 2D periodic array of patch antennas. Specifically, Bloch boundary conditions are applied to a unit cell of the impedance surface, and eigen surface wave modes and their eigen wavevectors are determined for a range of frequencies.
\item Calculating the surface impedance for a given operation frequency $\omega_0$, $Z(\omega_0) = \int_\mathrm{unit~cell}(E_\mathrm{x} / H_\mathrm{y}) \mathrm{d}x \mathrm{d}y$.
\end{enumerate}
A library that relates the surface impedances and patch antenna geometries can be created by repeating the above procedure for patch antennas of different sizes.

The distribution of surface impedance $Z(x,y)$ is determined by the following holographic technique. In the case of a scalar impedance surface, with a surface current $\mathbf{J}_\mathrm{surf}(x,y)$ produced by the electromagnetic source and the object far-field radiation $\mathbf{E}_\mathrm{rad}(x,y,z)$, the required surface impedance is
\begin{eqnarray}
Z(x,y) = j \left\{ X + M \rm{Re} \left[ (E_{rad,x}, E_{rad,y}) \left(
\begin{array}{c} 
J_\mathrm{surf,x}\\
J_\mathrm{surf,y}
\end{array} \right) \right] \right\}.\nonumber\\
\label{Eq_Z}
\end{eqnarray}
In the case of a tensor impedance surface, we have
\begin{eqnarray}
Z(x,y) &=& j \left( \begin{array}{c c}
X & 0\\
0 & X
\end{array} \right) \nonumber \\
 & & + j \frac{M}{2} \rm{Im} \left[
\left( \begin{array}{c}
E_\mathrm{rad,x}\\
E_\mathrm{rad,y}
\end{array} \right) (J_\mathrm{surf,x}, J_\mathrm{surf,y}) \right. \nonumber \\ 
& & \left. - \left(\begin{array}{c} 
J_\mathrm{surf,x}\\
J_\mathrm{surf,y}
\end{array} \right)  
(E_\mathrm{rad,x}, E_\mathrm{rad,y})
\right].
\label{Eq_Z_Tensor}
\end{eqnarray}

In the above two equations, $X$ represents the average impedance value, and $M$ spans the entire available impedance range. Using the holographic technique and the library of patch antennas, Sievenpiper and coworkers demonstrated a scalar impedance surface that scatters the current generated by a monopolar antenna into a linearly polarized plane wave propagating along $60^\circ$ from the surface normal (\fref{Figure_21}(c) and (e)). The surface current has a cylindrical distribution and can be described by $\mathbf{J}_\mathrm{surf} = \frac{1}{r^2}\exp(-j k_0 n_\mathrm{s} r) (x,y,0)$, where $r=(x^2+y^2)^{1/2}$, $k_0$ is the free space wavevector, and $n_\mathrm{s}$ is the effective index of the surface current, which is assumed to be a constant and is a function of the thickness and materials of the dielectric spacing layer between the metal patches and the metallic ground. They also experimentally demonstrated a tensor impedance surface that converts the current generated by a monopolar antenna to a circularly polarized far-field radiation propagating along $45^\circ$ direction (\fref{Figure_21}(d) and (f)).

\begin{figure}[htp]
\centerline{\includegraphics[width=3.2in]{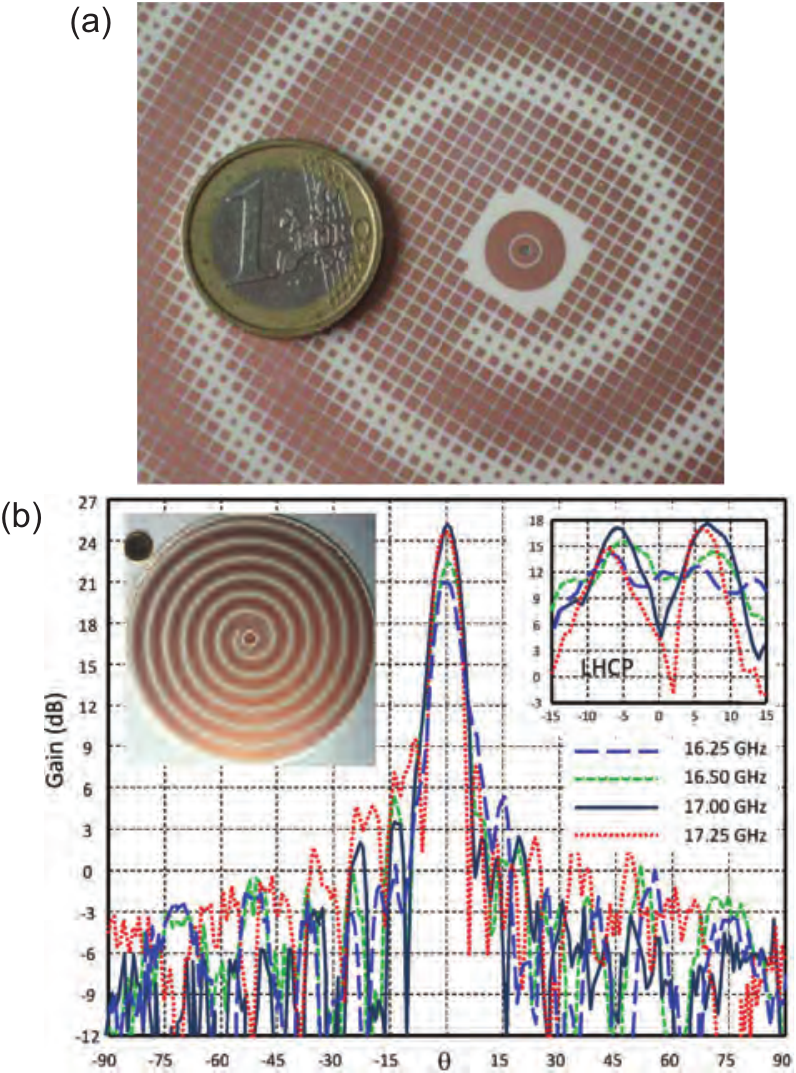}} \caption{(a) Section of an impedance surface near the central monopolar antenna. (b) Right-handed circularly polarized radiation profiles produced by the impedance surface antenna near 17 GHz. Inset is the entire antenna with a radius of 9.7 cm. Used with permission from~\cite{Maci_2011_IEEE}.}
\label{Figure_22}
\end{figure} 

\begin{figure}[htp]
\centerline{\includegraphics[width=3in]{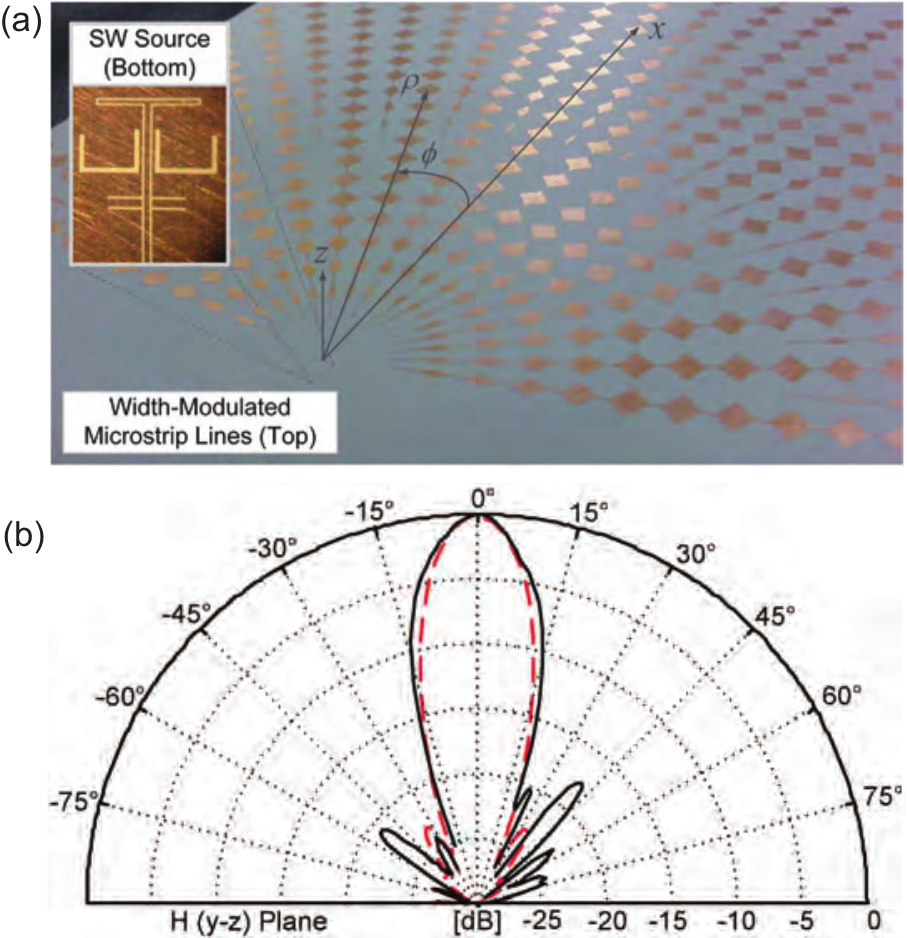}} \caption{(a) Planar 2D leaky-wave antenna consisting of radially directed and width-modulated microstrip lines and is able to transform a cylindrical surface wave into leaky waves. (b) Broadside beam pattern of the leaky-wave antenna. Solid and dashed curves are, respectively, measured and simulated beam patterns at $\sim 22$~GHz. Used with permission from~\cite{Podilchak_2013_IEEE}.}
\label{Figure_23}
\end{figure}

\begin{figure}[htp]
\centerline{\includegraphics[width=3.2in]{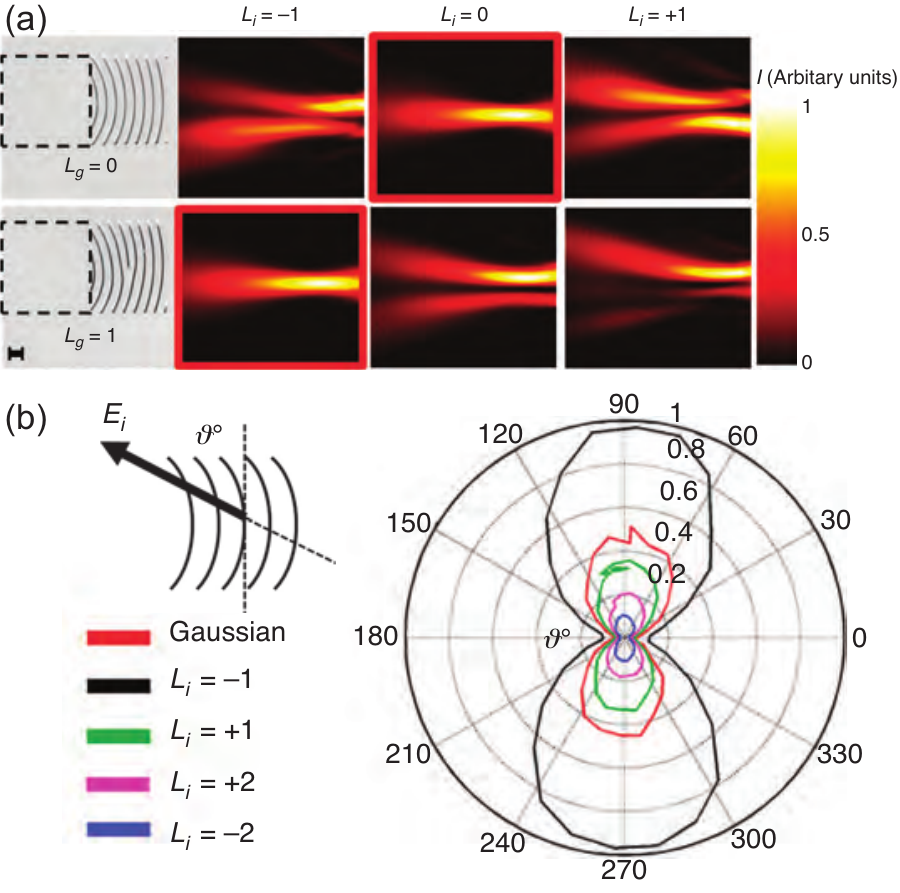}} \caption{(a) Left panel: metasurface holograms for detecting optical vortex beams. Right panel: simulation results of the intensity distribution of surface plasmon waves generated by illuminating the holograms at normal incidence with different optical vortex beams. (b) Photocurrent as a function of incident polarization measured for a metasurface hologram designed for detecting optical vortex beams with orbital angular momentum of $L_\mathrm{i} = -1$. Used with permission from~\cite{Genevet_Capasso_2012_NatCommun}.}
\label{Figure_24}
\end{figure}

Maci and colleagues used the same holographic principle to demonstrate metasurfaces with modulated surface impedance~\cite{Maci_2011_IEEE,Minatti_Maci_2011_IEEE}. They used square patch antennas of different sizes to create a spiral distribution of surface impedance that converts a surface current produced by a monopolar antenna to a collimated right-handed circularly polarized far-field radiation (\fref{Figure_22}). Podilchak and collaborators demonstrated experimentally~\cite{Podilchak_2013_IEEE} that width-modulated microstrip lines patterned on a grounded dielectric slab introduce a sinusoidally modulated surface impedance and provide appropriate conditions for leaky wave radiation (\fref{Figure_23}). 
\Fref{Figure_24} show a holographic metasurface that detects optical vortex beams with specific orbital angular momentum (OAM)~\cite{Genevet_Capasso_2012_NatCommun}. The nano-structured binary holograms shown in the left panel of \fref{Figure_24}(a) were created by calculating the interference pattern between a converging surface plasmon wave and an incident optical vortex beam. The simulated results in \fref{Figure_24}(a) show that a converging surface plasmon wave is generated only when an optical vortex beam with the correct OAM is scattered by the hologram. Experimental results in \fref{Figure_24}(b) show that a hologram can distinguish an optical vortex beam with OAM of $-1$ from optical vortex beams with other values of OAM. 

The major challenges in coupling an incident wave from free space into a surface wave with high efficiency are to suppress the reflection of the incident wave on the device surface and to prevent decoupling of the surface wave back into free space. In a series of work from the Zhou group~\cite{Sun_Zhou_2012_NatMater,Qu_Zhou_2013_EPL,Sun_Zhou_2016_LSA}, a few strategies were devised to address these challenges: (1) the entire surface of the metasurface coupler is designed to be impedance matched with free space to minimize direct reflection; (2) the lateral effective wavevector provided by the metasurface is designed to be sufficiently large so that a surface wave with a wavevector larger than the free space wavevector is excited; the surface wave becomes even more evanescent as it further interacts with the gradient metasurface, which prevents decoupling of the wave back to the free space; (3) the impedance mismatch between the supercells of the metasurface coupler is reduced to prevent scattering of the surface wave. Through these approaches, the authors were able to demonstrate coupling of an incident wave from free space into a surface wave with efficiencies of $\sim94\%$ in simulations and $\sim73\%$ in experiments using microwaves~\cite{Sun_Zhou_2016_LSA}.

\subsection{Control of surface waves}
\begin{figure*}[htp]
\centerline{\includegraphics[width=6in]{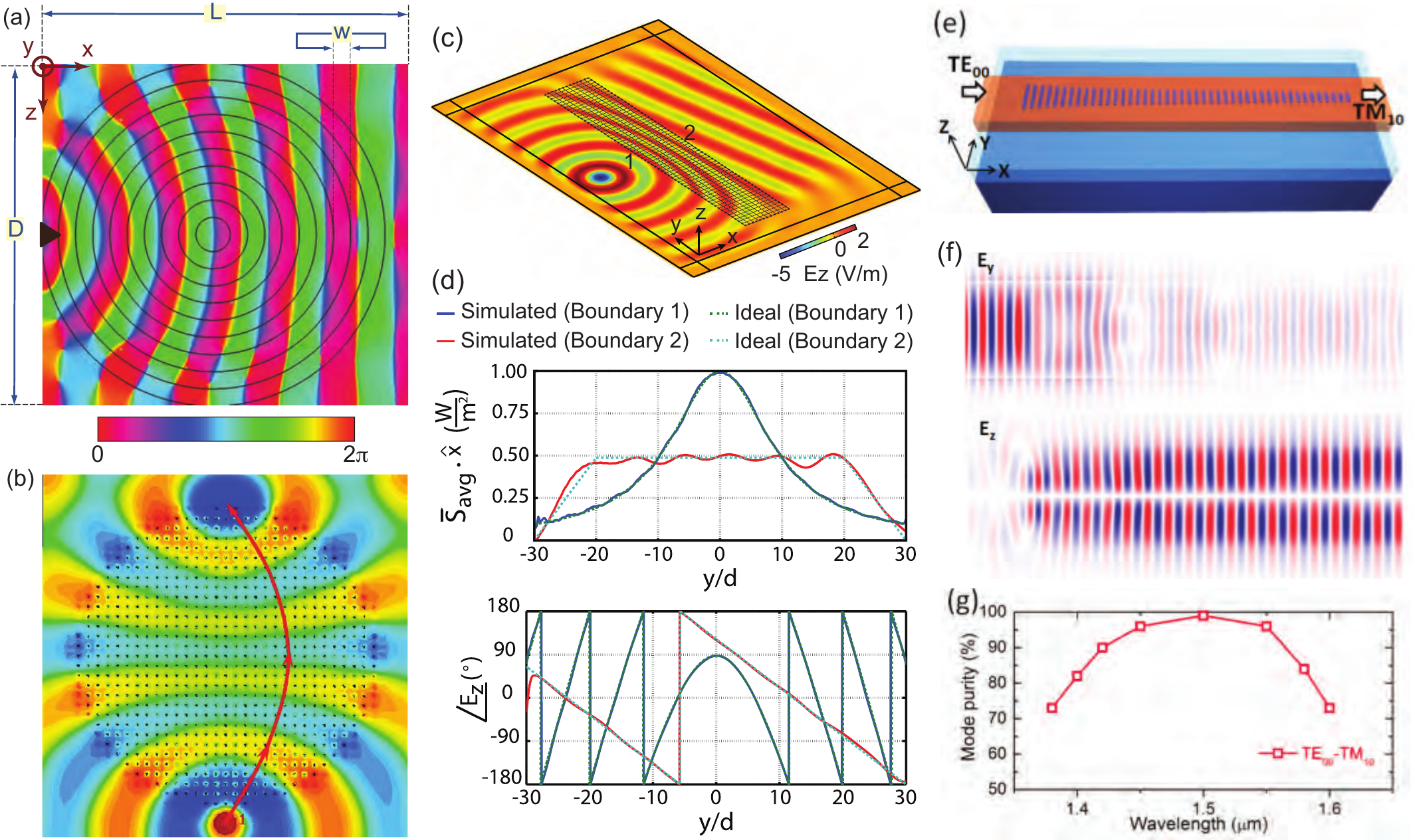}} \caption{(a) Luneburg lens based on graphene metasurface. Shown is the simulated phase of $E_\mathrm{y}$ of the surface plasmon at 30 THz on the graphene. $D = 1.5$~$\mu$m, $w = 75$~nm, and $L = 1.6$~$\mu$m. (b) Snapshot of the field in a metasurface Maxwell's fish-eye lens consisting of pins of different heights on a grounded slab and curvilinear trajectory of the real part of the Poynting vector. (c) Snapshot of simulated, vertical electric field ($E_\mathrm{z}$) of a metasurface that transforms a cylindrical surface wave into a surface wave with trapezoidal power density and linear phase progression. (d) Upper panel: simulated and ideal power densities along boundary 1 and boundary 2. Lower panel: phase profiles along boundary 1 and boundary 2. (e) Schematic of a telecom TE$_{00}$-to-TM$_{10}$ mode converter consisting of silicon phased array antennas patterned on a Si$_3$N$_4$ waveguide. The phase response is due to the optical Mie resonance in the silicon nanorod. (f) Simulated field evolution in the mode converter. (g) Purity of the converted TM$_{10}$ mode as a function of wavelength, showing that the mode converter works over a broad wavelength range. (a) used with permission from~\cite{Vakil_Engheta_2011_Science}, (b) used with permission from~\cite{Maci_2011_IEEE}, (c) and (d) used with permission from~\cite{Gok_Grbic_2013_PRL}, (e)-(f) used with permission from~\cite{Yu_2014_CLEO,Li_Yu_2014_CLEO}.}
\label{Figure_25}
\end{figure*} 

The examples that have been discussed in this section so far are all about coupling surface waves and waves propagating in free space using metasurfaces. The subject of controlling the propagation of surface waves confined to a 2D plane is a new frontier of metasurface research.

Vakil and Engheta proposed using graphene as an ultra-thin platform for controlling in-plane propagation of infrared electromagnetic waves~\cite{Vakil_Engheta_2011_Science}. They demonstrated theoretically that by designing and manipulating spatially inhomogeneous conductivity patterns on a sheet of graphene using the electric field effect, one can realize a number of transformation optical devices. The example of a graphene metasurface Luneberg lens is shown in \fref{Figure_25}(a). The research group of S. Maci used patch antennas of different sizes or metallic pins of different heights to demonstrate two types of in-plane planar lenses (Luneberg lenses and Maxwell's fish-eye~\cite{Maci_2011_IEEE}; the latter is shown in \fref{Figure_25}(b)). In their pioneering work, Gok and Grbic used the concept of transformation electromagnetics to demonstrate independent control of the power flow and phase progression of electromagnetic fields in a 2D space (\fref{Figure_25}(c) and (d))~\cite{Gok_Grbic_2013_PRL}. The resulting metasurface is a highly inhomogeneous, anisotropic media where each unit cell is characterized by a $2 \times 2$ permeability tensor in the plane and a scalar permittivity in the surface normal direction. These parameters were judicially chosen to create stipulated 2D distributions of wavevector and Poynting vector, as well as to ensure impedance matching between the adjacent unit cells so that there is no reflection and scattering of the surface wave as it propagates on the metasurface.

The concept of metasurfaces has been introduced into the field of integrated photonics where 1D phased array antennas patterned on optical waveguides enable the control of optical power flow and mode coupling in the waveguides. The 1D antenna array introduces a unidirectional phase gradient $\mathrm{d}\Phi / \mathrm{d}x$, where $\mathrm{d}\Phi$ is the difference in phase response between adjacent antennas that are separated from each other by a subwavelength distance of $\mathrm{d}x$. The phase gradient is equivalent to a unidirectional effective wavevector $\Delta \mathbf{k}$ along the waveguide, which leads to directional coupling of waveguide modes. That is, optical power couples preferentially from one waveguide mode to a second waveguide mode, whereas optical coupling from the second mode back to the first one is highly inefficient. As a result, the phase matching conditions are greatly relaxed, which enables the demonstration of extremely broadband and robust waveguide mode conversion (\fref{Figure_25}(e-g))~\cite{Yu_2014_CLEO,Li_Yu_2014_CLEO}.

\section{Active metasurfaces} \label{Active_Metasurfaces}

Active devices and components play a critical role in modern electromagnetic and photonic systems. Active control of metamaterials and metasurfaces extends their exotic passive properties by allowing fine resonance tuning to adapt to the operational conditions, and enabling a switchable resonant response, for instance, for signal modulation in communication and imaging. Furthermore, the concentration of optical power in metasurface resonators integrated with optical nonlinear materials can dramatically enhance the nonlinear response, as predicted in Pendry's original work on SRRs~\cite{Pendry_1999_IEEE}. As compared to bulk metamaterials, the planar configuration of metasurfaces facilitates the integration of active functional materials. A variety of functional materials providing tunable refractive indices through thermal excitation, voltage bias, magnetic field, optical pump, or mechanical deformation have been successfully incorporated into metasurfaces. In particular, semiconductors and graphene become the materials of choice for electrically tunable active metasurfaces. 

\subsection{Actively switchable and frequency tunable metal/semiconductor hybrid metasurfaces}
 \begin{figure*}[htp]
\centerline{\includegraphics[width=6in]{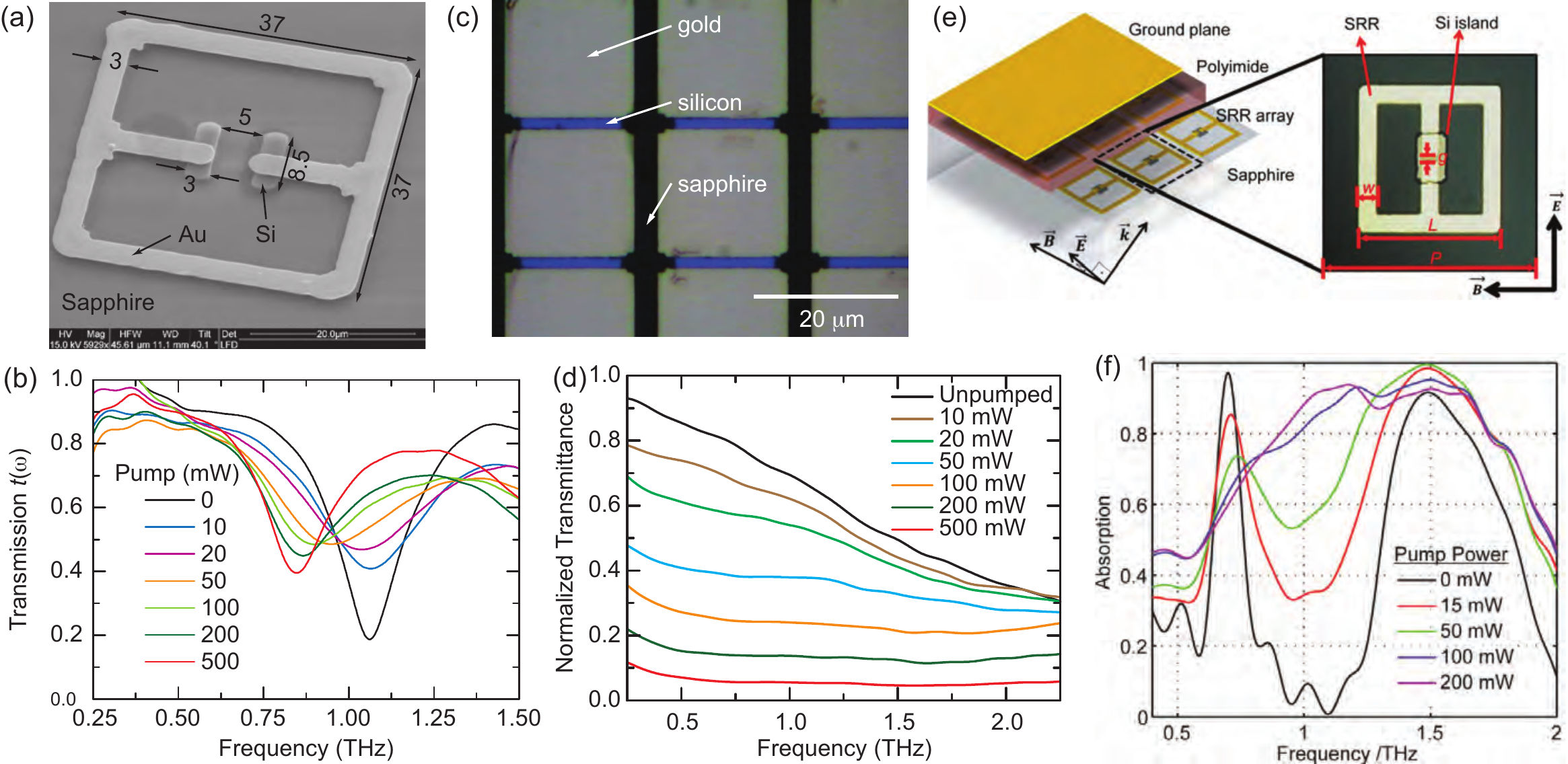}} \caption{Optically tunable THz metal/semiconductor hybrid metasurfaces. (a) SEM image of an electric SRR unit cell of a frequency tunable THz metasurface. (b) Upon photoexcitation of the silicon bars, the gap capacitance increases, which results in a lower resonance frequency. (c) Optical microscopy image of an ultra broadband THz modulator. (d) Without photoexcitation, the transmission is high at the low frequency side of the Lorentzian resonance of the gold grid; upon photoexcitation of the silicon region, it becomes effectively a wire grating showing low transmission. (e) Schematic (left panel) and optical microscopy image (right panel) of a THz metamaterial absorber consisting of silicon pads integrated at the gaps of SRRs. (f) Photoexcitation dramatically tunes the property from dual-band to a broadband absorption. (a) and (b) used with permission from~\cite{Chen_2008_NatPhoton}, (c) and (d) used with permission from~\cite{Heyes_Chen_2014_APL}, (e) and (f) used with permission from~\cite{Seren_Averitt_2014_AOM}.}
\label{Figure_26}
\end{figure*} 

The conductivity of semiconductors can be increased by orders of magnitude through doping, and thus semiconductors can be converted into plasmonic materials in the infrared and spectral ranges with longer wavelengths. Active tuning of the conductivity can be realized by carrier injection and depletion through photoexcitation and voltage bias. Such a unique capability makes semiconductors ideal materials for integration into metamaterial structures to accomplish active and dynamic functionalities, particularly in the microwave and THz frequency range. Varactor diodes have been widely used to realize frequency tunable and nonlinear response~\cite{Shadrivov_Kivshar_2006_OE,Zhu_Feng_2013_SciRep} in microwave metasurfaces. At THz frequencies, SRR arrays can be directly fabricated on top of semiconducting substrates such as intrinsic silicon and gallium arsenide, and the resonant response can be tuned through photoexcitation of free charge carriers at the substrate surface~\cite{Padilla_Averitt_2006_PRL}, resulting in an ultrafast switching speed~\cite{Chen_2007_OL}. Furthermore, semiconductors can be used as part of the resonant structure. In this case, photoexcitation dynamically modifies the structural geometry of the resonator, enabling switchable or frequency tunable response. As shown in \fref{Figure_26}(a), a pair of silicon bars form a part of the capacitive gap in an electric SRR unit cell. Under photoexcitation with near-infrared light, the silicon bars become metallic, which increases the SRR capacitance. Therefore, the frequency of the SRR \textit{LC} resonance is tuned to a lower frequency with the tuning range of about $20\%$~\cite{Chen_2008_NatPhoton}, as shown in \fref{Figure_26}(b). A variety of similar structures were demonstrated, resulting in a blue shift of the resonance frequency~\cite{Shen_Soukoulis_2011_PRL}. 

Optically modifying the metasurface geometric structure enables the transition between different types of resonances. In \fref{Figure_26}(c) silicon is integrated at the gaps of a metal patch array that exhibits a dipolar resonance without photoexcitation and allows high transmission below the resonance frequency. Under photoexcitation, the metallic silicon connects the metal patches, effectively forming a metal wire grating that blocks the low frequency THz waves, as shown in \fref{Figure_26}(d), and resulting in ultra broadband THz modulation~\cite{Heyes_Chen_2014_APL}. Recently, optically tunable THz metamaterial perfect absorbers~\cite{Seren_Averitt_2014_AOM} were demonstrated, as shown in \fref{Figure_26}(e) and (f), where silicon islands are located at the gaps of electric SRRs. Using such an approach, a variety of optical responses can be switched/tuned via photoexcitation, such as the handedness of chiral metasurfaces~\cite{Zhang_Zhang_2012_NatCommun,Zhou_OHara_2012_PRB} and plasmonic electromagnetically induced transparency (EIT)~\cite{Gu_Zhang_2012_NatCommun}.

\begin{figure*}[htp]
\centerline{\includegraphics[width=6in]{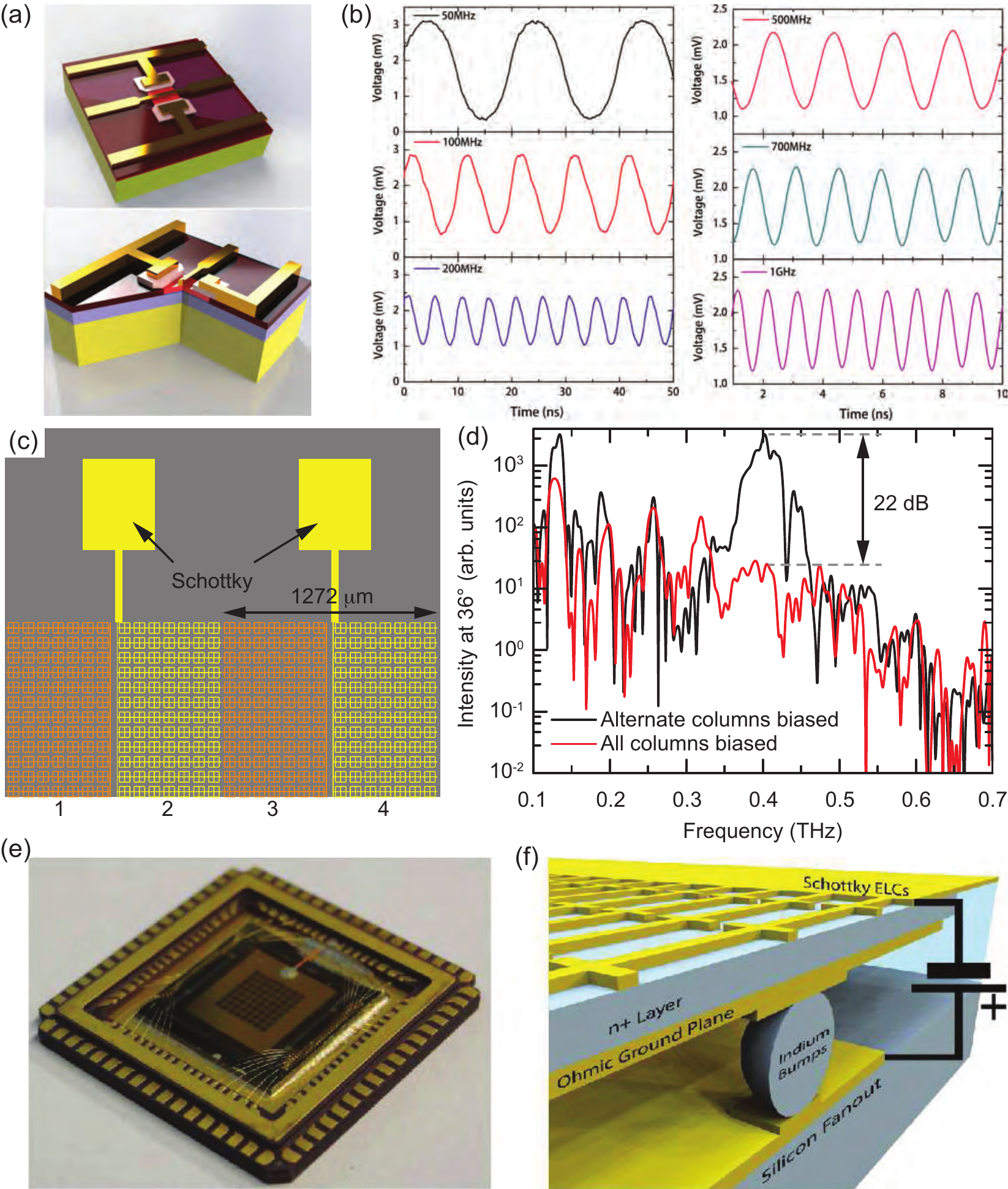}} \caption{Electrical modulation of metal/semiconductor hybrid metasurfaces. (a) Schematic of a unit cell of a high-speed THz metasurface modulator based on double-channel heterostructures. (b) Modulation performance at different frequencies. (c) Schematic of an active THz metasurface diffraction grating formed by 32 columns controlled by independent voltage biases, where different colors indicate different voltage biases. (d) Dynamic diffraction is enabled by applying reverse voltage biases to alternate columns. An unprecedented modulation depth of 22 dB is accomplished at the designed operation frequency of 0.4 THz. (e) THz spatial modulator with $8 \times 8$ pixels based on electrically switchable THz metasurface absorbers and used for THz compressive imaging, and (f) the corresponding device schematic consisting of a linked array of metallic resonators making Schottky contacts with an underlying n-doped semiconductor spacer, a metal ground plane serving as the ohmic contact, as well as the accessory structure enabling independent voltage biases to the pixels. (a) and (b) used with permission from~\cite{Zhang_Liu_2015_NL}, (c) and (d) used with permission from~\cite{Karl_Mittleman_2014_APL}, (e) and (f) used with permission from~\cite{Shrekenhamer_Padilla_2013_AOM}.}
\label{Figure_27}
\end{figure*}

Semiconducting hybrid metasurfaces feature electrical tuning of resonances via the application of a voltage bias, which is more convenient and practical for applications. The most prominent examples are the integration of varactor diodes for microwaves and Schottky junctions for THz frequencies. The first demonstration of an electrically switchable THz metasurface featured an unprecedented 50\% modulation depth~\cite{Chen_Averitt_2006_Nature}, which was further improved to 80\% through structural optimization~\cite{Chen_2009_NatPhoton}. Together with the causally connected phase modulation (up to 0.55 rad), this device allows broadband THz modulation~\cite{Chen_Averitt_2006_Nature} that can be used to replace a mechanical optical chopper in a lock-in THz detection scheme with modulation speed in the MHz range~\cite{Chen_2008_APL_FastModulation,Shrekenhamer_Padilla_2011_OE,Shrekenhamer_Padilla_2013_AOM}, limited either by the large device area accompanied by high stray capacitance or parasitic capacitance from the bonding electrodes and wires. Very recently, GHz electronic modulation speed has been demonstrated by using double-channel heterostructures supporting nanoscale 2DEGs with high carrier concentration and mobility~\cite{Zhang_Liu_2015_NL}, shown in \fref{Figure_27}(a). Through designing a composite hybrid metasurface structure to reduce the stray capacitance, 1 GHz modulation speed, 85\% modulation depth (\fref{Figure_27}(b)), and a phase shift of 1.19 rad were experimentally realized during real-time dynamic tests. Furthermore, a wireless free space modulation THz communication system based on this external THz modulator was tested using 0.2 Gbps eye patterns. This accomplishment opens an avenue toward the development of high performance THz wireless communication and imaging systems. 

In recent work, an electrically driven THz metasurface active diffraction grating was demonstrated to realize background-free THz modulation with an unprecedented 22 dB of dynamic range~\cite{Karl_Mittleman_2014_APL}. Each ``grating finger'' consists of an array of electrically connected and switchable SRRs forming a column that is controlled by an independent voltage bias, as shown in \fref{Figure_27}(c). The diffractive metasurface grating is created by applying a voltage bias to alternate columns within the 32-column metasurface structure, resulting in a frequency dependent diffraction angle for the incident broadband THz radiation. At the metasurface resonance frequency of 0.4 THz, the diffraction is strongest because of the largest transmission contrast between two neighboring columns. However, when the same voltage bias is applied to each of the columns, the structure behaves as a uniform metasurface with no observable diffraction. Therefore, application of an AC voltage to alternate columns results in background-free diffractive modulation of the incident THz radiation, as shown in \fref{Figure_27}(d). 

Spatial light modulators have been realized by pixelating the metasurface for independent control of reflection, transmission, or their phase. A prototype THz metasurface spatial light modulator with $4 \times 4$ pixels was realized to demonstrate reconfigurable interference patterns of double slits~\cite{Chan_Mittleman_2009_APL}. THz metasurface spatial light modulators with a larger number of pixels are possible, although the increasing number of electrical connecting wires makes them more complicated. One solution to this problem is a reflection-mode metasurface spatial modulator based on an electrically tunable metamaterial absorber structure~\cite{Shrekenhamer_Padilla_2013_AOM}. As shown in \fref{Figure_27}(f), a linked array of resonators and an underlying semiconductor layer create Schottky junctions, and a metal ground plane serves as the ohmic contact. Application of a reverse voltage bias enables tuning the frequency of the resonant absorption, with modulation speeds up to 10 MHz. This type of THz spatial modulators based on metamaterial absorbers, shown in \fref{Figure_27}(e), have been recently successfully employed in THz compressive imaging~\cite{Watts_Padilla_2014_NatPhoton}.

\subsection{Graphene hybrid metasurfaces}
\begin{figure*}[htp]
\centerline{\includegraphics[width=6in]{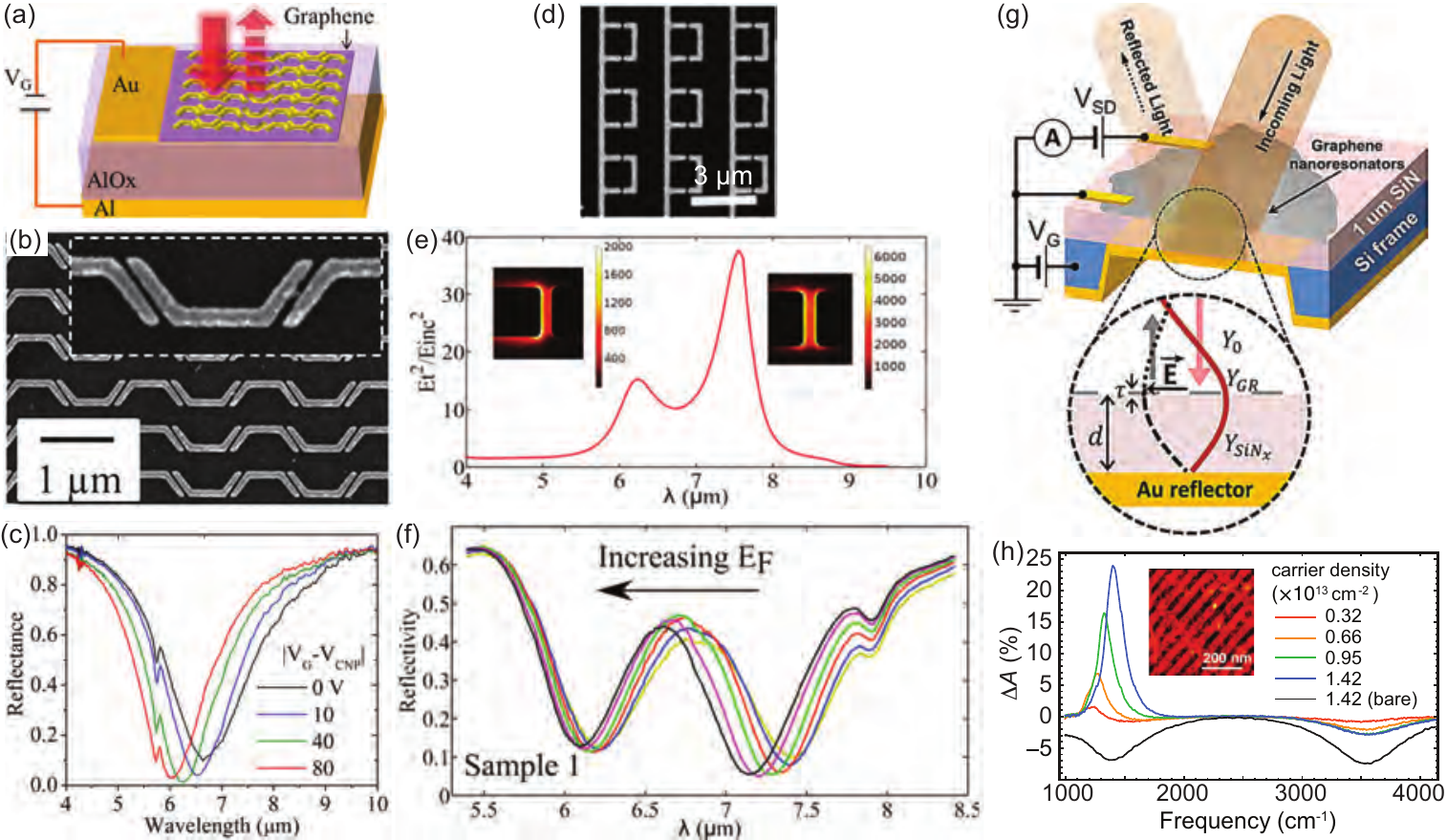}} \caption{Electrically tunable graphene-based metasurfaces. (a) Schematic of an ultrathin mid-infrared modulator based on a tunable metasurface absorber. (b) SEM image of the metasurface absorber. Inset: a zoomed-in view of a portion of the device. (c) Measured reflection spectra (normalized to the reflection spectrum of an aluminum mirror) of the metasurface absorber in (b) at different gate voltages $|V_\mathrm{G} - V_\mathrm{CNP}|$, where $V_\mathrm{CNP}$ is the gate voltage when the concentrations of electrons and holes in the graphene sheet are equal, i.e., charge neutral point (CNP). (d) SEM image of a metasurface structure exhibiting dual Fano resonances. (e) Average near-field intensity enhancement $\eta$ on graphene surface. Insets: spatial distribution of $\eta$ inside the gap for the two Fano resonances. (f) Measured reflection spectra of the device in (d) as a function of the Fermi energy. (g) Schematic of a metasurface modulator based on a graphene Salisbury screen. The inset illustrates the device with the optical waves at the resonance condition. (h) Change in absorption with respect to the absorption at CNP in 40-nm-wide graphene nanoapertures at various doping levels. The solid black curve corresponds to bare (unpatterned) graphene. The inset: AFM image of graphene nanoapertures with 40~nm width. (a)-(c) used with permission from~\cite{Yao_Capasso_2014_NL}, (d)-(f) used with permission from~\cite{Dabidian_Shvets_2015_ACSPhoton}, (g) and (h) used with permission from~\cite{Jang_Atwater_2014_PRB}. 
}
\label{Figure_28}
\end{figure*}

Except for fabrication of metallic metasurface structures directly on a substrate such as those shown in \fref{Figure_27}(a) and (c), integration of crystalline semiconductor films or islands into the critical regions of more complex metasurfaces (e.g., the structure shown in \fref{Figure_26} and \fref{Figure_27}(f) as well as other few-layer metasurfaces) poses significant fabrication challenges~\cite{Fan_Averitt_2013_IEEETHz} mainly due to the requirement of nano-lithography or transferring fragile semiconductor thin films. In this sense, the excellent mechanical properties and the tunable carrier density of graphene make it an excellent material to enable active metasurfaces~\cite{Emani_Boltasseva_2015_Nanophoton}. Graphene has largely tunable optical conductivity in the mid-infrared and THz frequency ranges. The doping of graphene can be adjusted through changing the bias voltage by a factor of ~10 at room temperature, which leads to a large change in its sheet conductivity $\sigma$ and therefore the in-plane electric permittivity $\epsilon_\parallel = 1 + i \sigma / (\epsilon_0 \omega t)$, where $t = 0.33$~nm is the thickness of single-layer graphene.

The resonant response of metasurfaces is of particular importance to enhance interactions between atomically thin graphene sheets and mid-infrared and THz radiation. Metallic plasmonic antennas are able to capture light from free space and concentrate optical energy into subwavelength spots. The electric field at these spots can be two to three orders of magnitude larger than the incident field. By placing graphene in the hot spots created by metallic plasmonic antennas and by tuning the optical conductivity of graphene, one can switch the resonance or tune the resonance frequency of the composite over a wide range. In the THz frequency range, intraband transitions in graphene sheets have been used to demonstrate broadband electrical modulation~\cite{Sensale-Rodriguez_Xing_2012_NatCommun}, and patterned graphene structures have been shown to exhibit resonant plasmonic response~\cite{Ju_Wang_2011_NatNano,Abajo_2014_ACSPhotonics}. Integrating graphene into metallic resonators has enabled the demonstration of THz metasurface electrical modulators~\cite{Lee_Zhang_2012_NatMater,Gao_Xu_2014_NL,Miao_Zhou_2015_PRX}. 

Mid-infrared metasurfaces with electrically tunable spectral properties have been experimentally demonstrated by controlling the carrier density of graphene~\cite{Yao_Capasso_2013_NL_Graphene}. Optimizing optical antenna designs has improved both the frequency tuning range and the modulation depth~\cite{Yao_Capasso_2013_NL}. In \fref{Figure_28}(a), the upper metasurface layer is separated from a back aluminum mirror by a thin aluminum oxide film. Such a reflect-array structure exhibits nearly perfect absorption~\cite{Yao_Capasso_2014_NL}. That is, at the OFF state the reflection is nearly zero, and the frequency at which near-zero reflection occurs can be tuned by applying a voltage bias that modifies the dispersion of the top graphene-antenna array (\fref{Figure_28}(c)). This approach provides a modulation speed in the tens of MHz range and an optical modulation depth close to 100\% with the latter defined as $1 - R_{min}(\lambda) / R_{max}(\lambda)$ where $R_{min}(\lambda)$ and $R_{max}(\lambda)$ are the minimum and maximum achievable reflectivity at a certain wavelength $\lambda$~\cite{Yao_Capasso_2014_NL}. 

A narrow spectral width is essential for realizing a high modulation depth based on resonance frequency tuning. For this purpose one may integrate graphene into metasurfaces that exhibit high Q-factor Fano resonances, such as the one shown in \fref{Figure_28}(d), which consists of an array of connected dipole and monopole resonators fabricated on top of a thin silicon dioxide layer on a silicon substrate~\cite{Dabidian_Shvets_2015_ACSPhoton}. This metasurface structure exhibits double plasmonic electromagnetically induced transparency (EIT) as illustrated by the two near-field intensity enhancement peaks in \fref{Figure_28}(e) and two reflection minima in \fref{Figure_28}(f). The graphene-SiO$_2$-silicon structure also enables back-gating to tune the graphene carrier density, which consequently tunes the Fano resonances (\fref{Figure_28}(f)). At a specific wavelength, it results in high reflection ``ON'' and low reflection ``OFF'' states with an experimentally measured modulation depth up to 90\%, though the insertion loss of 81\% is still rather high and the bandwidth is also rather small (a few percent of the operational wavelength)~\cite{Dabidian_Shvets_2015_ACSPhoton}. Phase modulation has been also shown recently in a similar graphene hybrid metasurface in the mid-infrared, which can be potentially used for motion sensing and tunable waveplates~\cite{Dabidian_Shvets_2015_arXiv}. 

Electrically tunable metasurfaces can be made of structured graphene sheets without utilizing metallic plasmonic antennas. \Fref{Figure_28}(g) shows a reflect-array mid-infrared modulator consisting of graphene nanoapertures (i.e., voids cut into a graphene sheet) separated from a metallic back mirror by a thin film of Si$_3$N$_4$~\cite{Mousavi_Shvets_2013_NL}. The width of the nanoapertures is chosen to be in the range of 20-60 nm, so that incident mid-infrared light can excite plasmonic resonances in the nanoapertures, leading to strongly enhanced optical absorption of up to 25\%. As the bias voltage changes the carrier doping of the perforated graphene sheet, the plasmonic resonances shift, giving rise to tunable amplitude and spectral position of the absorption peaks (\fref{Figure_28}(h)).

\subsection{Other resonance switchable and frequency tunable metasurfaces }
\begin{figure}[htp]
\centerline{\includegraphics[width=2.5in]{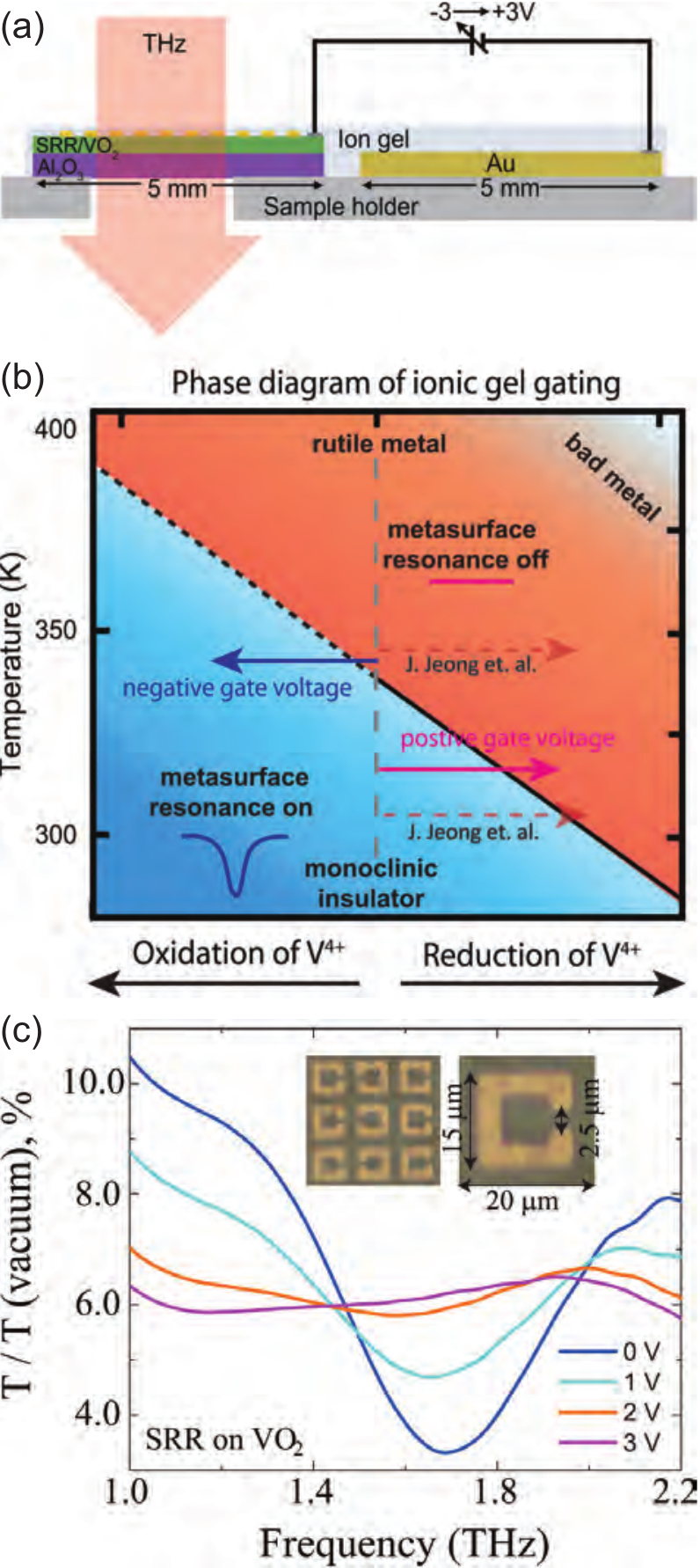}} \caption{(a) Schematic showing electrolyte gating of VO$_2$-based metasurfaces. (b) Phase diagram of VO$_2$. (c) Voltage dependent THz transmission spectra of the device in (a) at 315K. Insets: photos of gold SRRs sitting on VO$_2$ before the ionic gel is applied. Used with permission from~\cite{Goldflam_Basov_2014_APL}.
}
\label{Figure_29}
\end{figure}
While electrically tunable metasurfaces integrated with semiconductors and graphene are of utmost importance in applications, there are a variety of other functional materials and structures that have been successfully used to realize active metasurfaces. When a THz metallic SRR array was fabricated directly on top of a strontium titanate (STO) substrate (a phase transition material), the resonance frequency experiences a red-shift with decreasing temperature, due to the increasing refractive index of the STO substrate, although it suffers from significant insertion loss due to the high refractive index of the STO substrate~\cite{Singh_Chen_2011_OL}. Vanadium dioxide (VO$_2$) exhibits thermally driven insulator-to-metal phase transition and has attracted great interest in realizing thermally active metasurfaces at THz and infrared frequencies~\cite{Driscoll_Basov_2008_APL,Driscoll_Basov_2009_Science,Dicken_Atwater_2009_OE,Wen_2010_APL,Goldflam_Basov_2011_APL,Kats_Capasso_2013_OL}. The hysteresis of its phase transition has been utilized to demonstrate metasurface memory devices~\cite{Driscoll_Basov_2009_Science}. The resonance of metasurfaces based on VO$_2$ can be also switched through the application of a voltage bias~\cite{Goldflam_Basov_2014_APL}. In the latter case, a thin layer of ionic gel was applied on the surface of the metasurface (\fref{Figure_29}(a)). Application of positive (negative) voltage selectively tunes the metasurface resonance into the ``OFF'' (``ON'') state by inducing the VO$_2$ film into a more conductive (insulating) state. In particular, a positive voltage drives the following electrochemical reaction: VO$_2$ + $2x$ e$^-$ $\rightarrow$ VO$_{2-x}$ + $x$ O$^{2-}$, so that VO$_2$ goes across the boundary between the insulating and metallic states following the pink solid arrow in \fref{Figure_29}(b). As a result, application of positive voltages damps the SRR resonance, with 3 volts yielding complete suppression of the resonance (\fref{Figure_29}(c)). 

Liquid crystals can be also conveniently integrated with metasurface structures, and help realize electrically tunable spectral properties when the refractive index of liquid crystals is adjusted~\cite{Zhao_Zhou_2007_APL,Shrekenhamer_Padilla_2013_PRL}. The frequency tuning range is, however, quite limited and the operation speed is slow. THz superconducting metasurfaces consisting of resonant elements made of superconducting films instead of the typically used metals have shown outstanding switching and frequency tuning behaviors via thermal control~\cite{Chen_Taylor_2010_PRL_YBCO,Gu_Zhang_2010_APL,Wu_Jin_2011_OE,Kurter_Soukoulis_2011_PRL} or photoexcitation~\cite{Singh_Chen_2012_Nanophoton}, although this only applies to microwave and THz frequencies, limiting the applications of active superconducting metasurfaces. Last but not least, integration of micro-electro-mechanical systems (MEMS) into metasurfaces has enabled reconfigurable resonances by changing the geometry of the resonant elements through thermal or electrostatic actuation~\cite{Tao_Averitt_2009_PRL,Zhu_Kwong_2011_AdvMater,Ma_Lee_2014_LSA,HAN_Toshiyoshi_2014_OE,Kan_Shimoyama_2015_NatCommun}.

\subsection{Nonlinear metasurfaces}
\begin{figure}[htp]
\centerline{\includegraphics[width=3.2in]{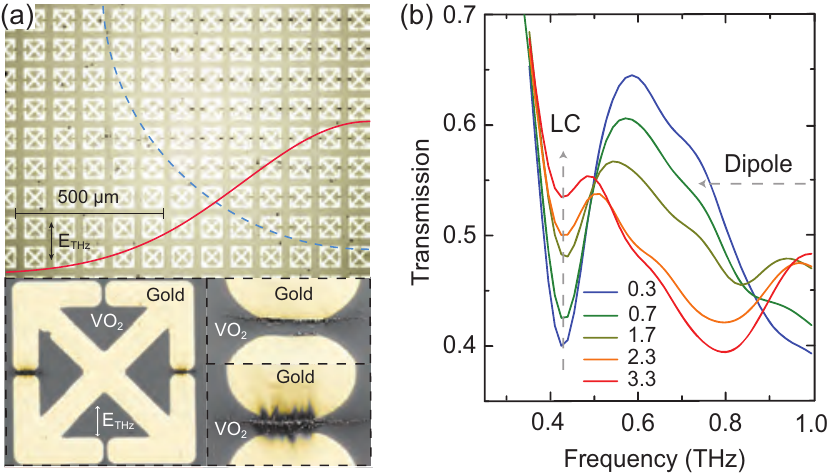}} \caption{(a) Top panel: optical image of an array of gold electric SRRs fabricated on top of a VO$_2$ film showing THz-field-induced damage illustrated by the black spots at the split gaps. The dashed blue circle approximates the THz beam waist, and the red curve approximates the THz intensity profile. Bottom panels: SEM images of a single SRR show that VO$_2$ is damaged by the vertically polarized THz field, with an expanded view of damage at the edge of the THz beam (right top) and near the beam centre (right bottom). (b) Experimental data showing incident field-dependent nonlinear transmission spectra of SRRs on VO$_2$ at 324 K, for in-gap fields ranging from 0.3 to 3.3 MV/cm. Used with permission from~\cite{Liu_Averitt_2012_Nature}. }
\label{Figure_30}
\end{figure}
The abilities of metasurfaces to promote light-matter interaction and manipulate local optical polarizations are ideally suited to enhance nonlinear optical effects. This is particularly significant in the THz frequency range due to the difficulties in generating high-power THz radiation. The concentration of incident THz waves relaxes the requirement of a strong THz source, and further reveals the ultrafast dynamics of electronic responses initiated by the intense THz pulses. An excellent example is the THz-field-induced insulator-to-metal transition in metal/VO$_2$ hybrid metasurfaces, where an array of electric SRRs were fabricated on top of a VO$_2$ film (\fref{Figure_30}(a))~\cite{Liu_Averitt_2012_Nature}. It was shown that the transmission spectra depend on the incident field strength, as shown in \fref{Figure_30}(b), due to the phase transition of VO$_2$ that is initiated by Poole-Frenkel electron liberation, followed by lattice equilibration on a picosecond timescale. The incident few hundreds kV/cm THz field is resonantly enhanced to the MV/cm level within the split gap, which is then sufficient to induce irreversible damage of the VO$_2$ film, as shown in \fref{Figure_30}(a). Such a methodology has been further applied to investigate nonlinear metasurfaces integrated with semiconductors such as gallium arsenide and indium arsenide~\cite{Fan_Averitt_2013_PRL,Seren_Averitt_2015_arXiv}, where the electric field induces intervalley scattering or impact ionization, resulting in a reduced carrier mobility or increased carrier density, thereby either damping or strengthening the metasurface resonant response.  Strong THz nonlinear response was also observed in superconducting metasurfaces under intense THz radiation~\cite{Grady_Chen_2013_NJP,Zhang_Jin_2013_APL}. Although the energy of a THz photon is well below that required to directly break a Cooper pair upon absorption and the applied THz pulses do not significantly raise the sample temperature, the transmission measurements reveal significant field-strength-dependent transmission spectra at various temperatures. It would be expected that the intense THz field can accelerate electrons that gain sufficiently high kinetic energy to induce Cooper pair breaking, which damps the resonance similar to the cases of resonance switching and frequency tuning under thermal and optical excitation.

\begin{figure*}[htp]
\centerline{\includegraphics[width=6in]{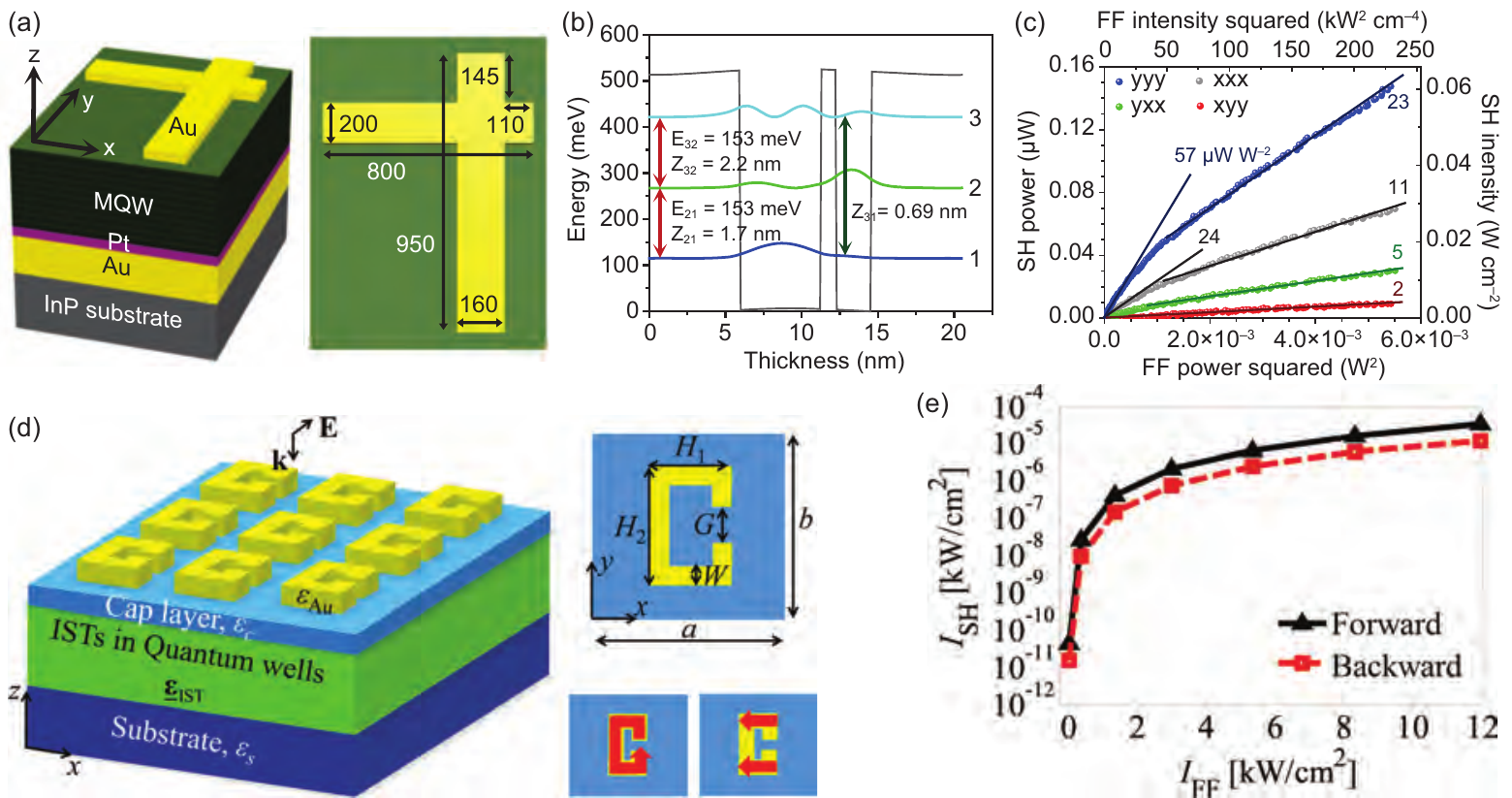}} \caption{(a) Unit cell of the SHG metasurface. Dimensions of the gold nanocross are given in nm, and the unit cell has a dimension of 1000 nm $\times$ 1300 nm. (b) Conduction band diagram of one period of an In$_{0.53}$Ga$_{0.47}$As/Al$_{0.48}$In$_{0.52}$As coupled quantum well structure designed as the nonlinear media for highly efficient SHG. The moduli squared of the electron wavefunctions for subbands 1, 2 and 3 are shown and labelled accordingly. Transitions between pairs of electron subbands are marked with double-headed red arrows, and the values of the transition energies (E$_{21}$ and E$_{32}$) and dipole moments (Z$_{21}$, Z$_{32}$ and Z$_{31}$) are shown next to each arrow. (c) SHG from metasurfaces based on (a) and (b). Shown are SH peak power (left axis) and intensity (right axis) as a function of pump peak power squared (bottom axis) or peak intensity squared (top axis) at a pump wavenumber of 1240 cm$^{-1}$ for different input/output polarization combinations. (d) Left panel: Schematic of a metasurface consisting of SRRs on top of a stack of MQWs. Upper right panel: Top view of one SRR. Lower right panel: Schematic showing the two main resonant modes of the SRR at the pump and SH frequency, respectively. (e) Intensity of the SH signal propagating in the forward and backward directions with respect to the nonlinear metasurface in (d) as a function of pump intensity. (a)-(c) used with permission from~\cite{Lee_Belkin_2014_Nature}, (d) and (e) used with permission from~\cite{Campione_Brener_2014_APL}.}
\label{Figure_31}
\end{figure*}
Conventionally, the phase matching condition in nonlinear processes, such as second harmonic generation (SHG), has to be satisfied in bulk nonlinear crystals to achieve efficient nonlinear optical generation. Under the condition of perfect phase matching, nonlinearly generated optical signals constructively build up, and optical power is continuously transferred from the pump(s) to the nonlinear optical signal. Metasurfaces greatly relax the requirement for phase matching as nonlinear processes occur within metasurfaces that have significantly reduced thicknesses. Giant second-harmonic (SH) response (\fref{Figure_31}(a-c)) has been experimentally demonstrated in plasmonic metasurfaces integrated with nonlinear media~\cite{Lee_Belkin_2014_Nature}. Specifically, InGaAs/AlInAs multiple quantum wells were used as the nonlinear media, which exhibit giant and electrically tunable nonlinear coefficients in the mid-infrared~\cite{Ahn_Chuang_1987_IEEE,Capasso_1994_IEEE,Gmachl_Cho_2003_IEEE}. The plasmonic metasurfaces were designed to not only enhance the local fields of both the pump and SH signal, but also to manipulate the near-field polarization, as the relevant field components involved in the harmonic generation are the ones normal to the quantum wells (due to the selection rules for intersubband transitions within quantum wells~\cite{Helm_2000}). The nonlinear metasurfaces achieved a nonlinear conversion efficiency of $\sim 2 \times10^{-6}$ using a pump intensity of only 15~kW/cm$^2$~\cite{Lee_Belkin_2014_Nature}, corresponding to an effective second-order nonlinear coefficient of $\chi^{(2)} \sim 30$ nm/V, about three orders of magnitude larger than that of LiNbO$_3$. Even larger $\chi^{(2)} \sim 250$ nm/V has been experimentally demonstrated in a nonlinear metasurface consisting of an array of SRRs and InGaAs/AlInAs multiple quantum wells (\fref{Figure_31}(d,e))~\cite{Campione_Brener_2014_APL}. The two plasmonic resonances of the SRRs enhance, respectively, the pump and SH signal (\fref{Figure_31}(d)).

These demonstrations have followed the original proposal of using SRRs by Pendry in 1999, where the resonance would localize electromagnetic energy within the small split-gaps and dramatically enhance nonlinear response in materials being introduced~\cite{Pendry_1999_IEEE}. Most experimental demonstrations of nonlinear metamaterials have been mainly focused in the microwave frequency range using packaged nonlinear electronic elements, such as varactor diodes, to introduce nonlinearity into the gaps of the metal resonators, resulting in nonlinear functionalities such as bistability~\cite{Powell_Kivshar_2007_APL,Wang_Soukoulis_2008_OE}, resonance tunability~\cite{Shadrivov_Kivshar_2006_OE,Huang_Smith_2010_APL}, and harmonic generation~\cite{Shadrivov_Kivshar_2008_APL,Rose_Smith_2011_PRL}. In the optical frequency regime, in addition to the difficulty in packaging nonlinear materials, it is challenging to fabricate bulk metamaterials consisting of complex three-dimensional nanostructures, and utilize their thickness to enhance nonlinear response. Experimental work has been scarce as compared to its microwave counterpart. One of the focuses has been harmonic generation using single layer metal SRR arrays (tens of nanometer thickness) excited at their magnetic resonance~\cite{Klein_Wegener_2006_Science,Klein_Wegener_2007_OE}, where the nonlinearity is associated with the dynamics of free and bound charges, particularly at the metal surface~\cite{Ciraci_Smith_2012_PRB}. This seems to be also responsible for the recently observed broadband THz generation under femtosecond near-infrared laser excitation in an array of gold SRRs~\cite{Luo_Soukoulis_2014_NatCommun}. It was also shown that the geometry of resonators, particularly the asymmetry ratio, plays a critical role as it governs the spatial overlap of the resonant modes at the pump and harmonic frequencies~\cite{OBrien_Zhang_2015_NatMater}. Although the nonlinear coefficients were often enhanced by orders of magnitude compared to common nonlinear crystals, the absolute conversion efficiency is still rather low. Further improvement is non-trivial and cannot be realized by simply stacking multi-layers for larger interaction thickness due to the impedance and propagation phase mismatches. 

\begin{figure*}[htp]
\centerline{\includegraphics[width=6in]{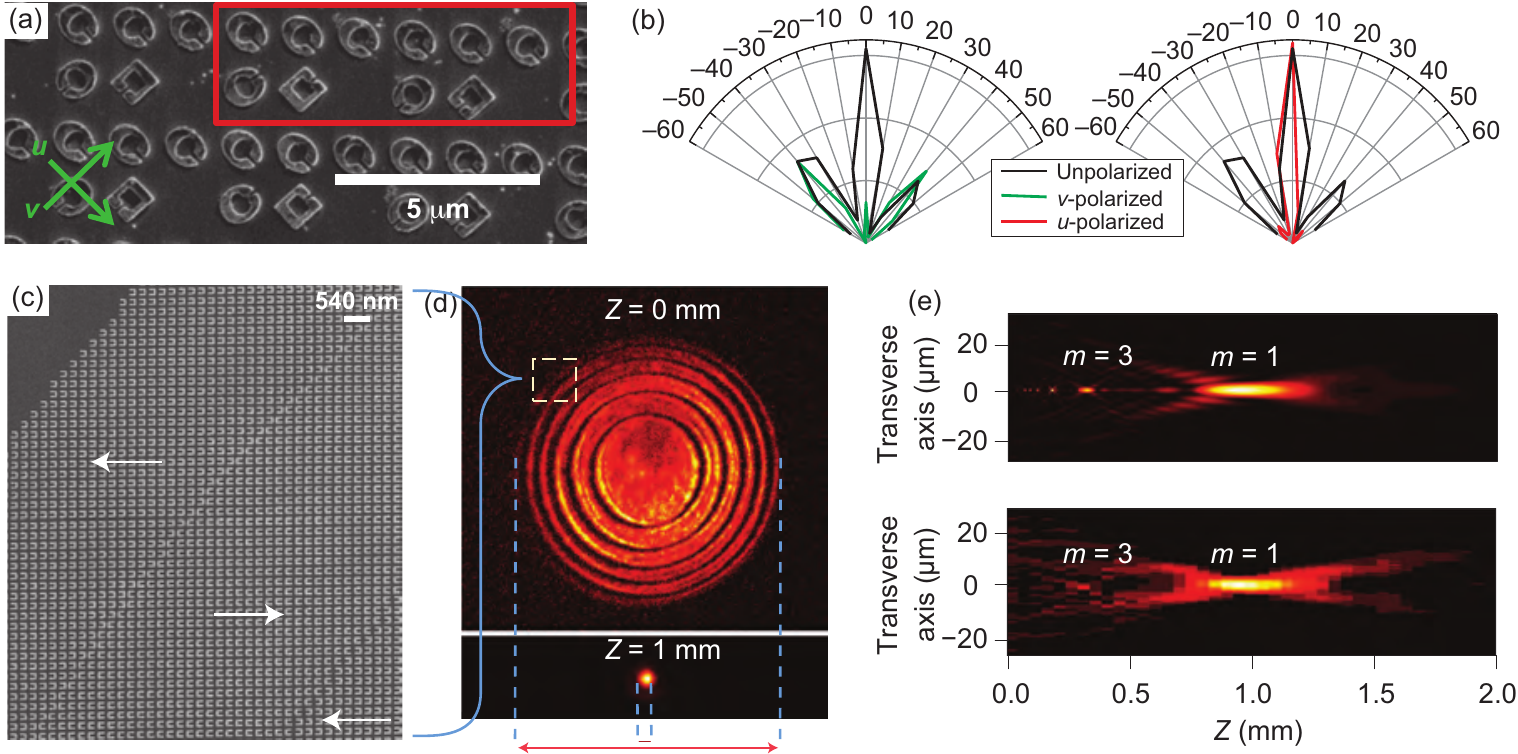}} \caption{(a) SEM image of a portion of a nonlinear metasurface that radiates SH signal of different polarizations into different directions (i.e., polarizing beam splitter for SH signal). A unit cell of the metasurface is denoted by the red rectangle. (b) Measured far-field profiles for the metasurface in (a) for two orthogonal polarizations of the SH radiation when the pump beam is polarized along the vertical direction. (c) SEM image of a portion of a nonlinear metasurface Fresnel zone plate (FZP) showing mirror inversion of SRRs in adjacent zones that radiate SH waves with opposite phases. Arrows mark the effective $\chi^{(2)}$ direction. (d) Recorded images of SH signal at a distance of $Z = 0$ and $Z = 1$ mm from the metasurface in (c). (e) Upper and lower panels are, respectively, simulation and measurement results showing the focusing of SH signal by the FZP ($m$ denotes focusing order). (a) and (b) used with permission from~\cite{Wolf_Brener_2015_NatCommun}, (c)-(e) used with permission from~\cite{Segal_Ellenbogen_2015_NatPhoton}.}
\label{Figure_32}
\end{figure*}
A step further is the demonstration of nonlinear phased arrays that radiate generated SH signals into different directions depending on their polarization states~\cite{Wolf_Brener_2015_NatCommun}. In the metasurface structure shown in \fref{Figure_32}(a), the top row of six identical resonators within the unit cell generate a single u-polarized broadside beam, shown in the right panel of \fref{Figure_32}(b), at the SH frequency. In the bottom row of four resonators within the unit cell, the left two resonators have a $\pi$ phase difference as compared to the right two resonators, as the local effective second-order nonlinear coefficient $\chi^{(2)}$ changes sign when the orientation of a SRR rotates by $180^\circ$. Together they generate two v-polarized beams at $\pm40^\circ$ as shown in the left panel of \fref{Figure_32}(b), where the radiation angles are determined by the period of the metasurface structure. The same effect was used to demonstrate complex wavefront engineering of the SH signal generated from metasurfaces consisting of gold SRRs (\fref{Figure_32}(c-e))~\cite{Segal_Ellenbogen_2015_NatPhoton}. Here the surface second-order nonlinearity of gold leads to SHG.  In experiments a nonlinear Fresnel zone plate (FZP) was demonstrated, which focuses the SH signal to the focal spots of the plate, leading to a large enhancement of the SH intensity.

\section{Summary and outlook}

Metamaterials and metasurfaces have led to the realization of novel electromagnetic properties and functionalities through tailoring subwavelength structures and integrating functional materials. In this paper we have reviewed the recent development of two-dimensional metamaterials -- metasurfaces -- by introducing the fundamental concepts, physical realization, and their promising applications in the control and manipulation of electromagnetic waves at frequencies ranging from microwave to visible light. One of our focuses is on the creation of an arbitrary phase profile for wavefront control and beam forming using both metallic and dielectric metasurfaces. Another focus is on the few-layer metasurfaces that address the efficiency issue encountered during the earlier development of metasurfaces. Active and nonlinear metasurfaces represent an important research direction that will greatly expand metasurface functionalities and applications. As a rapidly developing research field that has attracted world wide interest, it would be impossible (and not necessary) to include every aspect of its past success. For instance, we have not included metasurfaces for antireflection coatings~\cite{Chen_2010_PRL_ARC,Zhang_Guo_2014_APL}, photonic spin Hall effects in metasurfaces~\cite{Yin_Zhang_2013_Science} and ultrathin invisibility cloaks~\cite{Ni_Zhang_2015_Science}.

We see a number of promising areas in fundamental research and practical applications where metasurfaces could have an important impact: 

(1) Dispersionless flat lenses. Flat lenses that are able to correct chromatic aberration over a broad wavelength range, and reduce spherical aberration, coma, and other monochromatic aberrations, could revolutionize optical instrumentation. Substantially shrinking the complexity and size of optical instruments (e.g., replacing the entire set of compound lenses in a camera lens with a few dispersionless and aberration-corrected flat lenses) seems feasible in view of recent developments of metasurface lenses.  

(2) Optical modulators and spatial light modulators (SLMs) in the mid-infrared and THz spectral range. The lack of compact and fast modulators and SLMs has been a big challenge that prevents the wide applications of mid-infrared and THz technology in free-space communications, imaging, LIDAR (light detection and ranging), and homeland security (e.g., remote sensing, surveillance, and navigation in severe environments, such as foggy and dusty weather). Metasurfaces provide an ideal platform to create flat modulators in the mid-infrared and THz regimes as they enable a strong interaction between light and materials with tunable optical properties, and allow for introducing spatially-varying optical response. Strong light-material interactions enabled by metasurfaces allow for reducing the amount of tunable materials used so that one can increase the modulation speed.

(3) Radiative cooling metasurfaces. Metasurfaces that possess exceptional thermoregulatory properties have been an emerging field of research and have the potential to make an important technological impact. Fan and colleagues are pioneering the research on radiative cooling metasurfaces~\cite{Rephaeli_Fan_2013_NL,Zhu_Fan_2014_Optica,Raman_Fan_2014_Nature}, which have strong reflectivity in the solar radiation spectrum and enhanced emissivity in the thermal radiation spectrum. Metasurfaces based on multilayered thin films have demonstrated in experiments passive cooling of objects to a few degrees below the ambient air temperature under direct sunlight~\cite{Raman_Fan_2014_Nature}. Chen and colleagues recently proposed a fabric that blocks sunlight and provides passive cooling via the transmission of thermal radiation emitted by the human body~\cite{Tong_Chen_2015_ACSPhotonics}. It is interesting to note that radiative cooling has always been essential for the survival of animals living in harsh environmental conditions. Yu and co-workers recently reported the thermoregulatory strategies that enable Saharan silver ants to forage in the midday sun on the desert surface where temperatures can reach $70^\circ$C (which is not survivable by their primary predators). It was found that a monolayer of densely packed hairs with triangular cross-sections, in some sense a biological ``metasurface'', enhances not only the ant body's reflectivity in the visible and near-infrared, where solar radiation culminates, but also its emissivity in the mid-infrared~\cite{Shi_Yu_2015_Science}. The combined effect enables the ants to minimize absorption from solar radiation, and to efficiently dissipate heat back to the surroundings via blackbody radiation. Animals and plants living in extreme environments could provide us valuable scientific and engineering lessons on optical design and thermal management. In general, by designing the structural hierarchy, compositional heterogeneity, and local anisotropy of metasurface structures, one could create coatings that are optically thin and have desired spectral properties (reflectivity, absorptivity, transmissivity, and emissivity) over an extremely broad electromagnetic spectral range. Such ultra-thin and ultra-broadband metasurfaces will open doors to a variety of new applications, including control of radiative heat transfer, infrared camouflage and structural coloration.

(4) New material platforms for metasurfaces. Investigations of materials with low losses, tunability, high melting point, and CMOS compatibility for metamaterials and metasurfaces have been very active in recent years. Transition-metal nitrides such as TiN show comparable optical properties as gold in the visible and infrared but have much higher melting points~\cite{Naik_Boltasseva_2011_OME,Boltasseva_Atwater_2011_Science,Guler_Shalaev_2014_Science}, a property that can be explored for metasurface applications involving high optical intensity. Transparent conducting oxides (TCOs) such as indium-tin-oxide enable one to control the spectral location of the epsilon-near-zero point, which is associated with enhanced optical near-fields; the resulting strong interaction between light and TCOs can be exploited for optical modulation~\cite{Yi_Cubukcu_2013_APL,Park_Brongersma_2015_SciRep} and nonlinear optics. Phase-change materials such as chalcogenide alloys that have been used in rewritable CDs, DVDs, and Blu-ray discs, can be switched between the amorphous and crystalline states by laser or electrical current pulses with controlled duration and intensity~\cite{Hudgens_Johnson_2004_MRS,Wuttig_Yamada_2007_NatMater}. This material system has recently been used to demonstrate all-optical, non-volatile, metasurface switch~\cite{Gholipour_Zheludev_2013_AdvMater}, and high-resolution solid-state displays~\cite{Hosseini_Bhaskaran_2014_Nature}. SmNiO$_3$, a prototypical phase-transition perovskite nickelate, exhibits non-volatile and reversible large refractive index changes over an ultra-broad spectral range, from the visible to the long-wavelength mid-infrared. The super broadband performance is due to strong electron correlation effects~\cite{Shi_Ramanathan_2014_NatCommun}, and this new mechanism can be exploited to create a variety of active photonic devices.

\ack
H.T.C acknowledges support in part from the Los Alamos National Laboratory LDRD Program. N. Y. acknowledges support from NSF (grant ECCS-1307948), the AFOSR Multidisciplinary University Research Initiative program (grant FA9550-14-1-0389), and DARPA Young Faculty Award (grant D15AP00111). This work was performed, in part, at the Center for Integrated Nanotechnologies, a U.S. Department of Energy, Office of Basic Energy Sciences Nanoscale Science Research Center operated jointly by Los Alamos and Sandia National Laboratories. Los Alamos National Laboratory, an affirmative action/equal opportunity employer, is operated by Los Alamos National Security, LLC, for the National Nuclear Security Administration of the U.S. Department of Energy under Contract No. DE-AC52-06NA25396.

\section*{References}

\bibliography{ChenLANL}

\end{document}